\documentclass[twocolumn,prd,aps,longbibliography,superscriptaddress,preprintnumbers,tightenlines,showpacs,nofootinbib,amsfonts,amsmath]{revtex4-2}

\pdfoutput=1

\usepackage{epsfig}
\usepackage{graphics}
\usepackage{graphicx}
\usepackage{amsmath,amssymb,mathrsfs}
\usepackage{amsfonts}
\usepackage[usenames,dvipsnames]{xcolor}
\usepackage{xcolor}
\usepackage{wasysym}
\usepackage{times}
\usepackage{newtxmath}
\usepackage{gensymb}
\usepackage{appendix}
\usepackage{listings}
\usepackage{url}
\usepackage[normalem]{ulem}
\usepackage{alltt}
\usepackage[colorlinks=true,
urlcolor=blue]{hyperref}
\usepackage{cleveref}
\usepackage{longtable}
\usepackage{enumitem}
\setlist{nosep}
\usepackage{color}
\usepackage{calc}
\usepackage{tensor}
\usepackage{bm}
\usepackage{times}
\usepackage{multirow}
\usepackage{float}
\usepackage{dcolumn}
\usepackage[nolist,nohyperlinks]{acronym}
\usepackage{xspace}
\usepackage[english]{babel}
\usepackage[abs]{overpic}
\usepackage{pict2e}
\allowdisplaybreaks[1]
\usepackage[utf8]{inputenc}
\usepackage{gensymb}
\usepackage{bm}
\usepackage{stackengine}
\usepackage{boldline,multirow}
\usepackage{braket}
\usepackage{longtable}
\usepackage{orcidlink}

\usepackage{tabularx}
\usepackage{rotating}

\Crefname{figure}{Fig.}{Figs.}

\def\mR{\mathcal{R}}
\def\mP{\mathcal{P}}
\def\mN{\mathcal{N}}
\DeclareMathOperator{\Order}{\mathcal{O}}

\newcommand{\eg}{e.\,g.\@}

\newcommand{\ibar}{{\bar{\imath}}}
\newcommand{\jbar}{{\bar{\jmath}}}
\newcommand{\Lbar}{{\bar{L}}}
\newcommand{\Kbar}{{\bar{K}}}

\definecolor{dodgerblue}{HTML}{1E90FF}
\definecolor{viennared}{HTML}{DA0A14}

\hypersetup{citecolor=dodgerblue}

\begin{document}

\title{
\hspace*{-0.4cm}Worldtube excision method for intermediate-mass-ratio inspirals: scalar-field model in 3+1 dimensions \hspace*{-0.4cm}
}
\newcommand{\cornell}{\affiliation{Cornell Center for Astrophysics and Planetary
		Science, Cornell University, Ithaca, New York 14853, USA}}
\newcommand{\caltech}{\affiliation{Theoretical Astrophysics, Walter Burke
		Institute for Theoretical Physics, California Institute of Technology,
		Pasadena, California 91125, USA}}
\newcommand{\aei}{\affiliation{Max Planck Institute for Gravitational Physics
		(Albert Einstein Institute), Am M{\"u}hlenberg 1, 14476 Potsdam, Germany}}
\newcommand{\fullerton}{\affiliation{Nicholas and Lee Begovich Center for
		Gravitational-Wave Physics and Astronomy, California State University
		Fullerton, Fullerton, California 92831, USA}}
\newcommand{\southampton}{\affiliation{School of Mathematical Sciences and STAG Research Centre, University of Southampton, Southampton, SO17 1BJ, United Kingdom}}
\newcommand{\tata}{\affiliation{International Centre for Theoretical Sciences, Tata Institute of Fundamental Research, Bangalore 560089, India}}
\newcommand{\coimbra}{\affiliation{CFisUC, Department of Physics, University of Coimbra, 3004-516 Coimbra, Portugal}}

\author{Nikolas A. Wittek \orcidlink{0000-0001-8575-5450}} \aei
\author{Mekhi Dhesi \orcidlink{0000-0003-0017-4302}} \southampton
\author{Leor Barack \orcidlink{0000-0003-4742-9413}} \southampton
\author{Harald P. Pfeiffer \orcidlink{0000-0001-9288-519X}} \aei
\author{Adam Pound \orcidlink{0000-0001-9446-0638}} \southampton
\author{Hannes R. Rüter \orcidlink{0000-0002-3442-5360}} \coimbra
\author{Marceline S. Bonilla \orcidlink{0000-0003-4502-528X}} \fullerton %
\author{Nils Deppe \orcidlink{0000-0003-4557-4115}} \caltech  %
\author{Lawrence E.~Kidder \orcidlink{0000-0001-5392-7342}} \cornell  %
\author{Prayush Kumar \orcidlink{0000-0001-5523-4603}}
\tata  %
\author{Mark A. Scheel\,\orcidlink{0000-0001-6656-9134}} \caltech  %
\author{William Throwe \orcidlink{0000-0001-5059-4378}} \cornell  %
\author{Nils L.~Vu \orcidlink{0000-0002-5767-3949}} \caltech \aei  %
\date{\today}

\begin{abstract}
	Binary black hole simulations become increasingly more computationally expensive with smaller mass ratios, partly because of the longer evolution time, and partly because the lengthscale disparity dictates smaller time steps. The program initiated by Dhesi {\it et al.} [Phys.~Rev~D {\bf 104}, 124002 (2021)] explores a method for alleviating the scale disparity in simulations with mass ratios in the intermediate astrophysical range ($10^{-4}\lesssim q \lesssim 10^{-2}$), where purely perturbative methods may not be adequate. A region (``worldtube'') much larger than the small black hole is excised from the numerical domain, and replaced with an analytical model approximating a tidally deformed black hole. Here we apply this idea to a toy model of a scalar charge in a fixed circular geodesic orbit around a Schwarzschild black hole, solving for the massless Klein-Gordon field. This is a first implementation of the worldtube excision method in full 3+1 dimensions. We demonstrate the accuracy and efficiency of the method, and discuss the steps towards applying it for evolving orbits and, ultimately, in the binary black-hole scenario. Our implementation is publicly accessible in the SpECTRE numerical relativity code.
\end{abstract}

\maketitle

\acrodef{NR}{numerical relativity}
\acrodef{GW}{gravitational wave}
\acrodef{BBH}{binary black hole}
\acrodef{BH}{black hole}

\newcommand{\NR}[0]{\ac{NR}\xspace}
\newcommand{\BBH}[0]{\ac{BBH}\xspace}
\newcommand{\BH}[0]{\ac{BH}\xspace}

\newcommand{\citeme}[0]{{\color{purple}{Citation!}}}

\section{Introduction} \label{sec:Intro}

Inspiraling binary black holes (BBHs) are the most numerous source of
gravitational wave signals detected by the LIGO and Virgo observatories
\cite{%
  LIGOScientific:2016dsl,%
  LIGOScientific:2018mvr,%
  LIGOScientific:2020ibl,%
  LIGOScientific:2021djp}.  The mass ratio is one of the most
important characteristics of these binaries, and observations so far
\cite{%
  LIGOScientific:2018jsj,%
  Venumadhav:2019lyq,%
  LIGOScientific:2020ibl,%
  Nitz:2021zwj,%
  LIGOScientific:2021psn} predominantly find mass ratios close
to unity.  However, GW190814 and GW200210\_092254 have mass ratios
$q\equiv m_2/m_1\sim
0.11$~\cite{LIGOScientific:2020zkf,LIGOScientific:2021djp}, and
GW191219\_163120---where the secondary's mass suggests it is a
neutron star---is estimated to have $q\sim
0.04$~\cite{LIGOScientific:2021djp}.

It is likely that upcoming observing runs by ground-based detectors
will continue to record binaries with small mass-ratios.  Future
ground-based detectors like the Einstein
Telescope~\cite{Maggiore:2019uih} and Cosmic
Explorer~\cite{Evans:2021gyd}, featuring an improved low-frequency sensitivity,
will be able to detect the capture of stellar-mass black holes (BHs) by 
intermediate-mass BHs, with mass-ratios down to $q\sim 10^{-3}$~\cite{Jani:2019ffg}.
Moreover, space-borne detectors, like the LISA
observatory~\cite{eLISA:2013xep,LISA:2017pwj}, will be sensitive to
binaries with mass ratios in the entire range from $q\sim 1$ to extreme mass-ratio inspirals
with $q\sim 10^{-5}$~\cite{Jani:2019ffg,Katz:2019qlu,Gair:2010bx,Volonteri:2020wkx}.

In anticipation of this remarkable expansion in observational reach,
it is important to develop accurate theoretical waveform templates
that reliably cover the entire relevant range of mass ratios. Standard
Numerical Relativity (NR) methods~\cite{BauSha10} work well for mass
ratios in the range $ 0.1 \lesssim q \leq 1$ (see
\eg{}~\cite{Boyle:2019kee}).  However, simulations become
progressively less tractable at smaller $q$, and few numerical
simulations have been performed at $q < 0.1$ so far. The root cause is
a problematic scaling of the required simulation time with
$q$. Fundamentally, one expects the required simulation time to grow
in proportion to $q^{-2}$, where one factor of $q^{-1}$ is associated
with the number of in-band orbital cycles, and the second factor
$q^{-1}$ comes from the Courant-Friedrich-Lewy (CFL) stability limit
on the time step of the numerical simulation, arising from the
requirement to resolve the smaller black hole.  The state of the art
in small-$q$ NR is represented by the recent simulations performed at
RIT of the last 13 orbital cycles prior to merger of a black-hole
binary system with $q=1/128$~\cite{Lousto:2020tnb, Rosato:2021jsq}.
Head-on simulations, where the needed evolution time is orders of
magnitudes shorter than for inspirals, are possible at even smaller
mass-ratios~\cite{Sperhake:2011ik,Lousto:2022}. While these
simulations represent an important proof of concept, their
computational cost is extremely high, and it is presently impossible
to explore the full parameter space including spin and eccentricity.

Binaries with extreme mass-ratios, say $q\lesssim 10^{-4}$,
corresponding to a compact object orbiting a massive black hole in a
galactic nucleus, can be modeled with a perturbative expansion in $q$.
This ``gravitational self-force'' (GSF) approach~\cite{Barack:2018yvs,
  Pound:2021qin} incorporates order-by-order in $q$ the small
deviations of the motion of the small body away from the geodesic
motion that applies for test-bodies.  The GSF approach is the only
method for modeling extreme-mass-ratio inspirals, and development is
ongoing towards waveform models suitable for signal identification and
interpretation with
LISA~\cite{vandeMeent:2017bcc,Chua:2020stf,Hughes:2021exa,Pound:2019lzj,Warburton:2021kwk,Wardell:2021fyy}.
With NR being well-suited to comparable masses and the GSF approach to
extreme mass-ratios, the question arises of how to model the
intermediate mass-ratio regime.  For simple binary systems (of nonspinning
black holes in quasi-circular or eccentric inspirals) NR simulations
suggest~\cite{vandeMeent:2020xgc,Ramos-Buades:2022lgf}
that GSF calculations may be sufficiently accurate even at mass-ratios
reaching the NR regime.  ``Post-adiabatic'' GSF
waveforms~\cite{Wardell:2021fyy} for non-spinning, quasi-circular binaries have shown those predictions were somewhat over-optimistic in the $q>0.1$ range~\cite{Albertini:2022rfe}, but they have borne out the prediction for smaller mass ratios $\lesssim 0.1$.  However, it remains unclear
whether the two methods, separately applied, can achieve reliable
waveform models of intermediate-mass-ratio inspirals over the full
astrophysically relevant parameter space.

In this paper we continue the work of \cite{Dhesi:2021yje} to develop
a new approach to the simulation of intermediate-mass-ratio systems,
combining NR techniques with black hole perturbation theory. The
general idea is to excise a large region around the smaller black
hole.  Inside this region—a “worldtube” in spacetime—an approximate
analytical solution is prescribed for the spacetime metric, arising
from the perturbation theory of compact objects in a tidal
environment. An NR simulation is set up for the binary, in which the
worldtube’s interior is excised from the numerical domain, and
replaced with the analytical solution. At each time step of the
numerical evolution, the numerical solution (outside the worldtube)
and analytical solution inside are matched across the worldtube’s
boundary, in a process that fixes a priori unknown tidal coefficients
in the analytical solution, gauge degrees of freedom, and also
provides boundary conditions to the NR evolution.  The intended effect
of this construction is to partially alleviate the scale disparity
that thwarts the efficiency of the numerical evolution at small $q$:
The smallest length scales on the numerical domain is now that of the
worldtube-radius $R$, rather than the scale $m_2$ of the smaller
body. As a result, the CFL limit is expected to increase by a factor
$R/m_2\gg 1$, with a comparable gain in computational efficiency.

In Ref.~\cite{Dhesi:2021yje}, as also in the present work, we 
consider a linear scalar-field toy model where the small black hole is
replaced with a pointlike scalar charge moving on a circular geodesic
around a Schwarzschild black hole.  Instead of tackling the full
Einstein’s equations, one solves the less complicated massless linear
Klein-Gordon equation for a scalar field.  Our previous work
\cite{Dhesi:2021yje} decomposed the scalar field into spherical harmonics
and solved the resulting 1+1-dimensional (1+1D) partial differential equation for each
mode separately.  Such a modal decomposition will not be
possible in the fully nonlinear BBH case.  As a step towards 
the BBH case, in this paper, we derive and implement a generalized matching scheme in full 3+1D.  
Our implementation is publicly accessible as part of the SpECTRE platform \cite{spectrecode},
a new general-relativistic code developed by
the SXS collaboration, which employs a nodal discontinuous Galerkin method 
with task-based parallelism. The input file for the simulations presented in this paper is given as supplemental material. (An evolution of the  scalar field equation in 3+1D with a point source was performed in \cite{Vega:2009} using a different method.)

The paper is organized as follows. In Section~\ref{sec: Evolution} we
describe our scalar-field model, and formulate it as an
initial-boundary evolution problem suitable for implementation on
SpECTRE.  Section~\ref{sec: model} describes the construction of the
approximate analytical solution inside the worldtube. In
Section~\ref{sec: matching method n=2}, we show how the unknown
parameters of this local solution can be continuously determined from
the evolution data on the worldtube boundary, using a set of ordinary differential equations (in
time) derived from the Klein-Gordon equations. The fully
specified solution inside the worldtube is then used to formulate
boundary conditions for the evolution system. We present the results
of our simulations in Section~\ref{sec: Results}, and demonstrate a
good agreement with both analytical solutions in limiting cases, and
numerical results from other simulations. We explore the convergence
of our numerical solutions with worldtube size, and show that its rate
matches our theoretical expectations. Finally, in
Section~\ref{sec:Conclusions}, we summarize our findings and discuss
the next steps in our program.  We use geometrized units throughout
the text with $G = c = 1$.

\section{Numerical field evolution outside the worldtube}\label{sec: Evolution}

We place a pointlike particle with scalar charge $q$ on a fixed, geodesic circular orbit around a Schwarzschild black hole of mass $M$. The evolution of the scalar field $\Psi$ is governed by the massless Klein-Gordon equation,
\begin{equation}\label{eq: KG with source}
	g^{\mu \nu} \nabla_\mu \nabla_\nu \Psi = -4 \pi q \int \frac{\delta^4 ( x^{\alpha} - x^{\alpha}_p (\tau))}{\sqrt{-g}} d\tau.
\end{equation}
Here $g^{\mu \nu}$ is the inverse Schwarzschild metric, and $\nabla_\mu$ is the covariant derivative compatible with it. $x^{\alpha}_p(\tau)$ is the particle's geodesic worldline parameterized in terms of proper time $\tau$. In Kerr-Schild coordinates $x^\alpha = (t,x^i)$, parametrized by coordinate time $t$, the worldline with orbital radius $r_p$ and angular velocity $\omega=(M/r_p^3)^{1/2}$ is given by
\begin{equation}
	x_p^\alpha(t) = \Big(t, \, r_p \cos(\omega t),\, r_p \sin(\omega t), \, 0\Big),
\end{equation}
where we have fixed the orbital plane and phase without loss of generality. 

   We excise the interior of a sphere with constant Kerr-Schild radius $R = \sqrt{\delta_{ij}(x^i - x_p^i) (x^j - x_p^j)}$, centered on the particle's position, from the numerical domain. We refer to this excision region as the worldtube and elaborate in Sec.~\ref{sec: matching method n=2} how boundary conditions are provided to the evolution domain.

Outside the worldtube, the numerical evolution of the scalar-field variable $\Psi^{\mN}$ (`$\mN$' for `numerical', to contrast with the analytical solution inside the worldtube, to be introduced below) is governed by the source-free Klein-Gordon equation on the fixed background spacetime:
\begin{equation}\label{eq: KG without source}
	g^{\mu \nu} \nabla_\mu \nabla_\nu \Psi^{\mN} = 0.
\end{equation}
The background spacetime is given in the usual 3+1 split,
\begin{equation}
	ds^2 = -\alpha^2 dt^2 + \gamma_{ij}(dx^i + \beta^i dt)(dx^j + \beta^j dt),
\end{equation}
where $\alpha$ is the lapse, $\beta^i$ is the shift and $\gamma_{ij}$ is the spatial metric on $t =\text{const.}$ hypersurfaces. The background spacetime of our simulations is a single Schwarzschild black hole in Kerr-Schild coordinates. 

The Klein-Gordon equation is transformed into the standard first-order form by introducing the auxiliary variables~\cite{Scheel:2004}
\begin{subequations}\label{eq: evolution reduction}
	\begin{align}
	\Pi &= - \alpha^{-1} (\partial_t \Psi^{\mN} - \beta^i \partial_i \Psi^{\mN}), \\
	\Phi_i &= \partial_i \Psi^{\mN}.
	\end{align}
\end{subequations}
This introduces two constraint fields \cite{Holst:2004},
\begin{align}
\label{eq:NR-C1}
	C_i &= \partial_i \Psi^{\mN} - \Phi_i, \\
\label{eq:NR-C2}
	C_{ij} &= \partial_{i}\Phi_{j}-\partial_{j}\Phi_{i},
\end{align}
which must vanish for any solution to the original, second-order evolution equation.
Following \cite{Lindblom:2006}, we write the first-order evolution equations for the vacuum Klein-Gordon equation \eqref{eq: KG without source} as
\begin{subequations}
	\label{eq: DG evolution equations}
	\begin{align}
	  \partial_t \Psi^{\mN} - (1 + \gamma_1)  \beta^i \partial_i \Psi^{\mN}
    &=
    -\alpha \Pi - \gamma_1 \beta^i \Phi_i, \label{eq: dt Psi} \\
	  \partial_t \Pi
    - \beta^k \partial_k \Pi + \alpha \gamma^{ik} \partial_i \Phi_k - \gamma_1 \gamma_2 \beta^i \partial_i \Psi^{\mN}
    &= \nonumber \\
    \alpha K \Pi + \alpha \Gamma^i \Phi_i - \gamma^{ij}\Phi_i \partial_{\!j}& \alpha  -\gamma_1 \gamma_2 \beta^i \Phi_i,
     \label{eq: dt Pi} \\
	  \partial_t \Phi_i - \beta^k \partial_k \Phi_i + \alpha \partial_i \Pi - \gamma_2 \alpha \partial_i \Psi^{\mN}
    &=\nonumber\\
    - \Pi \partial_i \alpha +\Phi_{\!j} &\partial_i \beta^j - \gamma_2 \alpha \Phi_i. 
	\end{align}
\end{subequations}
The lapse $\alpha$, shift $\beta^i$, spatial metric $\gamma_{ij}$, inverse spatial metric $\gamma^{ij}$, trace of the extrinsic curvature $K := \gamma^{ij} K_{ij}$ and trace of the spatial Christoffel symbol $\Gamma^i := \gamma^{jk}\Gamma^i_{jk}$ appearing in Eqs.~(\ref{eq: DG evolution equations}) depend only on the background Schwarzschild spacetime. Explicitly, they read as follows in Kerr-Schild coordinates:
\begin{subequations}
\begin{align}
	\alpha &= \left(1+\frac{2M}{r}\right)^{-1/2}, \\
	\beta^i &= \frac{2M \alpha^2}{r^2}x^i, \\
	\gamma_{ij} &= \delta_{ij}+\frac{2M}{r^3} x^k x^l \delta_{ik} \delta_{jl}, \\
	\gamma^{ij} &= \delta^{ij}-\frac{2M \alpha^2}{r^3} x^i x^j, \\
	K &= \frac{2 M \alpha^3}{r^2} \left( 1 + \frac{3M}{r} \right), \\
	\Gamma^i &= \frac{8M^2 + 3Mr}{\left(2Mr+3r^2\right)^2} x^i,
\end{align}
\end{subequations}
where $r = \sqrt{\delta_{ij} x^i x^k}$ is the areal radius from the central black hole. The variables $\gamma_1$ and $\gamma_2$ appearing in Eqs.~\eqref{eq: DG evolution equations} are constraint damping parameters.  Compared to the first-order reduction presented in \cite{Holst:2004}, the additional term $\gamma_1 \gamma_2 \beta^i C_i$ in Eq.~\eqref{eq: dt Pi} ensures that the system is symmetric hyberbolic for any values of $\gamma_1$ and $\gamma_2$ \cite{Lindblom:2006}.
For $\gamma_2$, we found that a central Gaussian profile $\gamma_2 = A e^{-(\sigma r)^2 } + c$ with $A = 10$, $\sigma = 10^{-1}/M$ and $c = 10^{-4}$ results in a long-term stable evolution for all tested systems. We choose $\gamma_1=0$ throughout.

The evolution equations \eqref{eq: DG evolution equations} are in the general symmetric hyperbolic form
\begin{equation}
	\partial_t \psi^a + A^{i a}{}_{b} \partial_i \psi^b = F^a ,
\end{equation}
with $\psi^a := (\Psi, \Pi, \Phi_i)$ representing the set of first-order variables, enumerated by the indices $a$ and $b$. For the imposition of boundary conditions at a boundary with normal co-vector $\hat n_i$, we solve the (left) eigenvalue problem
\begin{equation}\label{eq:eigenproblem}
	e^{\hat{a}}{}_a \;\hat{n}_i A^{i a}{}_{b} = v_{(\hat{a})} e^{\hat{a}}{}_b 
\end{equation}
for the eigenvalues $v_{(\hat a)}$ and eigenvectors $e^{\hat a}{}_b$, enumerated by the index $\hat a$.
The $v_{(\hat a)}$ are known as the characteristic speeds, and the parentheses indicate that there is no implicit sum convention on the right hand side of Eq.~(\ref{eq:eigenproblem}).  The co-vector $\hat n_i$ is normalized with respect to the three metric, i.e. $\gamma^{ij}\hat n_i\hat n_j=1$, and we define $\hat n^i=\gamma^{ij}\hat n_j$. The characteristic fields $\psi^{\hat{a}}$ are obtained by projecting the evolved variables $\psi^a$ onto the set of eigenvectors $e^{\hat{a}}{}_a$:
\begin{equation}
	\psi^{\hat{a}} =  e^{\hat{a}}{}_{a}\, \psi^a.
\end{equation}
For the evolution system~\eqref{eq: DG evolution equations}, the characteristic fields are $\psi^{\hat a}=(Z^1, Z^2_i, U^+, U^-)$ with
\begin{subequations}\label{eq: char fields}
	\begin{align}
		Z^1 &= \Psi^{\mN}, \\
		Z^2_i &= P^k_i \Phi_k,\\ %
		U^\pm &= \Pi \pm \hat{n}^i \Phi_i - \gamma_2 \Psi^{\mN}.\label{eq: U minus} 
	\end{align}
\end{subequations}

Here, $P^k_i=\delta^k_i - \hat n^k\hat n_i$ denotes the projection operator orthogonal to $\hat n^i$, so that $Z^2_i$ carries only two degrees of freedom.
The corresponding characteristic speeds are $v_{Z^1} = - \hat{n}_i \beta^i(1 + \gamma_1)$ , $v_{Z^2} = - \hat{n}_i \beta^i$ and $v_{U^\pm} = - \hat{n}_i \beta^i \pm \alpha$. We note that the fields $U^\pm$ reduce to the known physical retarded/advanced derivatives $\partial_t \Psi \pm \partial_r \Psi$ in flat space with $\gamma_2 = 0$. The other characteristic fields result from the reduction of the PDE system to first order.

Boundary conditions must be specified at the external boundaries of the domain for each characteristic field, if and only if it is flowing into the domain, specifically those with negative characteristic speeds. There are three external boundaries in our domain: one excision sphere within the central black hole, one excision sphere around the scalar charge (the surface of the worldtube), and the outer boundary.

At the black hole excision sphere, all characteristic fields are flowing out of the computational domain into the excised domain, so no boundary conditions need to be applied. For the outer boundary and at the worldtube boundary, the fields $Z^1$, $Z^2_i$ may require boundary conditions, while $U^-$ always requires ones and $U^+$ never requires ones.

Boundary conditions for the physical characteristic field $U^-$ at the outer boundary are derived from the second-order Bayliss-Turkel radiation condition \cite{Bayliss:1980}. These boundary conditions are applied with the method of Bjorhus~\cite{Bjorhus:1995}.
At the worldtube boundary, the local solution inside the worldtube is used to provide boundary conditions for $U^-$, as explained in detail in Section~\ref{sec: matching method n=2}.

Boundary conditions for $Z^1$ and $Z^2_i$ can be derived by requiring that there are no constraint violations flowing into the domain \cite{Kidder:2005}, as described in Appendix \ref{sec: constraint preserving BCs}. These constraint-preserving boundary conditions are applied with the method of Bjorhus \cite{Bjorhus:1995} at the worldtube boundary and at the outer boundary; see Eq.~(\ref{eq: Bjorhus}).

\begin{figure}
	\includegraphics[width=\columnwidth]{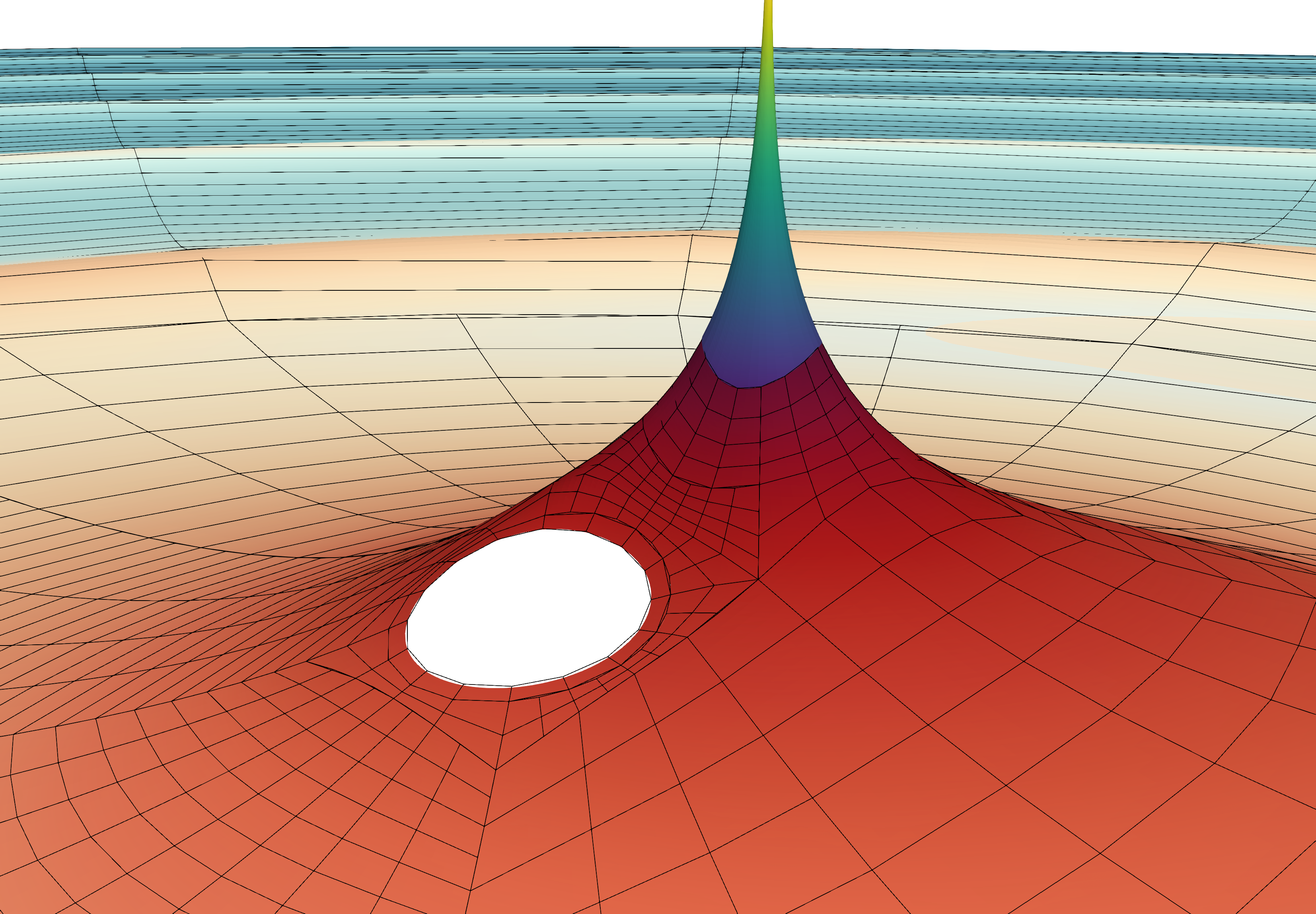}
	\caption{Illustration of the computational domain:  Shown is the equatorial plane, with height-deformation proportional to the value of the scalar field.  The grid lines correspond to the DG-element boundaries of the 3-D numerical evolution. The central blue/green peak represents the region inside the worldtube, where the approximate solution is dominated by the singularity of the scalar field at the point-charge. Left of the peak an excision region is cut out within the horizon of the central black hole. A zoomed-out view of the entire domain is shown in Figure~\ref{fig: top down}.}
	\label{fig: fancy plot}
\end{figure}

The evolution equations~(\ref{eq: DG evolution equations}) are solved with SpECTRE \cite{spectrecode}, which employs a nodal discontinuous Galerkin (DG) scheme in 3+1 dimensions. The domain is built up of several hundred DG elements, each endowed with a tensor product of Legendre polynomials using Gauss-Lobatto quadrature. The elements are deformed from unit cubes to fit the domain structure using a series of smooth maps as illustrated in Fig.~\ref{fig: fancy plot}. Discontinuous Galerkin methods require a choice of numerical flux that dictates how fields are evolved on element boundaries where they are multiply defined \cite{Hesthaven:2007}. Here we employ an upwind flux.

SpECTRE uses dual coordinate frames \cite{Scheel:2006} to solve the evolution equations. The components of the tensors in the evolution Eqs.~\eqref{eq: DG evolution equations} are constructed in Kerr-Schild coordinates $x^i$. We refer to these as the \textit{inertial} frame because the coordinates are not rotating with respect to the asymptotic frame at spatial infinity. The evolution equations for the inertial components are solved as functions of co-rotating coordinates $(\bar{t}, x^{\bar{\imath}}) = (\bar{t}, \bar{x}, \bar{y}, \bar{z})$ given by the transformation
\begin{subequations}
	\label{eq: grid coordinates}
	\begin{align}
		\bar{t} &= t, \\
		\bar{x} &= \;\; x \cos(\omega t) + y \sin(\omega t), \\
		\bar{y} &= - x \sin(\omega t) + y \cos(\omega t), \\
		\bar{z} &= z.
	\end{align}
\end{subequations}
Tensor components in this frame we denote with a bar, as in $g_{\bar \alpha \bar \beta}$.
For more demanding situations (e.g. binary black hole simulations), the transformation $x^i\to x^{\bar\imath}$ can take a
  much more complicated form~\cite{Hemberger:2012jz,Scheel:2014ina}.
The grid points of the DG domain, as well as the particle position $x_p^{\bar \imath}=(r_p,0,0)$ are constant in space in these coordinates, which we will refer to as \textit{grid} coordinates. The internal worldtube solution is evolved in the grid frame directly, which considerably simplifies the formulation of the matching scheme in Sec.~\ref{sec: matching method n=2}.

A Dormand-Prince time stepper is used to advance the solution of the numerical fields with a global time step. We apply a weak exponential filter to the evolution fields after every time step to ensure stability of the evolution.

The code is parallelized using the heterogeneous task-based parallelism framework Charm++ \cite{charm}. The inclusion of the worldtube does not adversely impact the parallel efficiency, as its computational cost is negligible compared to even a single DG element evaluation, and no additional communication between cores is introduced.

\section{Approximate solution inside the worldtube} \label{sec: model}

Inside the worldtube, the scalar field is given by an analytical expansion in powers of coordinate distance from the particle's worldline $x_p$. We use $\Psi^\mathcal{A}$ to denote this analytical solution, and we use a formal parameter $\epsilon=1$ to count powers of the separation between the worldline and the field point.

As in Ref.~\cite{Dhesi:2021yje}, we split the field $\Psi^\mathcal{A}$ into a puncture field $\Psi^\mathcal{P}$ and a regular field $\Psi^\mathcal{R}$:
\begin{equation}\label{psiA definition}
\Psi^\mathcal{A} = \Psi^\mathcal{P} + \Psi^\mathcal{R}.
\end{equation}
$\Psi^\mathcal{P}$ is an approximate particular solution to the inhomogeneous equation~\eqref{eq: KG with source}, and it will be fully determined in advance; $\Psi^\mathcal{R}$ is an approximate smooth solution to the homogeneous equation, and it will be determined dynamically through matching $\Psi^\mathcal{A}$ to $\Psi^\mathcal{N}$ at the worldtube boundary.%

We express both $\Psi^\mathcal{P}$ and $\Psi^\mathcal{R}$ in terms of the coordinate distance $\Delta x^\alpha := x^\alpha-\tilde x^{\alpha}$, where $\tilde x^\alpha$ is a reference point on $x_p$. For a given field point $x^\alpha$ at coordinate time $t$, we let $\tilde x^{\alpha}:=x_p^\alpha(t)$ be the point on $x_p$ at the same value of $t$, such that $\Delta t=0$. Tensors evaluated at $\tilde x^{\alpha}$ are written with a tilde, as in $\tilde g_{\mu \nu}$. To facilitate matching $\Psi^\mathcal{A}$ to $\Psi^\mathcal{N}$, we ultimately express both $\Psi^\mathcal{P}$ and $\Psi^\mathcal{R}$ in the co-rotating grid coordinates $(t,x^{\ibar})$ introduced in Eq.~\eqref{eq: grid coordinates}, but most of this section applies in both inertial and co-rotating coordinates. 

Unlike in Ref.~\cite{Dhesi:2021yje}, for $\Psi^\mathcal{P}$ we use an approximation to the Detweiler-Whiting singular field~\cite{Detweiler:2002mi}; this choice ensures that we can calculate the scalar self-force directly from the regular field $\Psi^\mathcal{R}$. Covariant expansions of the Detweiler-Whiting singular field are readily available to high order in $\epsilon$; see \cite{Heffernan:2012su, Haas:2006ne, Wardell:2011gb}, for example, with \cite{Heffernan:2012su} deriving the scalar singular field to the highest order in the literature, $\mathcal{O}(\epsilon^4)$. 
These covariant expressions contain several ingredients. First among them is Synge's world function $\sigma(x,\tilde{x})$~\cite{J.L.Synge:1960zz}, which is equal to half the squared geodesic distance between $x$ and $\tilde{x}$. Its gradient, $\tilde \sigma_{\alpha}:=\tilde \nabla_{\!\alpha}\sigma(x,\tilde{x})$, is a directed measure of distance from $\tilde x$ to $x$. The projection of $\tilde \sigma_{\alpha}$ tangent to the worldline has magnitude
\begin{equation}
\varrho:=\tilde \sigma_{\alpha}\tilde u^{\alpha},
\label{eq:rbar}
\end{equation}
and the projection normal to the worldline has magnitude
\begin{equation}
s:=\sqrt{(\tilde g^{\alpha \beta}+\tilde u^{\alpha}\tilde u^{\beta})\tilde \sigma_\alpha \tilde \sigma_{\beta}}\,,
\label{eq:s2}
\end{equation}
where $\tilde u^\alpha$ is the particle's four-velocity at time $t$.
In terms of these quantities, the covariant expansion of $\Psi^{\mathcal{P}}$ through order $\epsilon^2$ is given by~\cite{Heffernan:2012su, Wardell:2011gb}
\begin{align}
  \Psi^{\mathcal{P}} &=  q \bigg\{ \frac{1}{\epsilon s}+ \frac{\epsilon}{6 s^3}(\varrho^2-s^2) \tilde C_{u \sigma u \sigma} \nonumber \\
  &\quad + \frac{\epsilon^2}{24 s^3} \left[ (\varrho^2-3s^2) \varrho \tilde C_{u \sigma u \sigma | u} - (\varrho^2-s^2) \tilde C_{u \sigma u \sigma | \sigma} \right]\nonumber \\
  &\quad+\mathcal{O}(\epsilon^3) \bigg\}.
\label{eq:4Dsingular}
\end{align}
Here
\begin{align}
\tilde C_{u \sigma u \sigma} &:= \tilde C_{\alpha \beta \mu \nu} \tilde u^{\alpha} \tilde \sigma^{\beta} \tilde u^{\mu} \tilde \sigma^{\nu},\label{eq:C1}\\
\tilde C_{u \sigma u \sigma | \sigma} &:=\tilde\nabla_{\!\gamma}\tilde C_{\alpha \beta \mu \nu} \tilde u^{\alpha} \tilde \sigma^\beta \tilde u^\mu \tilde \sigma^\nu \tilde \sigma^\gamma\label{eq:C2}
\end{align}
are contractions of the Weyl tensor $C_{\alpha\beta\mu\nu}$ and its derivative  evaluated at the reference point $\tilde{x}$ on the particle's worldline.

We now express the covariant expansion~\eqref{eq:4Dsingular} in terms of Kerr-Schild coordinates. To achieve this we follow the method in~\cite{Wardell:2011gb}, %
which begins from an expansion of $\sigma(x,\tilde{x})$ in powers of $\Delta x^{\alpha}$, 
\begin{align}
  \sigma &= \frac{1}{2} \tilde g_{\alpha \beta}\Delta x^{\alpha}\Delta x^{\beta}+\tilde A_{\alpha \beta \gamma} \Delta x^{\alpha} \Delta x^{\beta} \Delta x^{\gamma} \nonumber\\
  & \quad +\tilde B_{\alpha \beta \gamma \delta} \Delta x^{\alpha} \Delta x^{\beta} \Delta x^{\gamma} \Delta x^{\delta} \nonumber\\
& \quad + \tilde C_{\alpha \beta \gamma \delta \rho}  \Delta x^{\alpha} \Delta x^{\beta} \Delta x^{\gamma} \Delta x^{\delta} \Delta x^{\rho}+ \dots\label{eq:sigma expansion}
\end{align}
Differentiating this with respect to $\tilde{x}^\alpha$, we obtain
\begin{equation}
\begin{split}
\tilde \sigma_{\alpha} &= - \tilde g_{\alpha \beta} \Delta x^{\beta} + (\frac{1}{2}\tilde g_{\beta\gamma,\alpha}-3 \tilde A_{\alpha \beta \gamma}) \Delta x^{\beta} \Delta x^{\gamma} \\ 
&\quad+(\tilde A_{\beta \gamma \delta,\alpha}-4 \tilde B_{\alpha\beta\gamma\delta}) \Delta x^{\beta} \Delta x^{\gamma} \Delta x^{\delta} \\
&\quad + (\tilde B_{\beta \gamma \delta \rho,\alpha}-5 \tilde C_{\alpha \beta \gamma \delta \rho})\Delta x^{\beta}\Delta x^{\gamma}\Delta x^{\delta}\Delta x^{\rho}+ \dots
\end{split}
\end{equation}
We then use the identity $2\sigma=\tilde \sigma_\alpha \tilde \sigma^\alpha$ to recursively determine the coefficients $\tilde A_{\alpha \beta \gamma}$, $\tilde B_{\alpha \beta \gamma \delta}$, $\tilde C_{\alpha \beta \gamma \delta \rho}$ and so on. This yields, for example, $\tilde A_{\alpha \beta \gamma}=\frac{1}{4}\tilde g_{(\alpha \beta, \gamma)}$. We now contract $\tilde \sigma_\alpha$ with the four-velocity, metric and Weyl tensor to get the coordinate expressions for $\varrho$, $s$, $\tilde C_{u \sigma u \sigma}$, and $\tilde C_{u \sigma u \sigma|\sigma}$ as per their definitions~\eqref{eq:rbar}, \eqref{eq:s2}, \eqref{eq:C1}, and \eqref{eq:C2}. Our final expression for $\Psi^\mathcal{P}$ is obtained by substituting all of these results into Eq.~\eqref{eq:4Dsingular} and re-expanding in powers of $\Delta x^\alpha$. %

We write the result in the style of \cite{Barack:2002mha}:%
\begin{equation}
\Psi^{\mathcal{P}}\! =q \bigg[  \frac{1}{\epsilon s_1} + \frac{\textrm{P}_3(\Delta x^\alpha)}{s_1^3} + \frac{\epsilon \textrm{P}_6(\Delta x^\alpha)}{s_1^5} + \frac{\epsilon^2 \textrm{P}_9(\Delta x^\alpha)}{s_1^7} +\mathcal{O}(\!\epsilon^3) \bigg].
\label{eq:puncture2}
\end{equation}
Here $s_1=\sqrt{ (\tilde g_{\alpha \beta}+\tilde u_\alpha \tilde u_\beta) \Delta x^\alpha \Delta x^\beta}$ is the leading coordinate approximation to $s$, and $\textrm{P}_n(\Delta x^\alpha)$ is a polynomial in $\Delta x^\alpha$ of homogeneous order $n$. 

The form~\eqref{eq:puncture2} is valid in any coordinate system. In the co-rotating grid coordinates, the reference point on $x_p$ is $\tilde x^{\bar{\alpha}}=x^{\bar\alpha}_p(t)=(t, r_p, 0, 0)$, and the coordinate separation $\Delta x^\ibar := x^\ibar - x_p^\ibar(t)$ is $\Delta x=\bar x-r_p$, $\Delta y=\bar y$, and $\Delta z=\bar z$. The distance $s_1$ then reduces to 
\begin{multline}
(s_1)^2 = \left(1+\frac{2M}{r_p}\right)\Delta x^2 + \Delta y^2 + \Delta z^2 \\ 
   + (\tilde u^t)^2\left(\frac{2M\Delta x}{r_p} + r_p\omega\Delta y\right)^2,
\end{multline}
where $\tilde u^t=(1-3M/r_p)^{-1/2}$. The polynomials $\textrm{P}_3(\Delta x^\alpha)$, $\textrm{P}_6(\Delta x^\alpha)$, and $\textrm{P}_9(\Delta x^\alpha)$  are too long to be included here. Instead we have made them available online as \textsc{Mathematica} code \footnote{\url{https://github.com/nikwit/Puncture-Field-KS-Coords}}.

We now turn to the regular field $\Psi^\mathcal{R}$. Because it approximates a smooth homogeneous solution,  %
we can write it as a Taylor series around $\tilde x^\alpha$. In the grid coordinates, such an expansion reads
\begin{equation}
\Psi^\mathcal{R}(t, x^\ibar) = \Psi^\mathcal{R}_0(t) + \epsilon \Psi^\mathcal{R}_{i}(t) \Delta x^\ibar + \epsilon^2 \Psi^\mathcal{R}_{\ibar \jbar}(t) \Delta x^\ibar \Delta x^\jbar + \mathcal{O}(\epsilon^3), 
\label{eq:3dregular}
\end{equation}
with the notation  $\Psi^\mathcal{R}_0(t) := \Psi^\mathcal{R}(t, x_p^\ibar)$, $\Psi^\mathcal{R}_{\ibar}(t) := \partial_\ibar \Psi^\mathcal{R}(t, x_p^\jbar)$, $\Psi^\mathcal{R}_{\ibar\jbar}(t) := \frac{1}{2}\partial_\ibar\partial_\jbar \Psi^\mathcal{R}(t, x_p^{\bar k})$, and so on. The coefficients $\Psi^\mathcal{R}_{\ibar_1\dots \ibar_k}$ in this series contain the full freedom in the approximate solution $\Psi^\mathcal{A}$. However, not all of these coefficients are independent; the field equation imposes relationships between them. As shown in Ref.~\cite{Pound:2012dk}, once the field equation is enforced, only the trace-free piece of each $\Psi^\mathcal{R}_{\ibar_1\dots \ibar_k}$ is left undetermined. An $n$th-order approximate solution $\Psi^\mathcal{R}$ contains $\sum_{k=0}^n(2k+1)=(n+1)^2$ of these undetermined functions. All other functions of $t$ in $\Psi^\mathcal{R}$ are related to these by ordinary differential equations (ODEs) that result from the field equations. In the next section we show how all the functions $\Psi^\mathcal{R}_{\ibar_1\dots \ibar_k}(t)$ can be  determined through the combination of (i) matching $\Psi^\mathcal{R}$ to $\Psi^\mathcal{N}$ and (ii) solving the ODEs in $t$ that follow from the field equation.

\section{Matching Method}\label{sec: matching method n=2}

The idea behind the matching method is straightforward. We numerically solve the scalar wave equation on a Schwarzschild background, excising the worldtube containing the scalar charge from the numerical domain. Inside the worldtube, the solution is given by the analytical approximation $\Psi^\mathcal{A}=\Psi^{\mathcal{P}} +\Psi^\mR$ described above. Outside the worldtube we have the numerical field $\Psi^\mN$. We demand
\begin{equation}\label{eq: field split}
	\Psi^\mN \overset{\Gamma}{=} \Psi^\mP + \Psi^\mR,
\end{equation}
where $\overset{\Gamma}{=}$ henceforth represents an equality that holds on the (2+1D) worldtube's boundary $\Gamma$.
We will show that this matching condition, together with the scalar wave equation, fully determine the regular field $\Psi^\mR$ inside the worldtube. This solution, in turn, provides boundary conditions for the evolution of the numerical field, specifically for $U^-$.

We formulate the matching scheme in the co-moving grid coordinates $x^{\bar \imath}$ introduced in Eq.~(\ref{eq: grid coordinates}).  %
The Euclidean distance to the particle is defined as $\rho := \sqrt{\delta_{\ibar \jbar} \Delta x^\ibar \Delta x^\jbar} = \sqrt{\delta_{i j} \Delta x^{i} \Delta x^{j}}$. The boundary of the worldtube is located at $\rho=R$, with normal vector $n^\ibar := \Delta x^\ibar / \rho$.
We note that $n^\ibar$ is normalized with respect to $\delta_{\ibar\jbar}$,
whereas $\hat n^i$ in Sec.~\ref{sec: Evolution} is normalized with
respect to the 3-metric $\gamma_{ij}$.

We now introduce the details of our matching scheme for order $n = 0$, $1$ and $2$, which are the expansion orders implemented numerically in this work. The matching scheme for an expansion of arbitrary order $n$ is given in Appendix~\ref{sec: matching method arbitrary n}. We start by re-writing the Taylor expansion in Eq.~\eqref{eq:3dregular} in terms of the quantities $\rho$ and $n^\ibar$, and we introduce an analogous expansion for the time derivative of the regular field:
\begin{subequations}
\label{eq: power series n=2}
\begin{equation}\label{eq: psi power series n=2}
  \Psi^\mR(t,x^\ibar)
=  \Psi^\mR_0(t) + \rho\Psi^\mR_{\ibar}(t) n^\ibar
+ \rho^2  \Psi^\mR_{\ibar \jbar}(t) n^\ibar n^\jbar
+ \Order(\rho^3),
\end{equation}
\begin{equation}
\dot{\Psi}^\mR(t,x^\ibar)
= \dot{\Psi}^\mR_0(t) + \rho \dot{\Psi}^\mR_{\ibar}(t) n^\ibar
+ \rho^2 \dot{\Psi}^\mR_{\ibar \jbar}(t) n^\ibar n^\jbar
+ \Order(\rho^3),\label{eq: dt psi power series n=2}
	\end{equation}
\end{subequations}
where we now drop the order-counting parameter $\epsilon=1$. The set of coefficients $\big\{ \Psi^\mR_0(t), \Psi^\mR_\ibar(t), \Psi^\mR_{\ibar \jbar}(t)\big\}$ have one, three and six independent components, respectively, for a total of ten. We will show that all of these can be uniquely determined at each time step from (i) the numerical field $\Psi^\mN(t, x^\ibar)$ at the worldtube boundary, and (ii) the Klein-Gordon equation \eqref{eq: KG without source}.

\subsection{Worldtube boundary data}
At each time step $t_s$ we enforce the continuity condition
\begin{align}\label{eq: continuity condition}
	\Psi^\mR(t_s, x^\ibar)  &\overset{\Gamma}{=} \Psi^\mN(t_s, x^\ibar) - \Psi^\mP(t_s, x^\ibar),
\end{align}
both for the field itself and its time derivative. In the following section we will omit explicit expressions which enforce continuity between the time derivative  of the regular field $\dot{\Psi}^\mR(t,x^\ibar)$ and the numerical field $\partial_t \Psi^\mN(t,x^\ibar)$ because they are completely analogous to the expressions for the fields themselves.

We will utilize  symmetric trace-free (STF) tensors, indicated with
  angular brackets, e.g.\ $A^{\langle k_1\ldots k_l\rangle}$.  
  Note that $A^{\langle k_1 \cdots k_l \rangle} B_{k_1 \cdots k_l} = A^{ k_1 \cdots k_l} B_{\langle k_1 \cdots k_l \rangle} = A^{\langle k_1 \cdots k_l \rangle} B_{\langle k_1 \cdots k_l \rangle}$; more details about STF tensors are given in Appendix~\ref{sec: matching method arbitrary n}. Transforming  Eq.~(\ref{eq: psi power series n=2}) to a STF basis using \eqref{eq: normal from STF} yields, at order $n=2$,
\begin{multline}\label{eq: regular field STF n=2}
	\Psi^\mR(t_s, x^\ibar) =  \Psi^\mR_0 (t_s) + \frac{1}{3} \rho^2 \delta^{\ibar \jbar} \Psi^\mR_{\ibar \jbar} + \rho \Psi^\mR_{\langle \ibar \rangle} n^{\langle \ibar \rangle} \\+ \rho^2 \Psi^\mR_{ \langle \ibar \jbar \rangle} n^{\langle \ibar} n^{ \jbar \rangle},
\end{multline}
with $n^{\langle \ibar \rangle} = n^\ibar$ and $n^{\langle \ibar} n^{\jbar \rangle} = n^\ibar n^\jbar - \frac{1}{3} \delta^{\ibar \jbar}$. Equation~(\ref{eq: regular field STF n=2}) will be used on the left-hand side in Eq.~(\ref{eq: continuity condition}).

The right-hand side of Eq.~\eqref{eq: continuity condition} is obtained
by evaluating the puncture field of Eq.~\eqref{eq:puncture2} and its time derivative at the coordinates of the DG collocation points on the worldtube surface, and subtracting them pointwise from the corresponding values of $\Psi^\mN(t_s, x^\ibar)$ and $\partial_t \Psi^\mN(t_s, x^\ibar)$. This expression is then projected numerically onto the set of spherical harmonics defined on the worldtube $\Gamma$ with constant radius $\rho$,
\begin{equation}\label{eq:spherical harmonic expansion}
	\Psi^\mN(t_s, x^\ibar) - \Psi^\mP(t_s, x^\ibar) \overset{\Gamma}{=} \sum_{l=0}^{n=2} \sum_{m= -l}^l a_{lm}^{\mN, \mR}(t_s) Y_{lm}(n^\ibar),
\end{equation}
where
\begin{equation}\label{eq:spherical harmonic coefficients}
	a_{lm}^{\mN, \mR}(t_s) = \oint_{\Gamma} \left[ \Psi^\mN(t_s, x^\ibar) - \Psi^\mP(t_s, x^\ibar) \right] Y^*_{lm}(n^\ibar) d \Omega
\end{equation}
are the spherical harmonic coefficients of the \textit{numerical, regular} field $\Psi^\mN(t_s, x^\ibar) - \Psi^\mP(t_s, x^\ibar)$ and $d \Omega$ is the area element of the flat-space unit 2-sphere. In practice we use real-valued spherical harmonics and evaluate the integral with the Gauss-Lobatto quadrature used by the DG method.

Both the spherical harmonics $Y_{lm}$ and the STF normal vector $n^{\langle \bar k_1} \cdots n^{\bar k_l \rangle}$ provide an orthogonal basis for functions on a sphere. They can be transformed into each other using Eqs.~\eqref{eq: STF Ylm relations},
\begin{multline}\label {eq: NR STF n=2}
	\sum_{l=0}^{n=2} \sum_{m= -l}^l a_{lm}^{\mN,\mR}(t_s) Y_{lm} (n^\ibar) \\
	= \Psi^{\mN, \mR}_{\langle 0 \rangle} (t_s) +\Psi^{\mN, \mR}_{\langle \ibar \rangle} (t_s) n^{\langle \ibar \rangle} + \Psi^{\mN, \mR}_{\langle \ibar \jbar \rangle} (t_s) n^{\langle \ibar } n^{\jbar \rangle}.
\end{multline}
We have thus expressed both sides of the continuity condition~\eqref{eq: continuity condition} in a basis of STF normal vectors, using Eqs.~\eqref{eq: regular field STF n=2} and \eqref{eq: NR STF n=2}. Orthogonality of the STF basis allows us to match order by order in the STF expansion:
\begin{subequations}\label{eq: matching all orders}
	\begin{align}
		\Psi^{\mN, \mR}_{\langle 0 \rangle}(t_s) &= \Psi^\mR_0(t_s) + \frac{1}{3} \rho^2 \delta^{\ibar \jbar} \Psi^\mR_{\ibar \jbar}(t_s), \label{eq: matching order 0} \\ 
		\Psi^{\mN, \mR}_{\langle \ibar \rangle}(t_s) &= \rho \Psi^\mR_ \ibar (t_s),\label{eq: matching order 1}  \\ 
		\Psi^{\mN, \mR}_{\langle \ibar \jbar  \rangle}(t_s) &= \rho^2 \Psi^\mR_{\langle \ibar \jbar \rangle}(t_s).\label{eq: matching order 2}
	\end{align}
\end{subequations}
We emphasise that Eqs.~(\ref{eq: matching all orders}) contain two distinct sets of coefficients:  The $\Psi^{\mN,\mR}$ on the left-hand-sides are expansion coefficients on the surface $\Gamma$, whereas the $\Psi^\mR$ on the right-hand-side are the Taylor expansion coefficients of the solution in the interior, Eq.~(\ref{eq: psi power series n=2}).
The continuity conditions for a field expanded to arbitrary order are given in Eq.~\eqref{eq: continuity STF}. For expansion orders $n = 0$ or $n = 1$ the second term in Eq.~\eqref{eq: matching order 0} falls away. The regular field inside the worldtube is then fully determined by the continuity condition and can directly be used to provide boundary conditions for the future evolution. In one dimension, this is equivalent to a linear polynomial in an interval being fully determined by its two endpoints.

For $n = 2$, Eqs.~\eqref{eq: matching all orders} provide only 9 equations for the 10 coefficients of $\Psi^\mR(t_s, x^\ibar)$ because the monopole of the regular numerical field $\Psi^{\mN, \mR}_{\langle 0 \rangle}$ in Eq.~\eqref{eq: matching order 0} contributes to both the zeroth-order coefficient $\Psi^\mR_0$ and the trace of the second-order coefficient, $\delta^{\ibar \jbar} \Psi^\mR_{\ibar \jbar}$. 
More generally, for arbitrary order, the STF expansion on the worldtube, Eq.~\eqref{eq: continuity STF}, provides only the trace-free components of $\Psi^{\mN,\mR}$, so that boundary-matching determines only the trace-free parts of the expansion $\Psi^\mR(t_s, x^\ibar)$ but not its traces.
Therefore, for expansions of order $n \geq 2$, additional equations are needed to fully determine the regular field inside the worldtube. These are provided by a series expansion of the Klein-Gordon equation, as we describe below in Sec.~\ref{sec:Klein-Gordon equation}.

The coefficients of the regular field's time derivative %
are determined completely analogously, with the continuity condition
\begin{align}\label{eq: continuity condition dt}
	\partial_t \Psi^\mR(t_s, x^i)  \overset{\Gamma}{=} \partial_t\Psi^\mN(t_s, x^i) - \partial_t\Psi^\mP(t_s, x^i). \
\end{align}
$\partial_t \Psi^\mN(t_s, x^i)$ is evaluated using its evolution equation~\eqref{eq: dt Psi} and then transformed into the co-moving grid frame by adding the advective term $v_g^i \partial_i \Psi^\mN$, where $v_g^i$ is the instantaneous local grid velocity. The matching conditions for the time derivative of the regular field $\dot{\Psi}^R(t_s)$ are then just the time derivative of the matching conditions for $\Psi^R(t_s)$, Eqs.~\eqref{eq: matching all orders}.

\subsection{Klein-Gordon equation}
\label{sec:Klein-Gordon equation}

We rewrite the Klein-Gordon equation~\eqref{eq: KG without source} in grid coordinates
\begin{equation}\label{eq: KG}
	0 = g^{\bar \mu \bar \nu} \partial_{\bar \mu} \partial_{\bar \nu} \Psi^\mR - \Gamma^{\bar \rho} \partial_{\bar \rho} \Psi^\mR, 
\end{equation}
where $\Gamma^{\bar\rho} \coloneqq g^{\bar\mu\bar\nu} \Gamma^{\bar\rho}_{\bar\mu\bar\nu}$. The metric quantities $g^{\mu \nu}$ and $\Gamma^\mu$ are expanded in the grid coordinates $x^\ibar$ to the same order $n$ as the regular field at each time step $t_s$. For $n = 2$ these expansions read
\begin{align}
	g^{\bar{\mu} \bar{\nu}}(t_s,x^\ibar) &= g^{\bar{\mu} \bar{\nu}}_0(t_s) + g^{\bar{\mu} \bar{\nu}}_\ibar(t_s) \Delta x^\ibar \nonumber\\
							&\quad + g^{\bar{\mu} \bar{\nu}}_{\ibar \jbar} (t_s)\Delta x^\ibar \Delta x^\jbar + \Order (\rho^3), \label{eq: metric expansion} \\ 
	\Gamma^{\bar{\mu}}(t_s,x^\ibar) &= \Gamma^{\bar{\mu}}_0(t_s) + \Gamma^{\bar{\mu}}_\ibar(t_s) \Delta x^\ibar \nonumber\\
							&\quad + \Gamma^{\bar{\mu}}_{\ibar \jbar}(t_s) \Delta x^\ibar \Delta x^\jbar + \Order (\rho^3). \label{eq: Christoffel expansion}
\end{align}
The expansion coefficients are given by $g_0^{\bar\mu\bar\nu}:=g^{\bar\mu\bar\nu}(t_s, x^\ibar_p)$, $g_\ibar^{\bar\mu\bar\nu}:=\partial_\ibar g^{\bar\mu\bar\nu}(t_s,x^\jbar_p)$, and
$g_{\ibar\jbar}^{\bar\mu\bar\nu}(t_s):=\frac{1}{2}\partial_\ibar\partial_\jbar g^{\bar\mu\bar\nu}(t_s, x^{\bar k}_p)$, and similarly for $\Gamma^{\bar\mu}_0$, $\Gamma^{\bar\mu}_\ibar$, and $\Gamma^{\bar\mu}_{\ibar\jbar}$.
Due to the spherical symmetry of the Schwarzschild spacetime, and our circular-orbit setup,  these expansion coefficients are in fact independent of $t_s$.

We now expand the Klein-Gordon equation in powers of $\rho$ by inserting the expansions for $\Psi^\mR$, $g^{\mu \nu}$ and $\Gamma^\mu$ from Eqs.~\eqref{eq: psi power series n=2}, \eqref{eq: metric expansion} and \eqref{eq: Christoffel expansion}, respectively, into Eq.~\eqref{eq: KG}. 
The $O(\rho^0)$ piece of the equation reads
\begin{align}
	g^{tt}_0 \ddot{\Psi}^\mR_0(t_s)  &+ 2 g_0^{t\ibar} \dot{\Psi}^\mR_{\ibar}(t_s) + 2 g_0^{\ibar \jbar}\Psi_{\ibar \jbar}^\mR(t_s)\nonumber \\
  \label{eq: expanded KG n=2}
  &-  \Gamma_0^t\dot{\Psi}_0^\mR(t_s) -  \Gamma_0^\ibar \Psi_{\ibar}^\mR(t_s) = 0.
\end{align}
This ODE provides an additional, independent relation between the expansion coefficients ${ \Psi}^\mR_0$, $\Psi^\mR_{\ibar}$ and $\Psi_{\ibar \jbar}^\mR$, which enables us to determine the remaining, trace degree of freedom of the regular field at $n\!=\!2$. 
Specifically, combining Eq.~\eqref{eq: expanded KG n=2} with the continuity conditions (\ref{eq: matching all orders}), and using 
$\Psi_{\ibar \jbar}^\mR = \Psi_{\langle \ibar \jbar \rangle}^\mR + \frac{1}{3} \delta^{\bar l \bar k} \Psi_{\bar l \bar k}^\mR \delta_{\ibar \jbar}$,
we obtain
\begin{multline}\label{eq: Phi0 ODE}
	 g^{tt}_0 \ddot{\Psi}^\mR_0(t_s) + 2 g_0^{t\ibar} \dot{\Psi}^{\mN, \mR}_\ibar(t_s) + 2 g_0^{\ibar \jbar} \Psi^{\mN, \mR}_{\langle \ibar \jbar \rangle}(t_s) 
	\\ + \frac{2 \delta_{\ibar \jbar}  g_0^{\ibar \jbar}}{\rho^2} \left(\Psi^{\mN, \mR}_0(t_s) - \Psi^\mR_0(t_s) \right) \\
	-  \Gamma_0^0\dot{\Psi}^\mR_0(t_s) - \Gamma_0^\ibar \Psi_\ibar^{\mN, \mR}(t_s) = 0 .
\end{multline}
We reduce this ODE to first order and use a Dormand-Prince time stepper to advance the zeroth-order coefficient $\Psi^\mR_0$ and its time derivative to the next time step $t_{s+1}$, taking the same global time step as the DG evolution. Together with the continuity conditions~\eqref{eq: matching all orders} at time step $t_{s+1}$, this completely determines all components of the second-order expansion of $\Psi^\mR(t, x^\ibar)$ in Eq.~\eqref{eq: psi power series n=2} at $t_{s+1}$. 

The coefficients of the numerical, regular field $\Psi^{\mN, \mR}_{\langle \bar k_0 \cdots \bar k_l \rangle}$ are updated each sub-step. As initial conditions of the ODE (\ref{eq: Phi0 ODE}) we take $\Psi^\mR_0(t_0) = \dot{\Psi}^\mR_0(t_0) = 0$.

In Appendix \ref{sec: matching method arbitrary n} we formulate the generalization of this method to an arbitrary order $n$, and in particular we derive the generalized form of the ODE on $\Gamma$. 

\subsection{Boundary conditions for $\Psi^\mN$} \label{sec: BC formulation}
Once the expansion of the regular field has been fully determined, it can be used to provide boundary conditions to the DG elements neighboring the worldtube. DG methods commonly formulate boundary conditions between elements using the numerical flux, and these conditions are applied to each of the characteristic fields defined in Eqs.~\eqref{eq: char fields}. We use the internal solution $\Psi^\mathcal{A}$ of the worldtube to provide boundary conditions for the characteristic field $U^-$ as if the interior of the worldtube were simply another DG element.
From the definition of $U^-$ in Eq.~\eqref{eq: U minus} and the definitions in Eq.~\eqref{eq: evolution reduction}, we obtain the boundary condition
\begin{equation} \label{eq: U minus BCs}
	U^-(t_s)\overset{\Gamma}{=} - \alpha^{-1} \partial_t\Psi^{\cal A}(t_s)+\left( \beta^\ibar - \hat{n}^\ibar \right) \partial_\ibar\Psi^{\cal A}(t_s)- \gamma_2 \Psi^{\cal A}(t_s).
\end{equation}
The analytical solution $\Psi^{\cal A}(t_s)$ was defined in Eq.~\eqref{psiA definition} as the sum of the regular field $\Psi^\mR$ and the puncture field $\Psi^\mP$, both of which are now fully determined. The time and spatial derivative are simply obtained from $\partial_t\Psi^{\cal A}(t_s) = \partial_t \Psi^\mN(t_s) + \partial_t \Psi^\mP(t_s)$ and $\partial_\ibar\Psi^{\cal A}(t_s) = \partial_\ibar \Psi^\mN(t_s) + \partial_\ibar \Psi^\mP(t_s)$. The fields $\Psi^\mR(t_s)$ and $\partial_t \Psi^\mR(t_s)$ are given by Eq.~\eqref{eq: power series n=2} and its time derivative.  The derivative normal to the worldtube boundary is similarly obtained by taking the appropriate spatial derivative of $\Psi^\mR$ in Eq.~\eqref{eq: psi power series n=2} analytically. The expression for the puncture field $\Psi^\mP(t_s)$ is given in Eq.~\eqref{eq:3dregular}, and its time and normal derivative are computed analytically. 
We evaluate all of these expressions at the grid coordinates $x^\ibar$ of all DG grid points that lie on element faces abutting the worldtube to formulate pointwise boundary conditions. The value of $U^-(t_s)$ at the boundary is used to apply a correction to the time derivative of the evolution equations using the upwind flux \cite{Hesthaven:2007}.

We initially tried to provide boundary conditions in the above fashion for all characteristic fields entering the numerical domain, including $Z^1$ and $Z^2_i$. However, we found that this caused substantial constraint violations entering the numerical domain at the worldtube boundary. Instead, we use constraint-preserving boundary conditions for $Z^1$ and $Z^2_i$ as described in Appendix~\ref{sec: constraint preserving BCs}.

\subsection{Roll-on function}\label{sec: roll-on}
The initial conditions we use for the simulations are $\Psi = \partial_t \Psi = 0$ for both the DG fields outside the worldtube and the regular field inside it. The puncture field $\Psi^\mP$ added to the regular field in Eq.~\eqref{eq: U minus BCs} initially creates a discontinuity at the worldtube boundary, due to the unphysical instantaneous appearance of the scalar charge source $t=0$. DG methods are very inefficient at resolving discontinuities within elements, due to the Gibbs phenomenon.

To alleviate this, we multiply the puncture field $\Psi^\mP$ with a roll-on function $w(t)$ that smoothly grows from 0 to 1 (up to double precision) between $t=0$ and $t=t_{\rm end}$. We found that this effectively stretches out the initial discontinuity and causes the fields to settle more smoothly to their final values.

We tested two different roll-on functions: $w(t) = \mathrm{sin} [\pi t / (2 t_{\rm end})]$ and $w(t) = \mathrm{erf}(12 t/ t_{\rm end} - 6)/2 + 1/2$, where $\mathrm{erf}$ is the Gaussian error function. There was little difference in the long term evolution between the two choices. 

The roll-on function ensures a smooth settling of the solution corresponding to the scalar charge slowly being turned on over its first 4 orbits. We found $t_{\rm end} = 300M$ to be a good choice for the simulations with orbital radius $r_p = 5 M$. 

\subsection{Error estimates}\label{sec: error estimates}

To estimate the errors that our matching method incurs, we apply the same analysis as we did for the $1+1$D case in~\cite{Dhesi:2021yje}. 
The estimates follow from a Kirchhoff representation of the scalar field. We first consider the field in the numerical domain, outside the tube $\Gamma$. Call this region $V$. Inside $V$, our field
$\Psi^\mathcal{N}$ satisfies the same homogeneous field equation as the exact solution
$\Psi$, $g^{\mu\nu}\nabla_{\mu}\nabla_{\nu}\Psi^\mathcal{N}=0$, but it inherits errors that
propagate out from $\Gamma$. We introduce a retarded Green's function $G(x,x')$
satisfying
\begin{equation}\label{eq:BoxG}
  \Box G(x,x') = \Box' G(x,x') = \delta^4(x,x')\, ,
\end{equation}
where $x$ and $x'$ denote any two points, primes denote quantities at $x'$, $\Box:=g^{\mu\nu}\nabla_{\mu}\nabla_{\nu}$, and $\delta^4(x,x'):=\frac{\delta^4(x^\mu-x^{\mu'})}{\sqrt{-g}}$. If we now take any point $x\in V$, then the equations~\eqref{eq:BoxG} and $\Box\Psi^\mathcal{N}=0$ imply the
identity
\begin{equation}
\Psi^\mathcal{N}(x') \delta^4(x,x') = \Psi^\mathcal{N}(x')\Box' G(x,x') - G(x,x')\Box'\Psi^\mathcal{N}(x').
\end{equation}
Integrating this equation over all $x'\in V$ and then
integrating by parts, we obtain the Kirchhoff representation
\begin{align}
  \Psi^\mathcal{N}\!(x) & =\! \int_V\! \left[\Psi^\mathcal{N}(x')\Box' G(x,x')
    - G(x,x')\Box'\Psi^\mathcal{N}(x')\right] dV' \nonumber\\
 & =\! \int_{\partial V}\!\! \left[\Psi^\mathcal{N}(x')\nabla_{\!\mu'} G(x,x')
   \!-\! G(x,x')\nabla_{\!\mu'}\Psi^\mathcal{N}\!(x')\right]d\Sigma^{\mu'}\!\!.
  \label{eq:kirchhoff_representation}
\end{align}
Here $d\Sigma^{\mu'}$ is the outward-directed surface element on $\partial V$. For us the relevant portion of $\partial V$ is the tube boundary $\Gamma$, where $d\Sigma^{\mu'}= \mathcal{O}(R^2) dt\, d\Omega$. As in Eq.~\eqref{eq:spherical harmonic coefficients}, here $d\Omega$ is the area element of the unit 2-sphere.

In the integral over $\Gamma$, we may replace $\Psi^\mathcal{N}$ with $\Psi^\mathcal{A}$. Our truncated expansion of $\Psi^\mathcal{A}$  introduces an inherent $\mathcal{O}(R^{n+1})$ error in $\Psi^\mN(x')$ and $\mathcal{O}(R^{n})$ error in $\nabla_{\mu'}\Psi^\mN(x')$ on the worldtube. Equation~\eqref{eq:kirchhoff_representation} implies that the $\mathcal{O}(R^{n})$ error in $\nabla_{\mu'}\Psi^\mN(x')$ dominates. Accounting for the $\mathcal{O}(R^2)$ surface element, we see that this creates an $\mathcal{O}(R^{n+2})$ error in $\Psi^\mathcal{N}(x)$.

An important takeaway from this analysis is that the error in the numerical domain is suppressed by the small spatial size of $\Gamma$. As a consequence, the error converges two orders faster than the analogous error in the $1+1$D problem in~\cite{Dhesi:2021yje}. 

However, we note that this analysis applies only at a fixed location $x$ outside the worldtube. At a point \emph{on} the worldtube boundary $\Gamma$, the errors in $\Psi^\mathcal{N}$ are inherently $\mathcal{O}(R^{n+1})$, and the errors in $\nabla_{\mu}\Psi^\mathcal{N}$ are inherently $\mathcal{O}(R^{n})$. There is no suppression due to the small spatial size of the worldtube in this case. The same is true of the errors at a point outside the worldtube if we consider a point $x$ that is at a fixed multiple of $R$ away from the worldline rather than at a fixed physical location.

We also note that in applications, we require outputs other than $\Psi^\mathcal{N}$: the regular field on the particle's worldline and the self-force, for example. The omitted terms in our expansion~\eqref{eq:3dregular} scale with a power of distance from the worldline, which might make us expect that we incur no error in $\Psi^\mathcal{R}(x_p)$ and $\partial_\mu\Psi^\mathcal{R}(x_p)$ (and therefore in the self-force). However, we can see this is incorrect by referring again to a Kirchhoff representation of the field. Our method enforces the field equation~\eqref{eq: KG with source} on $\Psi^\mathcal{A}$ up to an error $\sim R^{n-1}$ (two derivatives of the truncation error in $\Psi^\mathcal{A}$). If we momentarily ignore that error term in the field equation, and if we consider $V$ to be the interior of $\Gamma$ and repeat the steps that led to Eq.~\eqref{eq:kirchhoff_representation}, then we obtain the Kirchhoff representation
\begin{multline}
  \Psi^\mathcal{A}(x) = -4\pi q\int_\gamma G(x,x_p(\tau)) d\tau +\int_{\Gamma} \Big[\Psi^\mathcal{A}(x')\nabla_{\mu'} G(x,x') \\
         - G(x,x')\nabla_{\mu'}\Psi^\mathcal{A}(x')\Big]d\Sigma^{\mu'}\, .
  \label{eq:interior kirchhoff_representation}
\end{multline}
If we now take $x$ to be a point $x_p$ on the worldline and consider the integral over $\Gamma$, then we have $G(x,x')\sim 1/R$ and $\nabla_{\mu'}G(x,x')\sim 1/R^2$. We can combine this with $d\Sigma^{\mu'}\sim R^2$ and with the errors $\mathcal{O}(R^{n+1})$ in $\Psi^\mathcal{A}(x')$ and $\mathcal{O}(R^{n})$ in $\nabla_{\mu'}\Psi^\mathcal{A}(x')$ to deduce that the error in $\Psi^\mathcal{A}(x_p)$ is $\mathcal{O}(R^{n+1})$. If we take a derivative of Eq.~\eqref{eq:interior kirchhoff_representation}, we find that the error in $\partial_\mu\Psi^\mathcal{A}(x_p)$ is $\mathcal{O}(R^{n})$. These error estimates apply immediately to $\Psi^\mathcal{R}(x_p)$ as well. 

It is also straightforward to see that these estimates are not altered by the $\mathcal{O}(R^{n-1})$ error in the field equation, which we neglected in deriving Eq.~\eqref{eq:interior kirchhoff_representation}. That error contributes an error $\sim \int R^{n-1}G(x_p,x') dV'\sim R^{n+1}$ to $\Psi^\mathcal{R}(x_p)$, consistent with the error from the boundary integral.

In summary, we expect that for an $n$th-order analytical approximation, our method introduces the following errors:
\begin{align}
\text{Error in } \Psi^\mathcal{N}(x):&\quad \mathcal{O}(R^{n+2}), \label{eq: predicted erro psiN}\\
\text{Error in } \Psi^\mathcal{R}(x_p):&\quad \mathcal{O}(R^{n+1}),\label{eq: predicted erro psiR}\\
\text{Error in } \partial_\alpha\Psi^\mathcal{R}(x_p):&\quad \mathcal{O}(R^{n}),\label{eq: predicted erro dpsiR}
\end{align}
where $x$ is a point outside $\Gamma$ and $x_p$ is a point on the particle's worldline. Our numerical results in the next section will bear out these predictions. The error in $\partial_\alpha\Psi^\mathcal{R}(x_p)$, and hence in the self-force, will be particularly relevant when we allow the system to evolve (as opposed to keeping the particle on a fixed geodesic orbit). We defer discussion of this to the Conclusion.

Finally, before proceeding, we note that our error estimate for $\partial_\alpha\Psi^\mathcal{R}(x_p)$ might be too pessimistic in some instances. Specifically, time-antisymmetric components, linked to the dissipative pieces of the self-force, might converge more rapidly with $R$. This is because these components arise from the radiative piece of the field, equal to half the retarded solution minus half the advanced solution~\cite{Mino:2003yg}. For these pieces of the field, we can replace the Green's function $G$ in Eq.~\eqref{eq:interior kirchhoff_representation} with its radiative piece, $G^{\rm Rad}=\frac{1}{2}(G^{\rm Ret}-G^{\rm Adv})$. $G^{\rm Rad}$ is smooth when its two arguments coincide (because singularities cancel between $G^{\rm Ret}$ and $G^{\rm Adv}$), meaning it does not introduce the negative powers of $R$ that $G^{\rm Ret}$ introduces in Eq.~\eqref{eq:interior kirchhoff_representation}. We therefore might expect that errors scale with a higher power of $R$ in the dissipative components of $\partial_\alpha\Psi^\mathcal{R}(x_p)$. That could be extremely beneficial in practice because dissipative effects dominate over conservative ones on the long timescale of an inspiral~\cite{Hinderer:2008dm}, and dissipative effects must therefore be computed with higher accuracy. However, our numerical experiments in the next section do not entirely bear out this expectation of more rapid convergence, and we leave further investigation of it to future work.

\section{Results}\label{sec: Results}

\begin{figure}
	\includegraphics[width=\columnwidth]{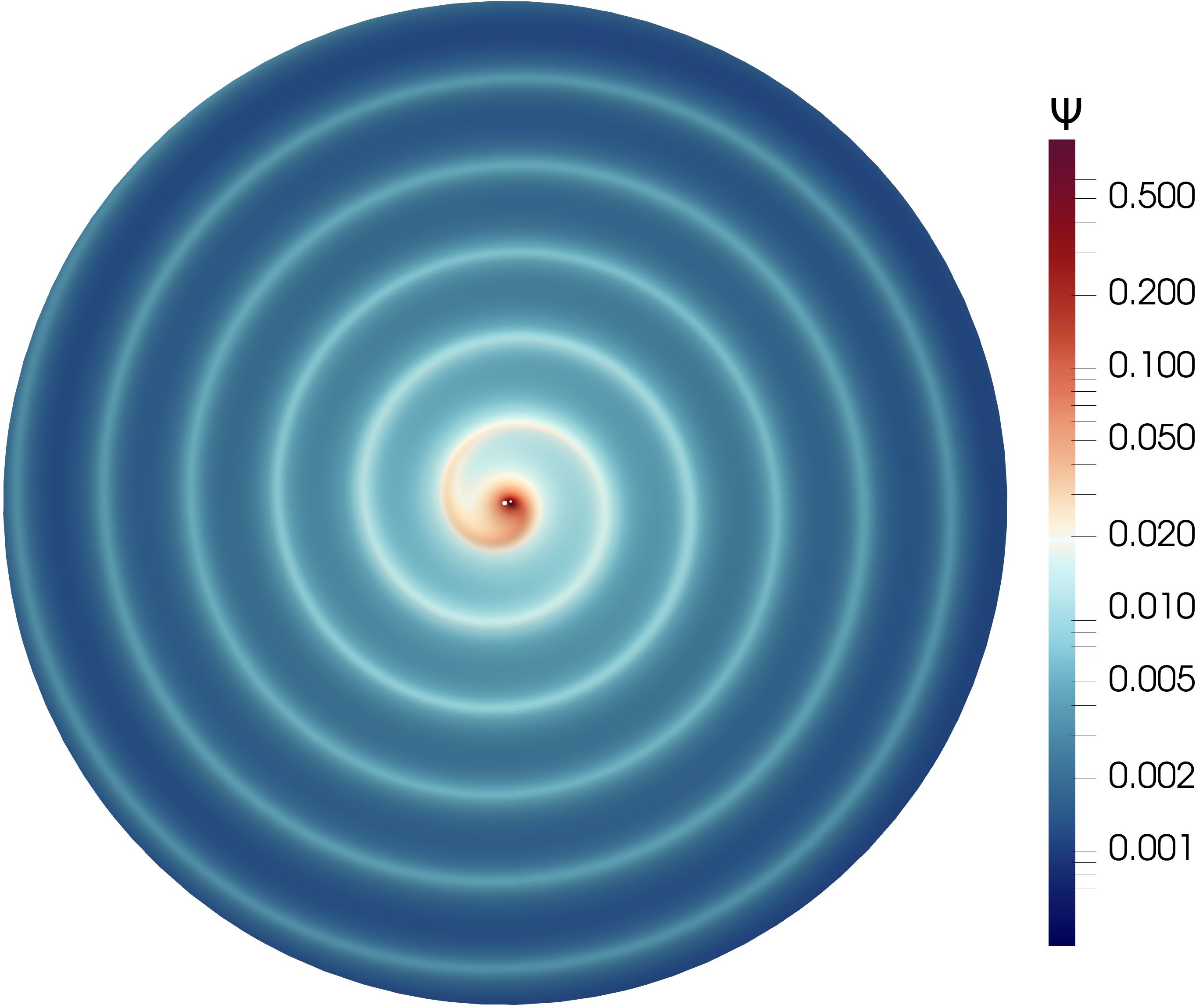}
	\caption{The equatorial plane of the domain, depicting the steady-state solution of the scalar field $\Psi^\mN$. The scalar charge creates an outward propagating spiral as it orbits the central black hole. Figure~\ref{fig: fancy plot} shows a tilted perspective zoomed into the center of the same plane with the spiral arms visible in the background.}
	\label{fig: top down}
\end{figure}

We use a central black hole of mass $M$ for the simulations.
The excision sphere inside the black hole has radius $1.9M$. The outer boundary is placed at $400M$. We use a CFL safety factor of 0.4. The scalar charge is placed on a circular orbit with radius $r_p = 5M$ with angular velocity $\omega = M^{1/2}r_p^{-3/2} \approx 0.09M^{-1}$.

The expansion terms of the puncture field converge more quickly with larger orbital radii of the scalar charge. The truncation error of the puncture field and hence of the worldtube solution is therefore particularly large at the relatively small orbital radius of $r_p = 5M$ used in our simulations. This ensures that the scheme is tested in an extreme region, comparable to binary black holes close to merger. Because the error due to the worldtube is comparatively large for a small $r_p$, it can be resolved with a lower resolution in the numerical domain, lowering the computational cost of the simulations.

We have implemented the worldtube scheme with the local solution expanded to orders $n = 0$, $1$ and $2$. The radius of the worldtube was varied between $0.2M$ and $1.6M$. The simulations were run until the field had settled to its steady state solution over the entire domain, which took between $3000M$ and $7000M$, depending on the magnitude of the settled error. %
Figure~\ref{fig: top down} shows a cut through the equatorial plane of the computational domain.

\begin{figure}
	\includegraphics[width=\columnwidth,trim=17 10 19 10]{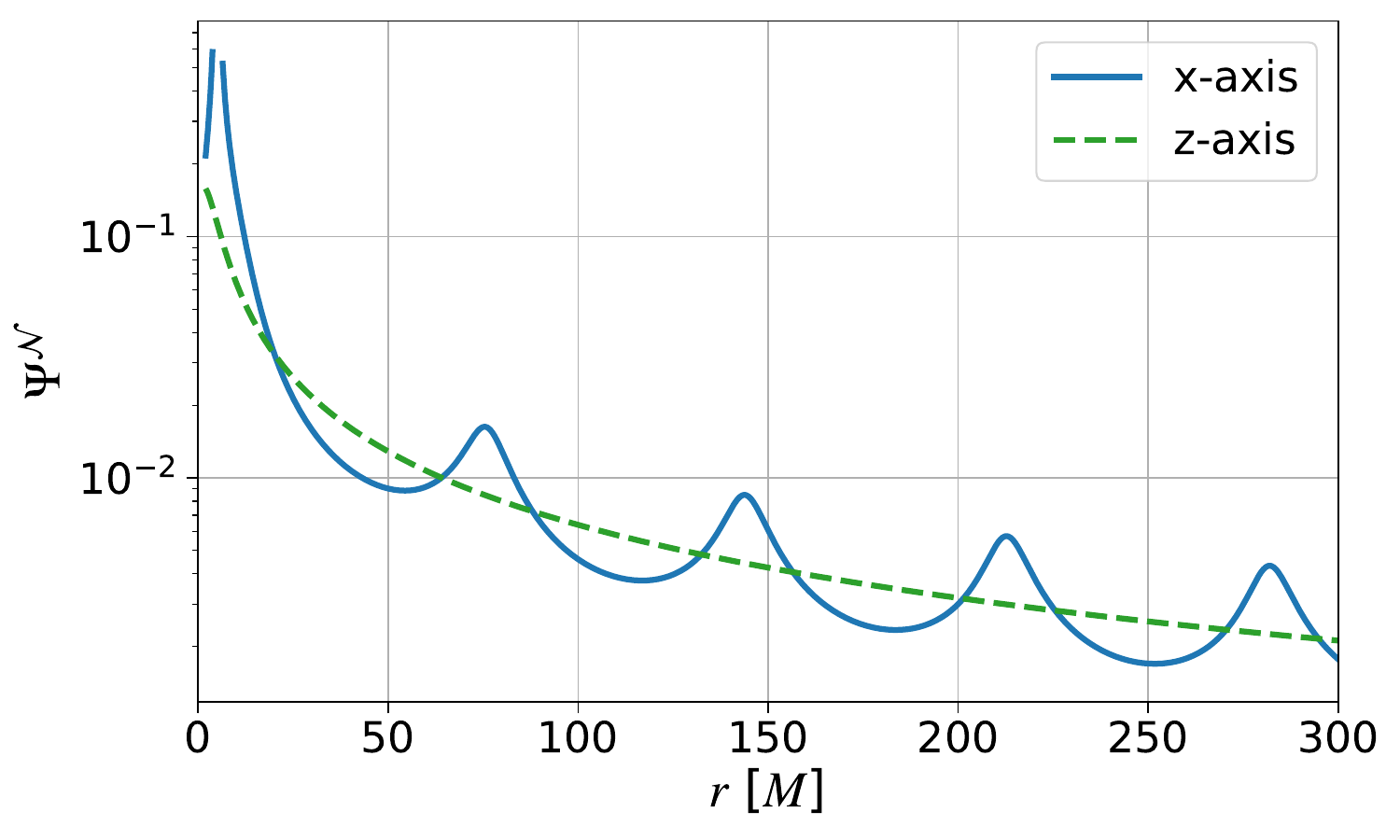}
	\caption{The steady-state solution of the scalar field $\Psi^\mN$ along the co-moving $x$-axis and $z$-axis of the domain.}
	\label{fig: fields along axes}
\end{figure}

Figure~\ref{fig: fields along axes} plots the steady state solution along two lines cut through the domain at late, constant Kerr-Schild times $t$: one along the co-moving $x$-axis connecting the central black hole center and the scalar charge and one along the $z$-axis normal to the charge's orbital plane. The undulations of the scalar field on the $x$-axis correspond to the arms of the spiral in Fig.~\ref{fig: top down}.
The missing part of the `$x$-axis' line corresponds to the worldtube; the field increases strongly near the worldtube, because of the scalar charge contained at the center of the worldtube.

We verify the validity and convergence of our simulations in three different ways:
First, in Sec.~\ref{sec: regular field at charge}, we compare the value of the regular field $\Psi^\mR$  and its spatial derivative $\partial_\ibar \Psi^\mR$ at the position of the charge to published numerical results obtained using frequency-domain self-force methods. 
Second, in Sec.~\ref{sec: z axis} we compare with the known axially symmetric analytical solution along the $z$-axis, given below in Eq.~\eqref{eq: z solution}. 
Finally, we perform an internal convergence test along the co-moving $x$-axis in Sec.~\ref{sec: x axis}.
For each simulation in the following sections, the resolution of the DG domain was increased until it no longer affected the steady-state solution. This guaranteed that the error measured was due to the worldtube, not the numerical evolution.

\subsection{Regular field $\Psi^\mR$ at the charge's position}\label{sec: regular field at charge}

The value of the regular field for a scalar charge in a circular geodesic orbit in Schwarzschild spacetime has been calculated in self-force literature. We compare the regular field of our simulations with the results of \cite{Diaz_Rivera:2004}, who quote the value $\Psi^\mR_{\rm ref}(0) = –0.01023418q/M$ for a circular orbit with radius $5M$.

For expansion orders $n = 0$ and $n = 1$, the regular field at the charge position is given directly by the monopole of the numerical field $\Psi^{\mN, \mR}_{\langle 0 \rangle}(t_s)$ in Eq.~\eqref{eq: matching order 0} (the second term on the right-hand side is absent for $n=0,1$). For $n = 2$, it is determined by solving the ODE~\eqref{eq: Phi0 ODE} inside the worldtube.

\begin{figure}
	\includegraphics[width=\columnwidth]{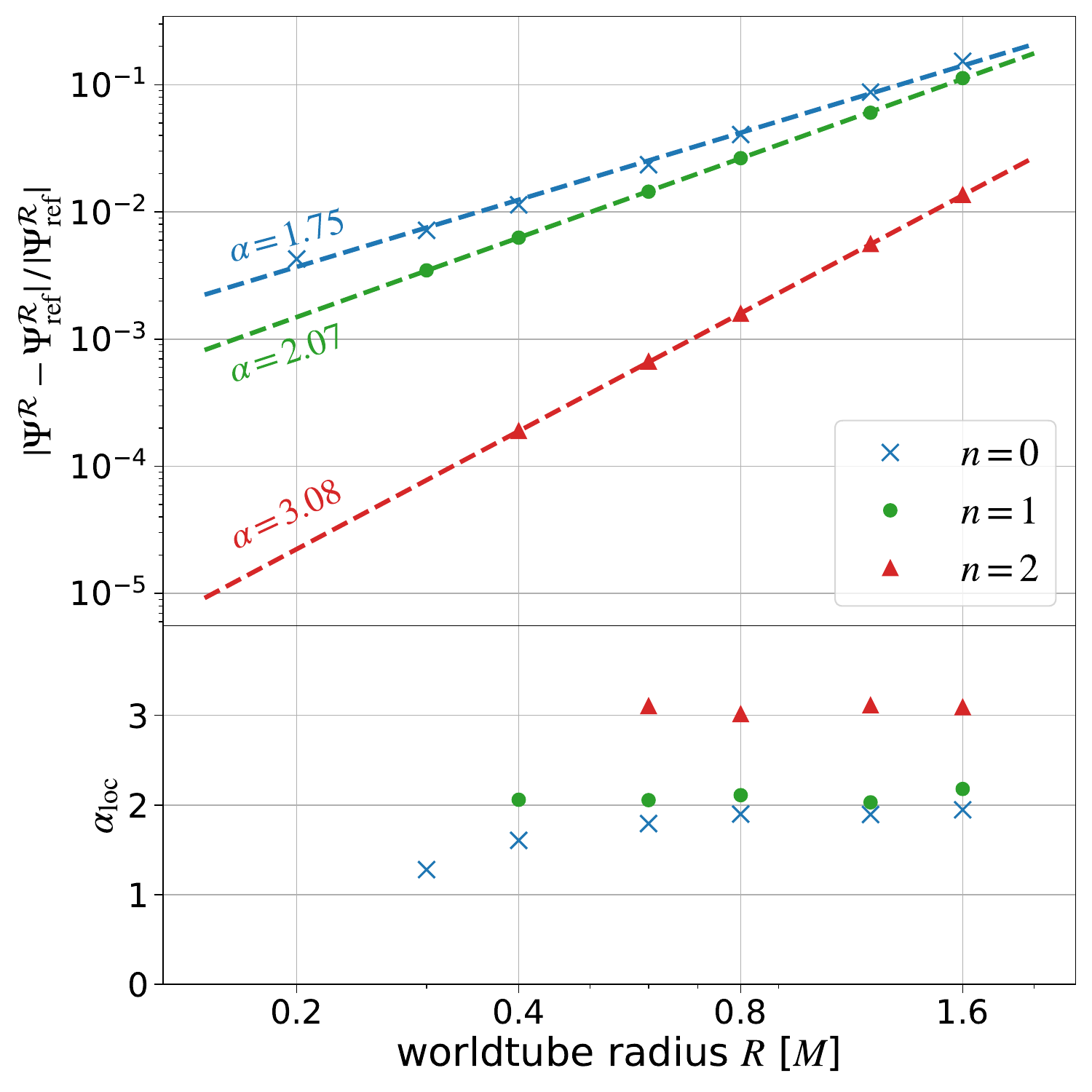}
	\caption{{\it Top panel:} The relative error of the regular field at the position of the charge compared to the value computed in \cite{Diaz_Rivera:2004}. Each cross represents the settled error at the final simulation time. The dashed lines are best fits for the relation $\varepsilon \propto R^\alpha$.
{\it Bottom panel:} The local convergence order between simulations of neighboring worldtube radii.
  }
	\label{fig: regular field}
\end{figure} 

The relative error $\varepsilon = |\Psi^\mR(0) - \Psi^\mR_{\rm ref}(0)| / |\Psi^\mR_{\rm ref}(0)|$, where $\Psi^\mR(0)$ is the final value of the regular field at the scalar charge, is shown in the top panel of Fig.~\eqref{fig: regular field}, with each marker representing a simulation. The dashed lines  show fits of the data, for each expansion order $n$, to a relation of the form $\varepsilon \propto R^\alpha$, where $R$ (recall) is the worldtube radius. The bottom panel displays the local convergence order, defined through
\begin{equation} \label{eq: local convergence order}
 \alpha_{{\rm loc}, i} = \frac{\log(\varepsilon_i) -  \log(\varepsilon_{i-1})}{\log(R_i) - \log(R_{i-1})},
\end{equation}
where $R_i$ are the worldtube radii in our sample, and $\varepsilon_i$ are the corresponding errors.  
We find that the error always decreases with smaller worldtube radius or higher order $n$ of the local solution as expected. Equation~\eqref{eq: predicted erro psiR} indicated a convergence order inside the worldtube of $\alpha = n+1$ at sufficiently small worldtube radii. For $n=1$ and $2$ we find that this prediction is confirmed quite well, with global convergence orders measured as $\sim\!2.07$ and $\sim\!3.08$, respectively, and local convergence order uniformly close to this value. At $n=0$ we measure a global convergence order of $\sim\!1.72$ and a local convergence order that appears to decrease with the worldtube radius. This suggests that for $n=0$ the scheme is not fully in the convergent regime for the values of $R$ we consider; rather, there are still significant contributions from higher-order terms.  At smaller worldtube radii $R$, these higher-order-in-$R$ contributions become less significant and the local convergence rate approaches the expected value of 1.

We also compare the gradient of $\Psi^\mR$ at the position of the particle, which enters the expression for the self-force acting on the particle due to back-reaction from the scalar field. The value of the radial derivative is given in \cite{Diaz_Rivera:2004} as $\partial_{r_s} \Psi^\mR_{\mathrm{ref}}(0) = 0.0004149937q/M^2$ using Schwarzschild coordinates $(t_s,r_s,\theta_s,\varphi_s)$.  We are not aware of any works which report the angular derivative of the scalar field at $r_p=5M$. Instead, it was computed for us to be $\partial_{\varphi_s} \Psi^\mR_{\mathrm{ref}}(0) = -0.01009125769q/M^2$ using the frequency domain code of \cite{Macedo_2022}. The coordinate transformation from Kerr-Schild time $t$ to Schwarzschild time $t_s$ is given by
\begin{equation}
	t_s = t + 2 M \ln \left( \frac{r}{2M} -1 \right),
\end{equation}
which makes the conversion between $\partial_{r_s}\Psi^\mR$ and the Kerr--Schild radial derivative $\partial_{r}\Psi^\mR$
\begin{align}
	\partial_{r_s} &= \partial_r + \frac{2M}{2M-r} \partial_t  \nonumber \\
	&= \partial_r + \frac{2M}{2M-r} (\partial_{\bar{t}} + \frac{\partial x^{\ibar}}{\partial t} \partial_\ibar ) ,
\end{align}
where in the second line we have transformed into the co-moving coordinate frame given in Eqs.~\eqref{eq: grid coordinates}. The reference values of the regular field's gradient at the particle's position are then given by
\begin{align}
 \partial_r \Psi^\mR_{\mathrm{ref}}(0) &= \partial_{\bar{x}} \Psi^\mR_{\mathrm{ref}}(0) = \partial_{r_s}\Psi^\mR_{\mathrm{ref}}(0) - \frac{2M \omega}{2M-r_p}  \partial_{\varphi_s} \Psi^\mR_{\mathrm{ref}}(0), \label{eq: dx ref}\\
   \partial_\varphi \Psi^\mR_{\mathrm{ref}}(0) &= r_p \partial_{\bar{y}}\Psi^\mR_{\mathrm{ref}}(0) = \partial_{\varphi_s} \Psi^\mR_{\mathrm{ref}}(0),  \label{eq: dy ref}
\end{align}
 which we use to compare to our simulation values.

  \begin{figure}
	\includegraphics[width=\columnwidth]{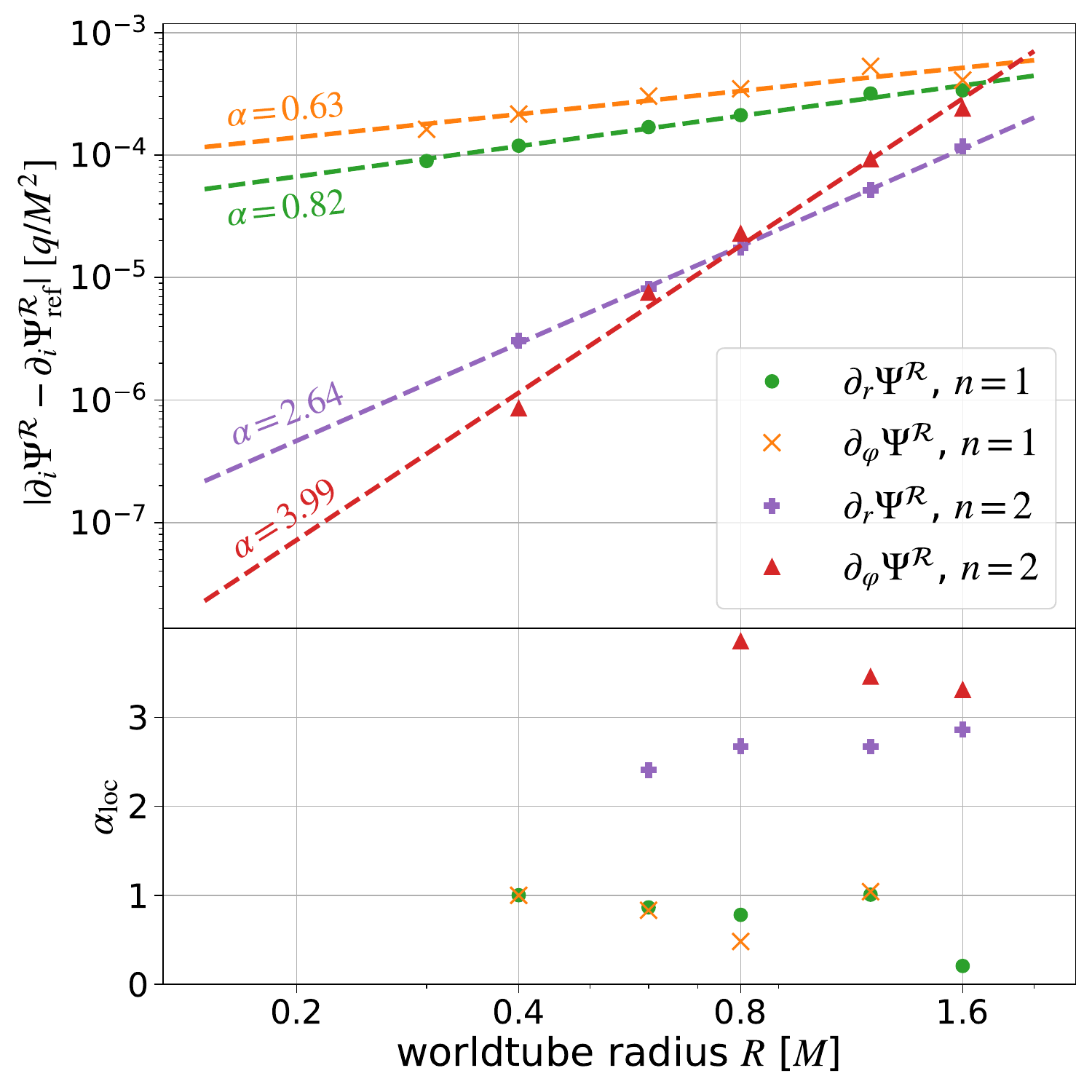}
	\caption{{\it Top panel:} The absolute difference between the radial and angular derivatives of the regular field at the position of the particle and the reference values of Eqs.~\eqref{eq: dx ref} and~\eqref{eq: dy ref}. Each cross represents the final value of a simulation. The dashed lines are best fits to the power-law relation $\propto R^{\alpha}$.	
		{\it Bottom panel:} The local convergence order between the simulations of adjacent worldtube radii as defined in Eq.~\eqref{eq: local convergence order}. The simulations at order $n = 1$ show a convergence order consistent with the predicted rate $\alpha = 1$. The $n = 2$ simulations show a higher convergence rate likely due to dominant higher order terms. Some anomalies are not visible.}
	\label{fig: derivatives regular field}
\end{figure} 

Figure~\ref{fig: derivatives regular field} compares the
  radial and azimuthal derivative of $\Psi^\mR$ obtained by our
  worldtube evolutions against these reference values.
  Shown are the differences from the reference values for orders $n=1$ and $2$,  with power-law fits  $\propto R^\alpha$. At zeroth order, the regular field is constant across the worldtube so the derivatives can not be computed.  The lower panel of Fig.~\ref{fig: derivatives regular field} plots the local convergence order $\alpha_{\mathrm{loc}}$ defined in \eqref{eq: local convergence order}.

 We argued in Eq.~\eqref{eq: predicted erro dpsiR} that the error of the regular field's derivatives at the particle position should scale with the worldtube radius as $\propto R^n$. For $n=1$, this behavior is confirmed by the local convergence $\alpha_{\mathrm{loc}}$ of our simulations with the exception of worldtube radius $R=1.6M$, which is anomalously lower than expected and skews the global convergence order. For $n=2$, the radial derivative $\partial_r \Psi^\mathcal{R}$ (linked to the conservative, time-symmetric piece of the self-force) shows a local convergence that approaches the expected order of $R^2$ at smaller worldtube radii. This suggests that the error is just entering the regime where the $\mathcal{O}(R^n)$ contribution becomes dominant. For the angular derivative $\partial_\varphi \Psi^\mathcal{R}$ (linked to the dissipative, time-antisymmetric piece of the self-force), the local convergence order is larger than 3 for all simulations. This could indicate that the error is still dominated by higher-order contributions at the sampled worldtube radii. Alternatively, it could indicate that dissipative quantities converge more rapidly with $R$ than conservative one, as suggested in Sec.~\ref{sec: error estimates}; however, the results for $n=1$ do not support that proposal, showing the same convergence rate for $\partial_\varphi \Psi^\mathcal{R}$ as for $\partial_r \Psi^\mathcal{R}$. We stress that in any case, the convergence is at least as rapid as predicted in Eq.~\eqref{eq: predicted erro dpsiR}.

\subsection{Solution along the $z$-axis}\label{sec: z axis}

The spherical symmetry of the Schwarzschild background allows for the Klein-Gordon Eq.~\eqref{eq: KG without source} to be decomposed into separately evolving spherical harmonic modes $\Psi_{l m}(r, t)$, where the spherical harmonic decomposition is centered on the black hole (different to the spherical harmonics introduced in Eq.~\eqref{eq:spherical harmonic expansion}, which are centered on the worldtube). On the polar axis ($x=y=0$) all modes vanish except the axially symmetric ones, i.e.~those with $m=0$. These modes are also static and admit simple analytical solutions \cite{Dhesi:2021yje}. Along the polar axis these solutions read 
\begin{equation}
	\begin{split}
		\Psi_{l0}(z) = &\frac{4 \pi \sqrt{1 - 3M/r_p}}{M} Y_{l0}(n^x)\times \\
		 &\;\;\Big(Q_l(r_p /M -1) P_l(z/M - 1) \Theta (r_p - z)   \\
		&\quad+Q_l(z/M - 1) P_l(r_p/M - 1) \Theta(z-r_p)\Big),
	\end{split}
\end{equation}
where $P_l$ and $Q_l$ are Legendre functions of the first and second kind, respectively, $n^x$ is the normal vector pointing in the direction of the $x$ coordinate axis, and $\Theta$ is the Heaviside function. The full solution along the $z$-axis is then given by
\begin{equation}\label{eq: z solution}
	\Psi_z(z) = q \sum_{l=0}^\infty \Psi_{l0}(z) Y_{l0}(n^z),
\end{equation}
where $n^z$ is normal vector pointing in the coordinate z direction. The expansion~(\ref{eq: z solution}) converges exponentially in $l$ everywhere except in the neighborhood of $z=r_p$, where the convergence is too slow to yield good results in practice. We therefore ignore this region and cut it out of plots when comparing with the analytical solution $\Psi_z$.

\begin{figure}
	\includegraphics[width=\columnwidth]{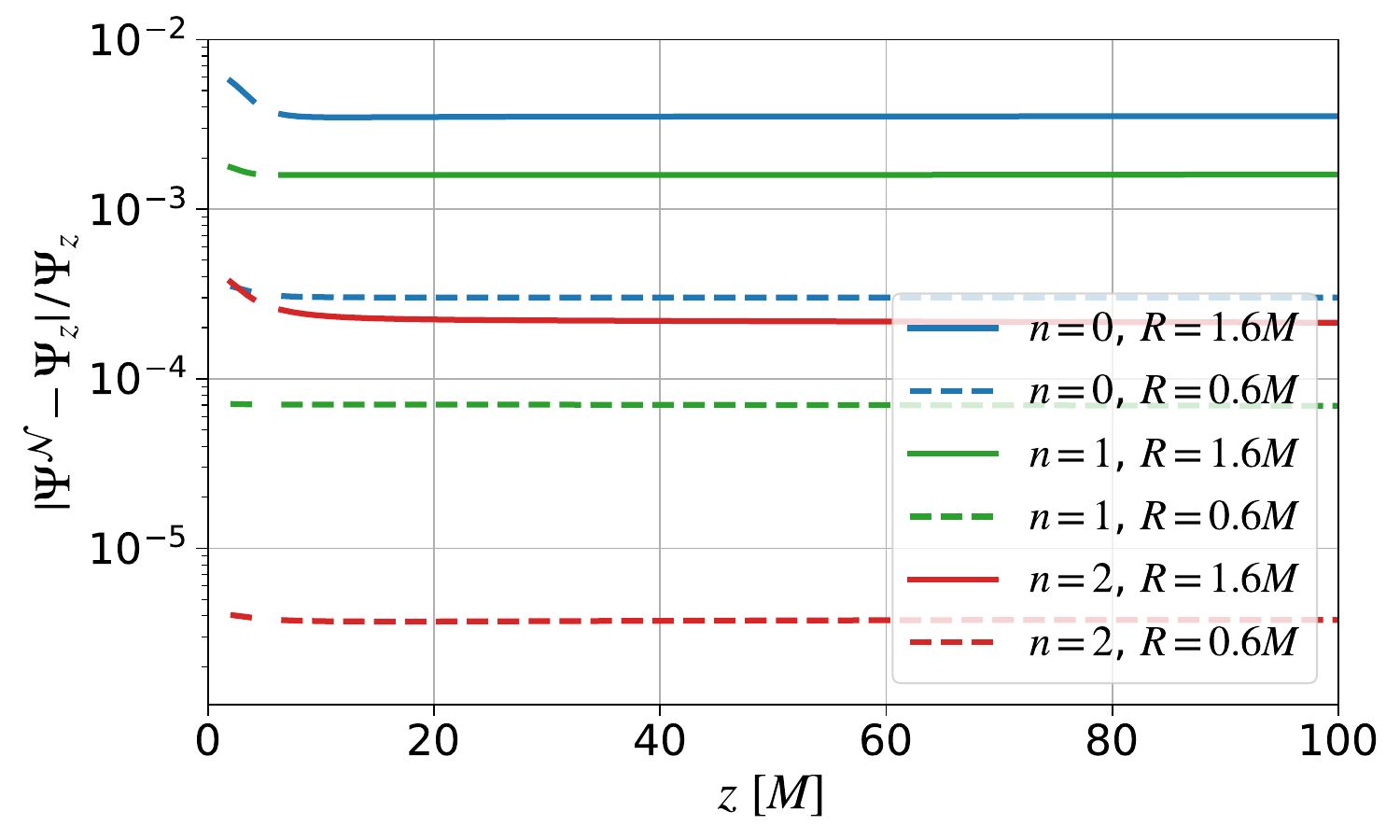}
	\caption{The relative error of the scalar field $\Psi$ along the $z$-axis compared to the analytical solution $\Psi_z$ given by Eq.~\eqref{eq: z solution}. We show two simulations with worldtube radii $1.6 M$ and $0.6 M$ for each order, 0, 1 and 2. The error decreases with higher order or smaller worldtube radius, as expected. A small region is cut out around $z=r_p=5M$, where Eq.~\eqref{eq: z solution} converges too slowly to be calculated to sufficient accuracy in practice. }
	\label{fig: z error}
\end{figure}

Figure~\ref{fig: z error} shows the relative error $|\Psi^\mN - \Psi_z| / \Psi_z$ between
  our numerical worldtube solutions $\Psi^\mN$ and Eq.~(\ref{eq: z solution}), computed at late evolution time, after $\Psi^\mN$ has settled into its steady state. The error is fairly constant along the axis. It is immediately clear that smaller $R$ and higher $n$ lead to improved agreement.

\begin{figure}
	\includegraphics[width=\columnwidth]{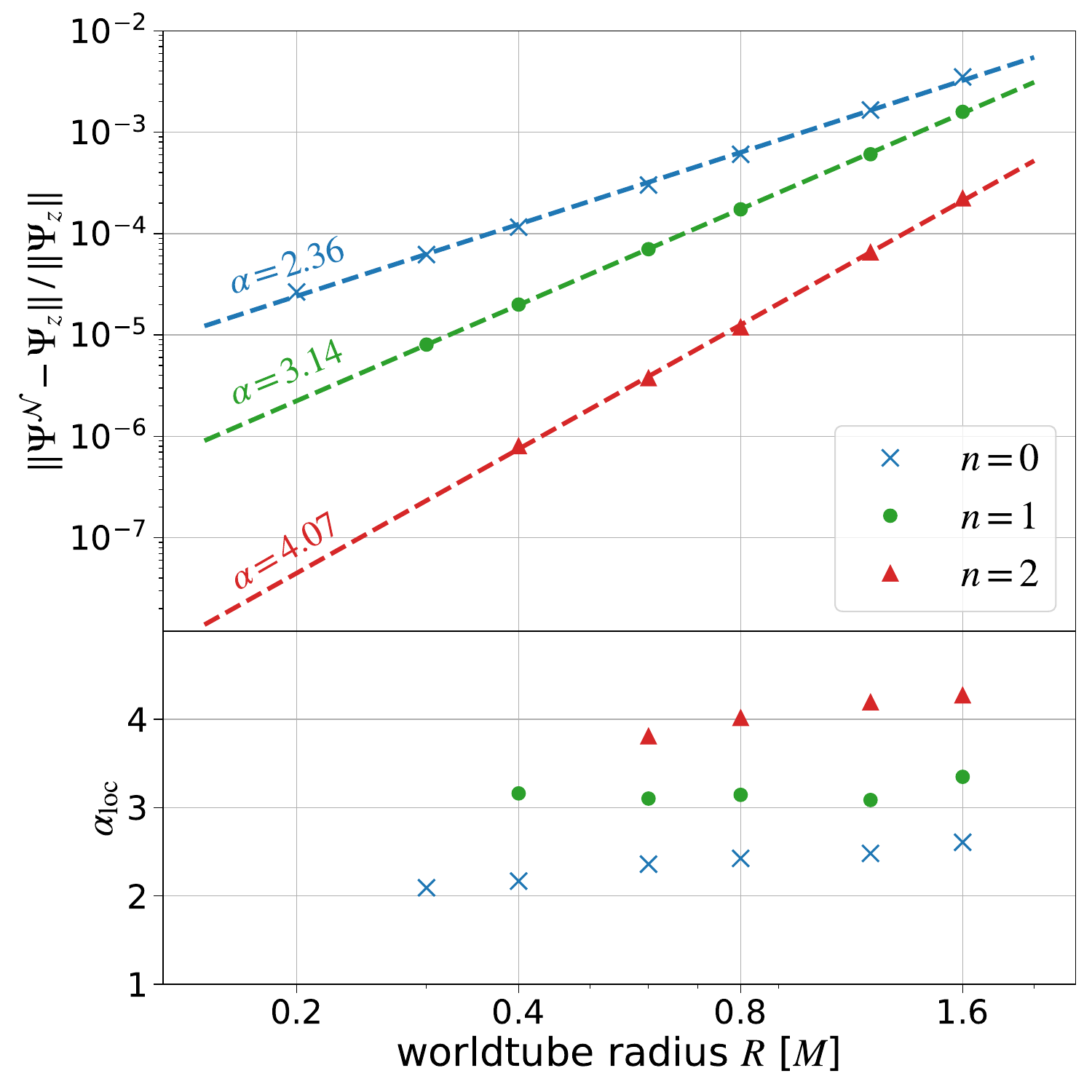}
	\caption{{\it Top panel:} The relative difference between the numerical, retarded field $\Psi^\mN$ along the $z$-axis and the analytical solution $\Psi_z$ given in Eq.~\eqref{eq: z solution}, integrated between $z = 10M$ and $z = 100M$ using the $L_1$-norm of Eq.~\eqref{eq: norm}. Each cross represents the final value of a simulation. The straight lines are a best fit to the power-law relation $\varepsilon \propto R^{\alpha}$.
	{\it Bottom panel:} The local convergence order between the simulations of adjacent worldtube radii as defined in Eq.~\eqref{eq: local convergence order}. The simulations with $n = 1$ and $n = 2$ show a constant convergence order consistent with the predicted rate $\alpha = n+2$. The $n = 0$ simulations show a higher convergence rate at larger worldtube radii but approach the expected value at smaller radii.}
	\label{fig: z integral}
\end{figure}

To investigate convergence with worldtube radius, we define the $L_1$-norm 
\begin{equation}\label{eq: norm}
	\Vert f(x) \Vert := \int_{10M}^{100M} |f(x)| dx,
\end{equation}  
which we use to integrate the relative error shown in Fig.~\ref{fig: z error} between $z = 10M$ and $z = 100M$ for each simulation.
Using this norm, the top panel of Fig.~\ref{fig: z integral} plots the relative differences
  between the analytical solution Eq.~(\ref{eq: z solution}) and numerical solutions $\Psi^\mN$ using various $R$ and $n$ and evaluated at late time in steady state.
Each symbol represents the integrated, relative error of a simulation's final value. Also plotted is a best fit of the error convergence $\propto R^{\alpha}$, where $R$ is the worldtube radius and $\alpha$ is the global convergence order. The lower panel of Fig.~\ref{fig: z integral} shows the local convergence order $\alpha_{\mathrm{loc}}$ as defined in Eq.~\eqref{eq: local convergence order}.

As explained in Section~\ref{sec: error estimates}, we expect the convergence order $\alpha = n+2$ in the volume outside the worldtube.
At order $n=2$, the global convergence order is best fit to $\alpha = 4.07$ which matches the predicted error. Order 1 has a fitted global convergence order of 3.14 and a local convergence order close to this value across the worldtube radii sampled. For the zeroth-order expansion, a global value of 2.33 is calculated, but the local order consistently decreases with smaller worldtube radii, which suggests that the error might still get contributions from higher-order terms at the larger worldtube radii, similar to the zeroth-order expansion in Fig.~\ref{fig: regular field}.

\subsection{Solution along the $x$-axis}\label{sec: x axis}

  The tests of our method so far compared to previously known
  data, either at the position of the charge or on the $z$-axis.  We now evaluate the
  convergence with worldtube radius in the volume, at locations where no analytic solution is available.
  To this end, we evaluate our numerical solutions $\varPsi^\mN$ along the co-rotating $\bar x$-axis,
  which passes through the center of the Schwarzschild black hole and the point charge. The settled field along this axis is shown as the blue curve in Figure~\ref{fig: fields along axes}. The simulation with $n = 2$ and $R = 0.4M$ is used as a reference solution, denoted $\Psi_{\mathrm{ref}}$, since it has the lowest error inside the worldtube and along the $z$-axis, as demonstrated above.

\begin{figure}
	\includegraphics[width=\columnwidth]{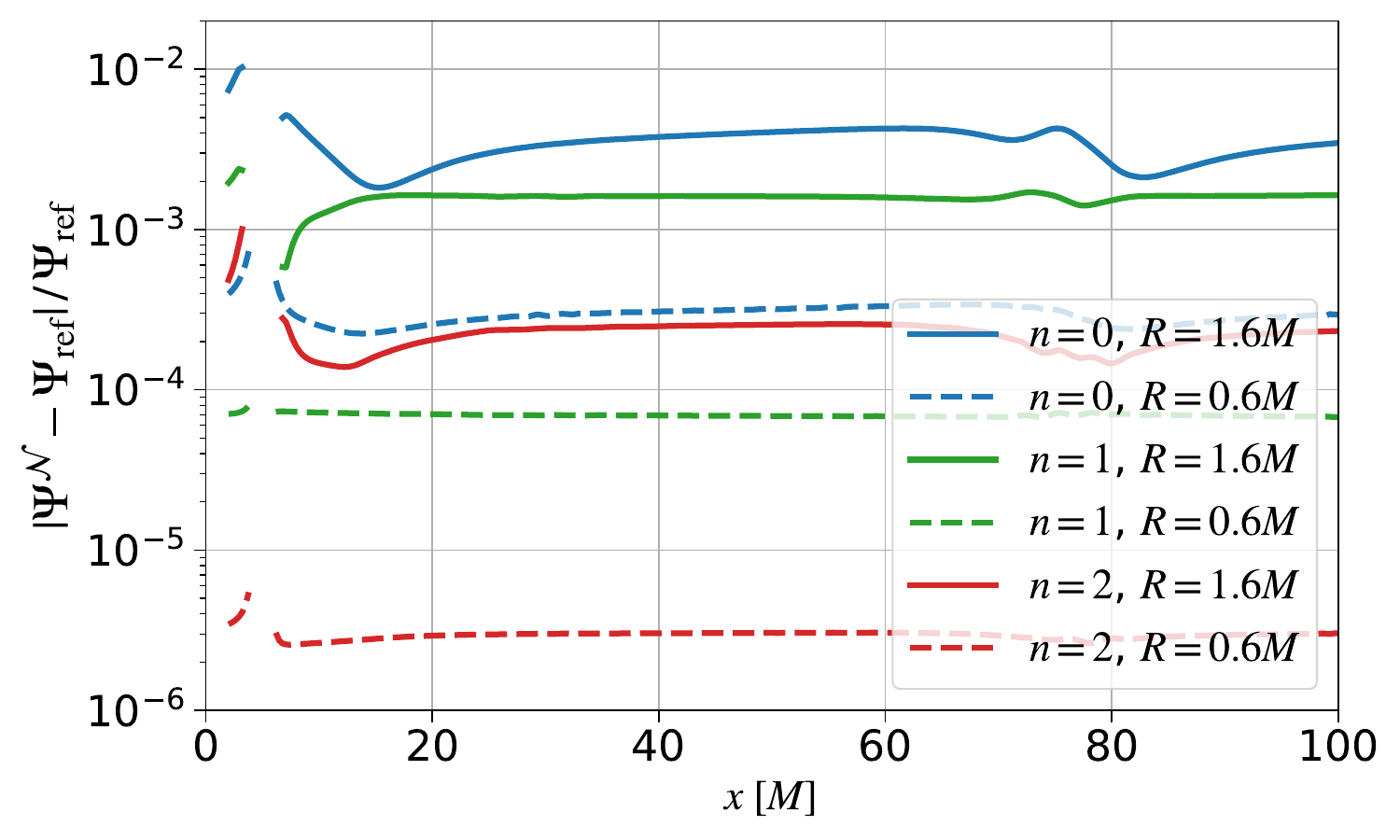}
	\caption{The relative error along the $x$-axis between $x = 10 M$ and $x = 100 M$ for two sample simulations of each order. The worldtube is centered at $x = 5 M$ and cut out from the plots.}
	\label{fig: x error}
\end{figure}

Figure~\ref{fig: x error} shows the relative difference with respect to the reference solution between $x = 1.9M$ and $x = 100M$ for two sample simulations at each order. The field along the $x$-axis has more features as it lies in the orbital plane of the charge. The error along the $x$-axis is therefore not quite as smooth as along the $z$-axis; for instance at $x\approx 80M$ some features are apparent, which coincide to a wave-crest in $\Psi^\mN$ (see Fig. \ref{fig: fields along axes}). The error decreases with both a higher expansion order $n$ and a decreasing worldtube radius $R$ as is expected.
\begin{figure}
	\includegraphics[width=\columnwidth]{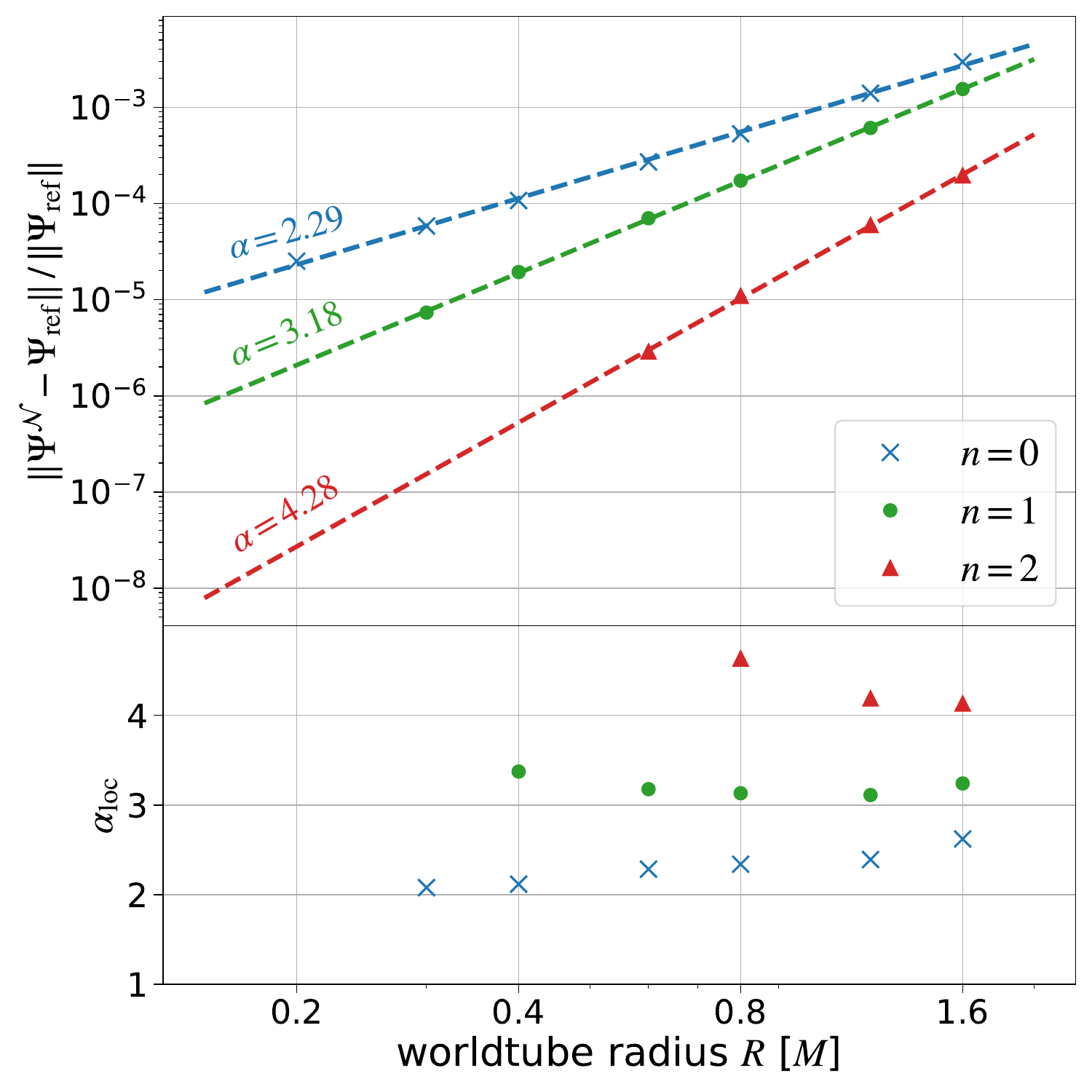}
	\caption{{\it Top panel:} The relative error $\varepsilon$ integrated along the co-moving x-axis compared to a reference solution $\Psi_{\mathrm{ref}}$ with $n = 2$ and $R = 0.4M$. Each cross represents a simulation, and the straight lines are a best fit of the relation $\varepsilon \propto R^\alpha$.
	{\it Bottom panel:} The local convergence order as defined in Eq.~\eqref{eq: local convergence order}. At first and second order, the simulations reproduce the expected convergence order of $\alpha = n+2$. At zeroth order, the local convergence approaches the expected order for smaller worldtube radii, likely indicating that the error still has contributions from higher-order terms in this regime.
}
	\label{fig: x integral}
\end{figure}

To quantify the convergence with respect to $R$ we
compute the norm Eq.~(\ref{eq: norm}) integrated along the co-moving x-axis, $\Vert \Psi^\mN(x) - \Psi_{\mathrm{ref}}(x) \Vert$.
This difference, normalized, is plotted in Fig.~\ref{fig: x integral}, where each marker represents an individual simulation. The straight lines show power law fits $\propto R^\alpha$. In the bottom panel we show the local convergence order $\alpha_{\mathrm{loc}}$ as defined in Eq.~\eqref{eq: local convergence order}. The convergence rates in worldtube radius $R$ are close to the expectation from Eq.~(\ref{eq: predicted erro psiN}), $\alpha = n+2$,  with global convergence order $\alpha$ equal to 2.25, 3.18, 4.29 for orders 0, 1 and 2, respectively. For $n=0$, the local convergence order $\alpha_{\mathrm{loc}}$ is steadily decreasing with the worldtube radius $R$ towards the expected value of $\alpha = 2$, which suggests it is just entering the convergent regime here. The local convergence rate of the $n = 2$ simulations appears to jump slightly at the smallest worldtube radius sampled, which we attribute to numerical error.

\section{Conclusions} \label{sec:Conclusions}
In this paper we continue the work of \cite{Dhesi:2021yje} and explore a novel approach to simulating high mass-ratio binary black holes. A large region (worldtube) is excised from the numerical domain around the smaller black hole, to alleviate the limiting CFL condition due to small grid spacing in this region. The solution inside the worldtube is represented by a perturbative approximation that is determined by the numerical solution on the boundary and in turn provides boundary conditions to the numerical evolution.

We test this method using the toy problem of a scalar charge in circular orbit around a central black hole. The simulations are carried out in 3+1D using SpECTRE, the new discontinuous Galerkin code developed by the SXS collaboration. In order to develop algorithms that generalize to the full GR problem, we do not decompose our solution into spherical harmonics as is the usual approach. We split the solution near the scalar charge into a puncture field, which is fully determined as a local expansion in Sec.~\ref{sec: model}, and a regular field, which is a smooth Taylor series with undetermined coefficients. The expansion coefficients in the regular field are determined by (i) the numerical solution on the worldtube boundary and (ii) the scalar wave equation as described in Sec.~\ref{sec: matching method n=2}. Our puncture is constructed from the Detweiler-Whiting singular field, allowing us to calculate the scalar self-force from our regular field.

We implement the described matching scheme for orders $n = 0, 1$ and $2$ and perform numerical simulations for a circular orbit of radius $r_p=5M$ for a variety of worldtube radii.   In Sec.~\ref{sec: error estimates} we make a theoretical argument for how the error introduced by the excision should converge with the excision radius in- and outside the worldtube. We confirm these results in Sec.~\ref{sec: Results} and show that the scheme solves the scalar wave equation with high accuracy even at relatively large worldtube radii. We further validate our method by comparing against known values of the Detweiler-Whiting regular field and its first derivatives on the particle's worldline.

\begin{figure}
	\includegraphics[width=\columnwidth]{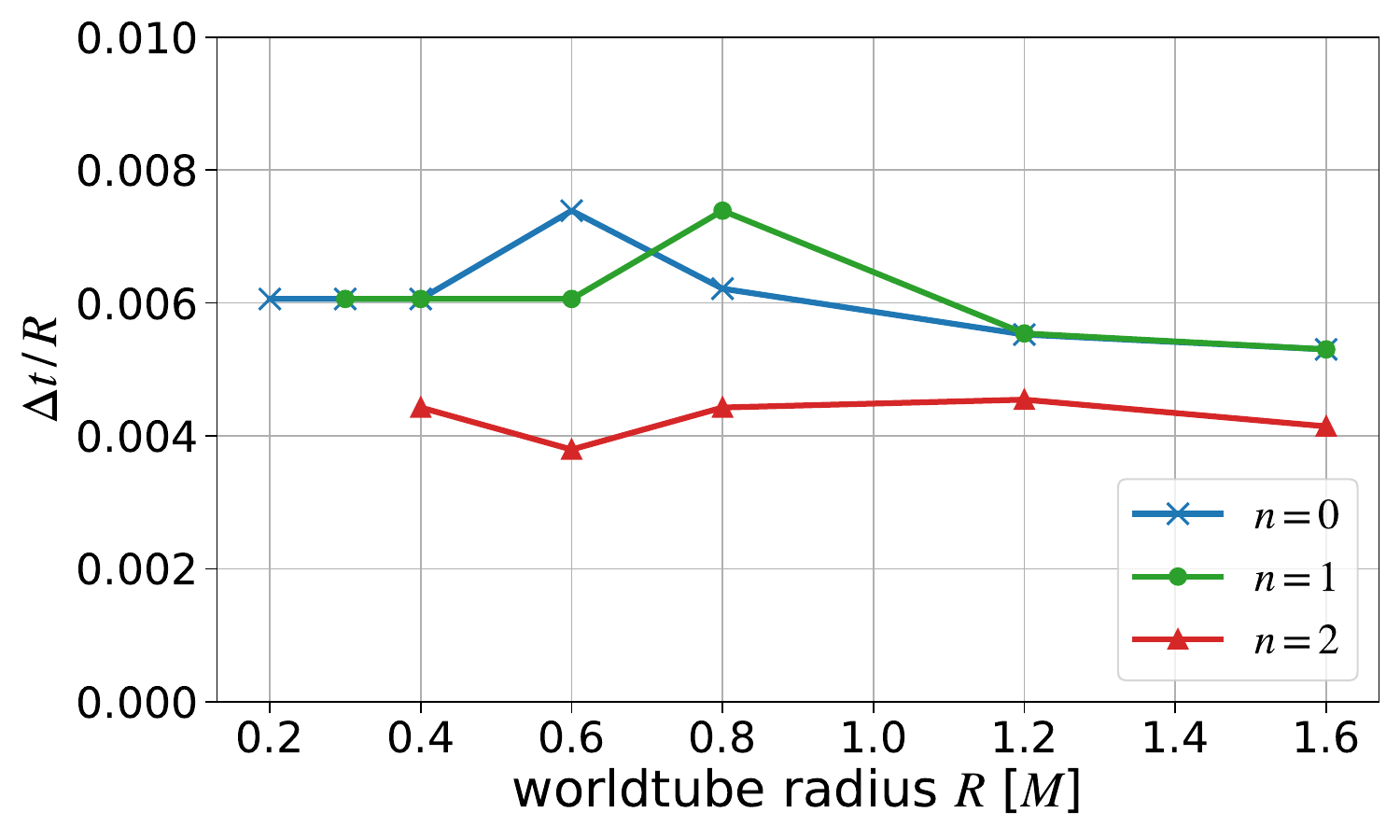}
  \caption{\label{fig:timestep}Time steps $\Delta t$ of the worldtube simulations presented here.}
\end{figure}
The ultimate goal of the worldtube method is to speed up BBH simulations at large mass-ratios by alleviating the CFL condition.  Figure~\ref{fig:timestep} considers the time steps sizes $\Delta t$ taken by our primary simulations.  Plotted is $\Delta t/R$ vs the worldtube radius $R$.  Note that the resolutions of our simulations were adjusted such that for each simulation, numerical truncation error is subdominant compared to the worldtube error, resulting in differences in $\Delta t$ for simulations with different expansion orders at the same worldtube radius.  Nevertheless, it is apparent that for fixed order, the time step is roughly proportional to the worldtube radius.  Therefore, the promise that larger worldtube radii allow larger time steps $\propto R$ indeed holds. Ideally, the worldtube error should be comparable to or somewhat smaller than the NR error. Our results show that this can be achieved by either decreasing the worldtube radius or by increasing the expansion order.  The former, of course, would lead to a smaller grid-spacing and a more significant CFL condition, whereas the latter has no noticeable performance cost. We have discussed the next steps towards tackling BBH simulations using the worldtube method in \cite{Dhesi:2021yje}.

Before tackling the full BBH problem, our next step will be to include the back-reaction of the scalar field onto the charged particle \cite{Diener:2012}. In the present work we have computed the first derivatives of the Detweiler-Whiting regular field, from which we can construct the scalar self-force, but we have so far ignored the effect of that force. Once it is accounted for, the equations of motion for the scalar charge $q$ of bare mass $\mu_0$ are given by \cite{Quinn_2000}
\begin{align}
	u^{\beta} \nabla_\beta(\mu u_\alpha) &= q \partial_\alpha \Psi^\mR, \\ 
	\mu &= -q \Psi^\mR + \mu_0.
\end{align}
Allowing the particle's trajectory to evolve dynamically in this way will be an important step toward the full gravity problem. SpECTRE uses a series of control systems which adjust certain time-dependent parameters of smooth coordinate maps to deform the grid~\cite{Scheel:2006}. While they usually ensure that excision spheres stay inside a black hole's apparent horizon, we will use these control systems to enable the worldtube to track the inspiraling scalar charge. The control systems are needed for the binary black hole case and such a scalar evolution will ensure they work as expected.

\begin{acknowledgments}
We thank Benjamin Leather for computing a reference value of the angular derivative of the regular field at the particle position. This work was supported in part by the Sherman Fairchild Foundation and by NSF Grants No.~PHY-2011961, No.~PHY-2011968, and No.~OAC-1931266 at Caltech, by NSF Grants No.~PHY-1912081 and No.~OAC-1931280 at Cornell, and by NSF Grants No.~PHY-1654359 and No.~PHY-2208014 at Cal State Fullerton. A.P. acknowledges the support of a Royal Society University Research Fellowship. P.K.’s research was supported by the Department of Atomic Energy, Government of India; and by the Ashok and Gita Vaish Early Career Faculty Fellowship at the International Centre for Theoretical Sciences.  SpECTRE uses \texttt{Charm++}/\texttt{Converse}~\cite{laxmikant_kale_2021_5597907,kale1996charm++}, which was developed by the Parallel Programming Laboratory in the Department of Computer Science at the University of Illinois at Urbana-Champaign. H. R. R. acknowledges support from the Fundação para a Ciência e Tecnologia (FCT) within the projects UID/04564/2021, UIDB/04564/2020, UIDP/04564/2020 and EXPL/FIS-AST/0735/2021. SpECTRE uses \texttt{Blaze}~\cite{Blaze1,Blaze2}, \texttt{HDF5}~\cite{hdf5}, the GNU Scientific Library (\texttt{GSL})~\cite{gsl}, \texttt{yaml-cpp}~\cite{yamlcpp}, \texttt{pybind11}~\cite{pybind11}, \texttt{libsharp}~\cite{libsharp}, and \texttt{LIBXSMM}~\cite{libxsmm}. The figures in this article were produced with \texttt{matplotlib}~\cite{Hunter:2007, thomas_a_caswell_2020_3948793}, \texttt{NumPy}~\cite{harris2020array}, and \texttt{ParaView}~\cite{paraview,paraview2}. The authors thank Charlie Vu for helpful discussion.

\end{acknowledgments}

\appendix
\section{Constraint-preserving boundary conditions}\label{sec: constraint preserving BCs}

The domain features two boundaries which require boundary conditions on the characteristic fields flowing into the domain: the worldtube boundary and the outer boundary. For the fields $Z^1$ and $Z^2_i$ (arising from reduction to first-order form), we use constraint-preserving boundary conditions which are formulated analogously to \cite{Kidder:2005} and applied using the Bjorhus condition. We begin by rewriting Eq.~(\ref{eq:NR-C1}) in terms of the characteristic fields with respect to a boundary with normal vector $\hat{n}^i$,
\begin{align}
	C_i &= \partial_i \Psi - \Phi_i \\
  \label{eq:C1-in-char-fields}
	&= \partial_i Z^1 - \frac{1}{2}(U^{+} - U^-) \hat{n}_i - Z_i^2.
\end{align}
The normal component of this constraint is given by
\begin{equation} \label{eq: normal one constraint}
	\hat{n}^i C_i = \hat{n}^i \partial_i Z^1 - \frac{1}{2}(U^{+} - U^-) - \hat{n}^i Z_i^2 .
\end{equation}
Vanishing of the constraints implies in particular that the normal component vanishes, $\hat{n}^iC_i=0$,
we interpret as a boundary condition on $Z^i$:
\begin{equation} \label{eq: one constraint BC}
	(\hat{n}^i \partial_i Z^1)_{BC}  = \frac{1}{2}(U^{+} - U^-) + \hat{n}^i Z_i^2.
\end{equation}
Applying the same procedure to Eq.~(\ref{eq:NR-C2}) yields
\begin{equation} \label{eq: normal two constraint}
	\hat{n}^i C_{ij} = \hat{n}^i \partial_i Z^2_j + \frac{1}{2}\hat{n}^i \hat{n}_{\!j}\, \partial_i(U^+\! - U^-) - \frac{1}{2}\partial_{\!j}\,(U^+\! - U^-) - \hat{n}^i \partial_j Z^2_i
\end{equation}
and
\begin{equation}\label{eq: two constraint BC}
	(\hat{n}^i \partial_i\, Z^2_j)_{BC} = - \frac{1}{2}\hat{n}^i \hat{n}_{\!j}\,\partial_i (U^+ \!- U^-) + \frac{1}{2}\partial_{\!j}\,(U^+\! + U^-) + \hat{n}^i \partial_j Z^2_i.
\end{equation}

In order to implement Eqs.~(\ref{eq: one constraint BC}) and (\ref{eq: two constraint BC}), we return to the evolution equations in first order form,
\begin{equation}\label{eq: non-conservative system}
	\partial_t \psi^\alpha + A^{i \alpha}_\beta \partial_i \psi^\beta = F^\alpha.
\end{equation}
Projecting onto the characteristic fields, one finds
\begin{align}
	e^{\hat{a}}{}_a (\partial_t \psi^a + A^{i a}_b \partial_i \psi^b) &= e^{\hat a}{}_a F^a, \\
	\partial_t \psi^{\hat{a}} + e^{\hat{a}}{}_a A^{i a}_b(P^k_i + \hat{n}^k \hat{n}_i) \partial_k \psi^b &=  e^{\hat a}{}_a F^a, \\
	\partial_t \psi^{\hat{a}} + v_{(\hat{a})} \hat{n}^k \partial_k \psi^{\hat{a}} +  e^{\hat{a}}{}_a A^{i a}_b P^k_i \partial_k \psi^b &=  e^{\hat a}{}_a F^a.
\end{align}

Boundary conditions are now applied by modifying the term
  $v_{(\hat a)}\hat{n}^k\partial_k\psi^{\hat{a}}$:
Iff $v_{(\hat a)}\!<\!0$ at a grid-point on the boundary, then the following modified evolution equation is used at that grid-point:
  \begin{align}\label{eq: Bjorhus}
	d_t \psi^{\hat{a}} = D_t \psi^{\hat{a}} + v_{(\hat{a})} \left(\hat{n}^i\partial_i \psi^{\hat{a}} - (\hat{n}^i\partial_i \psi^{\hat{a}})_{BC}\right).
\end{align}
Here
  \begin{equation}
	D_t \psi^{\hat{a}} \equiv - e^{\hat{a}}{}_a A^{i a}_b \partial_i \psi^b + e^{\hat a}{}_aF^a 
\end{equation}
represents the volume time-derivative of the characteristic fields.
  In other words, the time-derivative arising from the volume equations is corrected with a term that ensures the desired boundary condition. Summing Eqs.~\eqref{eq: normal one constraint} and ~\eqref{eq: one constraint BC} yields
\begin{equation}
	\hat{n}^i \partial_i Z^1 - (\hat{n}^i \partial_i Z^1)_{BC} = \hat{n}^i C_i.
\end{equation}
Analogously, combining Eqs.~\eqref{eq: normal two constraint} and~\eqref{eq: two constraint BC} results in
\begin{equation}
	\hat{n}^i \partial_i Z^2_j - (\hat{n}^i \partial_i Z^2_j)_{BC} = \hat{n}^i C_{ij}.
\end{equation}
Finally, we insert this into the Bjorhus condition~\eqref{eq: Bjorhus} to obtain
\begin{align}
	d_t Z^1 &= D_t Z^1 + v_{Z^1}\hat{n}^i C_i, \\
	d_t Z^2_i &= D_t Z^2_i + v_{Z^2}\hat{n}^i C_{ij},
\end{align}
where boundary corrections are only imposed when the corresponding characteristic speeds are negative.

\section{Matching method at arbitrary order}
\label{sec: matching method arbitrary n}

The main text in Sec.~\ref{sec: matching method n=2}
  develops our matching scheme up to order $n=2$. Here we show how this can be generalized to arbitrary order $n$. 

First, we introduce some notation and useful identities. We make use of multi-index notation according to \cite{Poisson:2014} where a capital index $L$ stands for a collection of $l$ indices,
\begin{equation}
	A_L = A_{k_1 k_2 \cdots k_l}.
\end{equation}
The tensor product of $l$ coordinate vectors or $l$ normal vectors is abbreviated as
\begin{align}
	x^L = x^{k_1} x^{k_2} \cdots n^{k_l}, \\
	n^L = n^{k_1} n^{k_2} \cdots n^{k_l},
\end{align}
and the tensor product of $l$ Kronecker symbols is written as
\begin{equation}
	\delta^{2L} = \delta^{k_1 k_2} \delta^{k_3 k_4} \cdots \delta^{k_{2l-1} k_{2l}}.
\end{equation}

Symmetric, trace-free (STF) tensors are written with angular brackets around the indices. The combination of $l$ STF normal vectors is defined as \cite{Thorne:1980}
\begin{align}
	n^{\langle L \rangle} &= n^{\langle k_1} \cdots n^{k_l \rangle} =  \sum_{k=0}^{\lfloor l/2\rfloor} \tilde{c}_k^{\,l} \delta^{(2K} n^{L - 2K)}, \label{eq:symmetrized-n}\\
  \intertext{where}
	\tilde{c}^{\,l}_k &= (-1)^k \frac{l! (2l - 2k - 1)!!}{(2l-1)!! (l-2k)! (2k)!!}.
\end{align}
The parentheses in Eq.~(\ref{eq:symmetrized-n}) indicate indices to be symmetrized and $\lfloor l/2\rfloor$ is the largest integer less than or equal to $l/2$.
The inverse expression is given by \cite{Blanchet:1986}
\begin{align}
	n^{L} &= n^{ k_1} \cdots n^{k_l} =  \sum_{k=0}^{{\lfloor l/2\rfloor}} c_k^l \delta^{(2K} n^{\langle L - 2K \rangle )} \label{eq: normal from STF} \\ 
\intertext{with}
	c^l_k &= \frac{l! (2l-4k+1)!!}{(2l-2k+1)!!(l-2k)! (2k)!!}.
\end{align}
The $n^{\langle L \rangle}$ provide an orthogonal basis for functions on a sphere, and each $n^{\langle L \rangle}$ is an eigenfunction of the Laplacian $\nabla^2:=\delta^{ab}\partial_a\partial_b$, satisfying $\nabla^2 n^{\langle L \rangle} = -\frac{\ell(\ell+1)}{\rho^2} n^{\langle L \rangle}$. For a fixed $l$, the STF tensors $n^{\langle L \rangle}$  span the same functions as the set of spherical harmonics $Y_{lm}(\theta, \phi)$ of rank $l$. This can be seen by expressing the normal vector as $n^i = (\sin \theta \cos \phi,\ \sin \theta \sin \phi,\ \cos \theta)$, leading to \cite{Poisson:2014}

\begin{subequations}
	\label{eq: STF Ylm relations}
	\begin{align}
		Y_{lm} &= \mathcal{Y}_{lm}^{*\langle L \rangle} n_{\langle L \rangle} \label{eq: Ylm to STF}, \\ 
		n^{\langle L \rangle} &= N_l \sum_{m=-l}^l \mathcal{Y}_{lm}^{\langle L \rangle} Y_{lm}, \\
    \intertext{where}
		\mathcal{Y}_{lm}^{\langle L \rangle} &\coloneqq \frac{1}{N_l} \int_{S^2} n^{\langle L \rangle} Y^*_{lm} d \Omega, \\
		N_l &\coloneqq \frac{4 \pi l!}{(2l+1)!!}.
	\end{align}
\end{subequations}

We start by expanding the regular scalar field $\Psi^{\mR}(t, x^i)$ and its time derivative in a power series around the charge's position to arbitrary order $n$ as shown in Eq.~\eqref{eq: power series}. We will show that all free components of the expansion can be uniquely determined at each time step from (i) numerical data from the worldtube boundary and (ii) the Klein-Gordon equation~\eqref{eq: KG without source}.

\subsection{Worldtube boundary data}

We carry the Taylor expansion of the regular field given in Eq.~\eqref{eq:3dregular} to $n$th order and give an analogous expansion for the time derivative
\begin{subequations}
	\label{eq: power series}
	\begin{align}
		\Psi^\mR(t,x^\ibar) &= \sum_{l=0}^{n}  \Psi^\mR_{\bar k_1 \cdots \bar k_n}(t) x^{\bar k_1} \cdots x^{\bar k_n} + \Order(\rho^{n+1}), \label{eq: power series psi} \\
		\partial_t \Psi^\mR(t,x^\ibar) &= \sum_{l=0}^{n} \dot{ \Psi}^\mR_{\bar k_1 \cdots \bar k_n}(t) x^{\bar k_1} \cdots x^{\bar k_n} + \Order(\rho^{n+1}), \label{eq: power series dt}
	\end{align}
\end{subequations}
 which has $\frac{1}{2}\sum_{i=0}^n (i+2)(i+1) = \frac{1}{6}(n+3)(n+2)(n+1)$ components. 

The continuity condition at the worldtube boundary is given by Eq.~\eqref{eq: continuity condition}. Both sides of this equation can be expressed in a basis of STF normal vectors.  The regular field $\Psi^\mR(t_s, x^\ibar)$ on the left is transformed using Eq.~\eqref{eq: normal from STF} to give
\begin{align}
	\Psi^\mR(t_s, x^i) &= \sum_{l=0}^{n} \rho^l \Psi^\mR_{\bar{L}} (t_s) \sum_{k=0}^{\lfloor l/2\rfloor} c_k^l \delta^{(2\bar{K}} n^{\langle \bar{L}-2\bar{K} \rangle )} \\
	&= \sum_{l=0}^{n} \sum_{k=0}^{\lfloor l/2\rfloor} \rho^l c_k^l \Psi^\mR_{\bar{L} - 2\bar{K} \ibar_1 \ibar_1 \cdots \ibar_k \ibar_k} (t_s) n^{\langle \bar{L} - 2\bar{K} \rangle }. \label{eq: regular field STF} 
\end{align}
As in Sec.~\ref{sec: matching method n=2}, the right-hand side of Eq.~\eqref{eq: continuity condition} is calculated by projecting the numerical, regular field onto spherical harmonics up to order $n$ to obtain the coefficients $a_{lm}^{\mN, \mR}$. This expansion is then transformed to STF normal vectors using Eq.~\eqref{eq: Ylm to STF}, 
\begin{align}
	\sum_{l=0}^n\Psi^{\mN, \mR}_{\langle \bar{L} \rangle} (t_s) n^{\langle \bar{L} \rangle}  &=  \sum_{l=0}^n \sum_{m= -l}^l a_{lm}(t_s) Y_{lm} (x^\ibar). \label{eq: NR STF}
\end{align}
The orthogonality of the STF normal vectors allows us to match Eqs.~\eqref{eq: regular field STF} and \eqref{eq: NR STF} order by order, yielding a system of algebraic equations:
\begin{align}\label{eq: continuity STF}
	\Psi^{\mN, \mR}_{\langle \bar{L} \rangle}(t_s) = \sum_{k=0}^{\lfloor \frac{n - l}{2} \rfloor} \rho^{l + 2k} c_k^{l + 2k} \Psi^\mR_{\bar{L} \ibar_1 \ibar_1 \cdots \ibar_k \ibar_k}(t_s), \quad 0 \leq l \leq n.
\end{align}
Each tensor component in the set $\left\{ \Psi^{\mN, \mR}_{\langle \bar{L}\rangle}\right\}_l$ with $0 \leq l \leq n$ fixes one degree of freedom of the Taylor coefficients $\left\{\Psi^\mR_{\bar{L}} \right\}_l$ for a total of $(n+1)^2$ equations. The matching equations for the time derivative coefficients $\left\{\dot{\Psi}^\mR_{\bar{L}} \right\}_l$ are derived completely analogously. 

\subsection{Klein-Gordon equation}
The remaining components of $\left\{\Psi^\mR_{\tilde{L}} \right\}_l$ are fixed by the Klein-Gordon equation~\eqref{eq: KG}. The metric quantities $g^{\mu \nu}$ and $\Gamma^\mu$ are expanded to the same order $n$ as the regular field in the grid frame $x^\ibar$,
\begin{align}
	g^{\bar \mu \bar \nu}(t_s,x^\ibar) &= \sum_{l=0}^{n} 	 g^{\bar \mu \bar \nu}_{\bar L}\, x^{\bar L} + \Order (\rho^{n+1}), \\ 
	\Gamma^{\bar \mu}(t_s,x^\ibar) &= \sum_{l=0}^{n}  \Gamma^{\bar \mu}_{\bar L} \,  x^{\bar L} + \Order (\rho^{n+1}),
\end{align}
where the expansion coefficients are given by
\begin{align}
	g^{\bar \mu \bar \nu}_{\bar L} \coloneqq \frac{1}{l!} \partial_{\bar L} g^{\bar \mu \bar \nu}(t_s,x_p^\ibar), \\
	\Gamma^{\bar \mu}_{\bar L}  \coloneqq \frac{1}{l!} \partial_{\bar L}  \Gamma^{\bar \mu}(t_s,x_p^\ibar).
\end{align}
Here, recall, $x_p^\ibar=(r_p,0,0)$. Substituting these expansions into the Klein-Gordon equation, we obtain
\begin{multline}
	0=\left( \sum_{l=0}^{n} g^{\bar \mu \bar \nu}_{\bar L}  x^{\bar L}  \right) \partial_{\bar \mu} \partial_{\bar \nu} \left( \sum_{l=0}^{n} \Psi^\mR_{\bar L} x^{\bar L} \right) \\
	+ \left(\sum_{l=0}^{n} \Gamma^{\bar \mu}_{\bar L}  x^{\bar L} \right) \partial_{\bar \mu} \left( \sum_{l=0}^{n} \Psi^\mR_{\bar L } x^{\bar L} \right).
\end{multline}
We now split the partial derivative into its time part $\partial_t$ and spatial part $\partial_\ibar$ and solve order by order in $\rho$. The $k$th-order equation reads
\begin{equation}\label{eq: perturbed KG}
	\begin{aligned}
		0 &= \sum_{l=0}^{k} \left( g_{\Kbar-\Lbar}^{tt} \ddot{\Psi}^\mR_\Lbar + 2 (l+1) g_{\Kbar-\Lbar}^{t \ibar} \dot{\Psi}^\mR_{\Lbar \ibar} \right.\\
		&\quad+ (l+2) (l+1) g_{\Kbar-\Lbar}^{\ibar \jbar} \Psi^\mR_{\Lbar \ibar \jbar} - \Gamma_{\Kbar-\Lbar}^t \dot{\Psi}^\mR_\Lbar \\
		&\quad \left. - (l+1) \Gamma_{\Kbar-\Lbar}^\ibar \Psi^\mR_{\Lbar \ibar} \right) \qquad 0 \leq k \leq n-2, 
	\end{aligned}
\end{equation}
where we have made use of the identity $\partial_\ibar a_{\bar{L}}(t) x^{\bar L} = l a_{\bar{ L} -1 \ibar}(t) x^{\Lbar-1}$ for $a_\Lbar$ completely symmetric $a_{(\Lbar)} = a_\Lbar$. The set of equations~\eqref{eq: perturbed KG} fixes $\frac{1}{6} (n+1)n(n-1)$ components of $\Psi^\mR$ which, when combined with equations Eq.~\eqref{eq: continuity STF}, fixes all components of the expansion~\eqref{eq: power series}. 

\bibliography{references}

\begin{thebibliography}{79}%
\makeatletter
\providecommand \@ifxundefined [1]{%
 \@ifx{#1\undefined}
}%
\providecommand \@ifnum [1]{%
 \ifnum #1\expandafter \@firstoftwo
 \else \expandafter \@secondoftwo
 \fi
}%
\providecommand \@ifx [1]{%
 \ifx #1\expandafter \@firstoftwo
 \else \expandafter \@secondoftwo
 \fi
}%
\providecommand \natexlab [1]{#1}%
\providecommand \enquote  [1]{``#1''}%
\providecommand \bibnamefont  [1]{#1}%
\providecommand \bibfnamefont [1]{#1}%
\providecommand \citenamefont [1]{#1}%
\providecommand \href@noop [0]{\@secondoftwo}%
\providecommand \href [0]{\begingroup \@sanitize@url \@href}%
\providecommand \@href[1]{\@@startlink{#1}\@@href}%
\providecommand \@@href[1]{\endgroup#1\@@endlink}%
\providecommand \@sanitize@url [0]{\catcode `\\12\catcode `\$12\catcode
  `\&12\catcode `\#12\catcode `\^12\catcode `\_12\catcode `\%12\relax}%
\providecommand \@@startlink[1]{}%
\providecommand \@@endlink[0]{}%
\providecommand \url  [0]{\begingroup\@sanitize@url \@url }%
\providecommand \@url [1]{\endgroup\@href {#1}{\urlprefix }}%
\providecommand \urlprefix  [0]{URL }%
\providecommand \Eprint [0]{\href }%
\providecommand \doibase [0]{https://doi.org/}%
\providecommand \selectlanguage [0]{\@gobble}%
\providecommand \bibinfo  [0]{\@secondoftwo}%
\providecommand \bibfield  [0]{\@secondoftwo}%
\providecommand \translation [1]{[#1]}%
\providecommand \BibitemOpen [0]{}%
\providecommand \bibitemStop [0]{}%
\providecommand \bibitemNoStop [0]{.\EOS\space}%
\providecommand \EOS [0]{\spacefactor3000\relax}%
\providecommand \BibitemShut  [1]{\csname bibitem#1\endcsname}%
\let\auto@bib@innerbib\@empty
%</preamble>
\bibitem [{\citenamefont {Abbott}\ \emph {et~al.}(2016)\citenamefont {Abbott}
  \emph {et~al.}}]{LIGOScientific:2016dsl}%
  \BibitemOpen
  \bibfield  {author} {\bibinfo {author} {\bibfnamefont {B.~P.}\ \bibnamefont
  {Abbott}} \emph {et~al.} (\bibinfo {collaboration} {LIGO Scientific,
  Virgo}),\ }\bibfield  {title} {\bibinfo {title} {{Binary Black Hole Mergers
  in the first Advanced LIGO Observing Run}},\ }\href
  {https://doi.org/10.1103/PhysRevX.6.041015} {\bibfield  {journal} {\bibinfo
  {journal} {Phys. Rev. X}\ }\textbf {\bibinfo {volume} {6}},\ \bibinfo {pages}
  {041015} (\bibinfo {year} {2016})},\ \bibinfo {note} {[Erratum: Phys.Rev.X 8,
  039903 (2018)]},\ \Eprint {https://arxiv.org/abs/1606.04856}
  {arXiv:1606.04856 [gr-qc]} \BibitemShut {NoStop}%
\bibitem [{\citenamefont {Abbott}\ \emph
  {et~al.}(2019{\natexlab{a}})\citenamefont {Abbott} \emph
  {et~al.}}]{LIGOScientific:2018mvr}%
  \BibitemOpen
  \bibfield  {author} {\bibinfo {author} {\bibfnamefont {B.~P.}\ \bibnamefont
  {Abbott}} \emph {et~al.} (\bibinfo {collaboration} {LIGO Scientific,
  Virgo}),\ }\bibfield  {title} {\bibinfo {title} {{GWTC-1: A
  Gravitational-Wave Transient Catalog of Compact Binary Mergers Observed by
  LIGO and Virgo during the First and Second Observing Runs}},\ }\href
  {https://doi.org/10.1103/PhysRevX.9.031040} {\bibfield  {journal} {\bibinfo
  {journal} {Phys. Rev. X}\ }\textbf {\bibinfo {volume} {9}},\ \bibinfo {pages}
  {031040} (\bibinfo {year} {2019}{\natexlab{a}})},\ \Eprint
  {https://arxiv.org/abs/1811.12907} {arXiv:1811.12907 [astro-ph.HE]}
  \BibitemShut {NoStop}%
\bibitem [{\citenamefont {Abbott}\ \emph
  {et~al.}(2021{\natexlab{a}})\citenamefont {Abbott} \emph
  {et~al.}}]{LIGOScientific:2020ibl}%
  \BibitemOpen
  \bibfield  {author} {\bibinfo {author} {\bibfnamefont {R.}~\bibnamefont
  {Abbott}} \emph {et~al.} (\bibinfo {collaboration} {LIGO Scientific,
  Virgo}),\ }\bibfield  {title} {\bibinfo {title} {{GWTC-2: Compact Binary
  Coalescences Observed by LIGO and Virgo During the First Half of the Third
  Observing Run}},\ }\href {https://doi.org/10.1103/PhysRevX.11.021053}
  {\bibfield  {journal} {\bibinfo  {journal} {Phys. Rev. X}\ }\textbf {\bibinfo
  {volume} {11}},\ \bibinfo {pages} {021053} (\bibinfo {year}
  {2021}{\natexlab{a}})},\ \Eprint {https://arxiv.org/abs/2010.14527}
  {arXiv:2010.14527 [gr-qc]} \BibitemShut {NoStop}%
\bibitem [{\citenamefont {Abbott}\ \emph
  {et~al.}(2021{\natexlab{b}})\citenamefont {Abbott} \emph
  {et~al.}}]{LIGOScientific:2021djp}%
  \BibitemOpen
  \bibfield  {author} {\bibinfo {author} {\bibfnamefont {R.}~\bibnamefont
  {Abbott}} \emph {et~al.} (\bibinfo {collaboration} {LIGO Scientific, VIRGO,
  KAGRA}),\ }\href@noop {} {\bibinfo {title} {{GWTC-3: Compact Binary
  Coalescences Observed by LIGO and Virgo During the Second Part of the Third
  Observing Run}}} (\bibinfo {year} {2021}{\natexlab{b}}),\ \Eprint
  {https://arxiv.org/abs/2111.03606} {arXiv:2111.03606 [gr-qc]} \BibitemShut
  {NoStop}%
\bibitem [{\citenamefont {Abbott}\ \emph
  {et~al.}(2019{\natexlab{b}})\citenamefont {Abbott} \emph
  {et~al.}}]{LIGOScientific:2018jsj}%
  \BibitemOpen
  \bibfield  {author} {\bibinfo {author} {\bibfnamefont {B.~P.}\ \bibnamefont
  {Abbott}} \emph {et~al.} (\bibinfo {collaboration} {LIGO Scientific,
  Virgo}),\ }\bibfield  {title} {\bibinfo {title} {{Binary Black Hole
  Population Properties Inferred from the First and Second Observing Runs of
  Advanced LIGO and Advanced Virgo}},\ }\href
  {https://doi.org/10.3847/2041-8213/ab3800} {\bibfield  {journal} {\bibinfo
  {journal} {Astrophys. J. Lett.}\ }\textbf {\bibinfo {volume} {882}},\
  \bibinfo {pages} {L24} (\bibinfo {year} {2019}{\natexlab{b}})},\ \Eprint
  {https://arxiv.org/abs/1811.12940} {arXiv:1811.12940 [astro-ph.HE]}
  \BibitemShut {NoStop}%
\bibitem [{\citenamefont {Venumadhav}\ \emph {et~al.}(2020)\citenamefont
  {Venumadhav}, \citenamefont {Zackay}, \citenamefont {Roulet}, \citenamefont
  {Dai},\ and\ \citenamefont {Zaldarriaga}}]{Venumadhav:2019lyq}%
  \BibitemOpen
  \bibfield  {author} {\bibinfo {author} {\bibfnamefont {T.}~\bibnamefont
  {Venumadhav}}, \bibinfo {author} {\bibfnamefont {B.}~\bibnamefont {Zackay}},
  \bibinfo {author} {\bibfnamefont {J.}~\bibnamefont {Roulet}}, \bibinfo
  {author} {\bibfnamefont {L.}~\bibnamefont {Dai}},\ and\ \bibinfo {author}
  {\bibfnamefont {M.}~\bibnamefont {Zaldarriaga}},\ }\bibfield  {title}
  {\bibinfo {title} {{New binary black hole mergers in the second observing run
  of Advanced LIGO and Advanced Virgo}},\ }\href
  {https://doi.org/10.1103/PhysRevD.101.083030} {\bibfield  {journal} {\bibinfo
   {journal} {Phys. Rev. D}\ }\textbf {\bibinfo {volume} {101}},\ \bibinfo
  {pages} {083030} (\bibinfo {year} {2020})},\ \Eprint
  {https://arxiv.org/abs/1904.07214} {arXiv:1904.07214 [astro-ph.HE]}
  \BibitemShut {NoStop}%
\bibitem [{\citenamefont {Nitz}\ \emph {et~al.}(2021)\citenamefont {Nitz},
  \citenamefont {Kumar}, \citenamefont {Wang}, \citenamefont {Kastha},
  \citenamefont {Wu}, \citenamefont {Sch\"afer}, \citenamefont {Dhurkunde},\
  and\ \citenamefont {Capano}}]{Nitz:2021zwj}%
  \BibitemOpen
  \bibfield  {author} {\bibinfo {author} {\bibfnamefont {A.~H.}\ \bibnamefont
  {Nitz}}, \bibinfo {author} {\bibfnamefont {S.}~\bibnamefont {Kumar}},
  \bibinfo {author} {\bibfnamefont {Y.-F.}\ \bibnamefont {Wang}}, \bibinfo
  {author} {\bibfnamefont {S.}~\bibnamefont {Kastha}}, \bibinfo {author}
  {\bibfnamefont {S.}~\bibnamefont {Wu}}, \bibinfo {author} {\bibfnamefont
  {M.}~\bibnamefont {Sch\"afer}}, \bibinfo {author} {\bibfnamefont
  {R.}~\bibnamefont {Dhurkunde}},\ and\ \bibinfo {author} {\bibfnamefont
  {C.~D.}\ \bibnamefont {Capano}},\ }\href@noop {} {\bibinfo {title} {{4-OGC:
  Catalog of gravitational waves from compact-binary mergers}}} (\bibinfo
  {year} {2021}),\ \Eprint {https://arxiv.org/abs/2112.06878} {arXiv:2112.06878
  [astro-ph.HE]} \BibitemShut {NoStop}%
\bibitem [{\citenamefont {Abbott}\ \emph
  {et~al.}(2021{\natexlab{c}})\citenamefont {Abbott} \emph
  {et~al.}}]{LIGOScientific:2021psn}%
  \BibitemOpen
  \bibfield  {author} {\bibinfo {author} {\bibfnamefont {R.}~\bibnamefont
  {Abbott}} \emph {et~al.} (\bibinfo {collaboration} {LIGO Scientific, VIRGO,
  KAGRA}),\ }\href@noop {} {\bibinfo {title} {{The population of merging
  compact binaries inferred using gravitational waves through GWTC-3}}}
  (\bibinfo {year} {2021}{\natexlab{c}}),\ \Eprint
  {https://arxiv.org/abs/2111.03634} {arXiv:2111.03634 [astro-ph.HE]}
  \BibitemShut {NoStop}%
\bibitem [{\citenamefont {Abbott}\ \emph {et~al.}(2020)\citenamefont {Abbott}
  \emph {et~al.}}]{LIGOScientific:2020zkf}%
  \BibitemOpen
  \bibfield  {author} {\bibinfo {author} {\bibfnamefont {R.}~\bibnamefont
  {Abbott}} \emph {et~al.} (\bibinfo {collaboration} {LIGO Scientific,
  Virgo}),\ }\bibfield  {title} {\bibinfo {title} {{GW190814: Gravitational
  Waves from the Coalescence of a 23 Solar Mass Black Hole with a 2.6 Solar
  Mass Compact Object}},\ }\href {https://doi.org/10.3847/2041-8213/ab960f}
  {\bibfield  {journal} {\bibinfo  {journal} {Astrophys. J. Lett.}\ }\textbf
  {\bibinfo {volume} {896}},\ \bibinfo {pages} {L44} (\bibinfo {year}
  {2020})},\ \Eprint {https://arxiv.org/abs/2006.12611} {arXiv:2006.12611
  [astro-ph.HE]} \BibitemShut {NoStop}%
\bibitem [{\citenamefont {Maggiore}\ \emph {et~al.}(2020)\citenamefont
  {Maggiore} \emph {et~al.}}]{Maggiore:2019uih}%
  \BibitemOpen
  \bibfield  {author} {\bibinfo {author} {\bibfnamefont {M.}~\bibnamefont
  {Maggiore}} \emph {et~al.},\ }\bibfield  {title} {\bibinfo {title} {{Science
  Case for the Einstein Telescope}},\ }\href
  {https://doi.org/10.1088/1475-7516/2020/03/050} {\bibfield  {journal}
  {\bibinfo  {journal} {JCAP}\ }\textbf {\bibinfo {volume} {03}},\ \bibinfo
  {pages} {050}},\ \Eprint {https://arxiv.org/abs/1912.02622} {arXiv:1912.02622
  [astro-ph.CO]} \BibitemShut {NoStop}%
\bibitem [{\citenamefont {Evans}\ \emph {et~al.}(2021)\citenamefont {Evans}
  \emph {et~al.}}]{Evans:2021gyd}%
  \BibitemOpen
  \bibfield  {author} {\bibinfo {author} {\bibfnamefont {M.}~\bibnamefont
  {Evans}} \emph {et~al.},\ }\href@noop {} {\bibinfo {title} {{A Horizon Study
  for Cosmic Explorer: Science, Observatories, and Community}}} (\bibinfo
  {year} {2021}),\ \Eprint {https://arxiv.org/abs/2109.09882} {arXiv:2109.09882
  [astro-ph.IM]} \BibitemShut {NoStop}%
\bibitem [{\citenamefont {Jani}\ \emph {et~al.}(2019)\citenamefont {Jani},
  \citenamefont {Shoemaker},\ and\ \citenamefont {Cutler}}]{Jani:2019ffg}%
  \BibitemOpen
  \bibfield  {author} {\bibinfo {author} {\bibfnamefont {K.}~\bibnamefont
  {Jani}}, \bibinfo {author} {\bibfnamefont {D.}~\bibnamefont {Shoemaker}},\
  and\ \bibinfo {author} {\bibfnamefont {C.}~\bibnamefont {Cutler}},\
  }\bibfield  {title} {\bibinfo {title} {{Detectability of Intermediate-Mass
  Black Holes in Multiband Gravitational Wave Astronomy}},\ }\href
  {https://doi.org/10.1038/s41550-019-0932-7} {\bibfield  {journal} {\bibinfo
  {journal} {Nature Astron.}\ }\textbf {\bibinfo {volume} {4}},\ \bibinfo
  {pages} {260} (\bibinfo {year} {2019})},\ \Eprint
  {https://arxiv.org/abs/1908.04985} {arXiv:1908.04985 [gr-qc]} \BibitemShut
  {NoStop}%
\bibitem [{\citenamefont {Seoane}\ \emph {et~al.}(2013)\citenamefont {Seoane}
  \emph {et~al.}}]{eLISA:2013xep}%
  \BibitemOpen
  \bibfield  {author} {\bibinfo {author} {\bibfnamefont {P.~A.}\ \bibnamefont
  {Seoane}} \emph {et~al.} (\bibinfo {collaboration} {eLISA}),\ }\href@noop {}
  {\bibinfo {title} {{The Gravitational Universe}}} (\bibinfo {year} {2013}),\
  \Eprint {https://arxiv.org/abs/1305.5720} {arXiv:1305.5720 [astro-ph.CO]}
  \BibitemShut {NoStop}%
\bibitem [{\citenamefont {Amaro-Seoane}\ \emph {et~al.}(2017)\citenamefont
  {Amaro-Seoane} \emph {et~al.}}]{LISA:2017pwj}%
  \BibitemOpen
  \bibfield  {author} {\bibinfo {author} {\bibfnamefont {P.}~\bibnamefont
  {Amaro-Seoane}} \emph {et~al.} (\bibinfo {collaboration} {LISA}),\
  }\href@noop {} {\bibinfo {title} {{Laser Interferometer Space Antenna}}}
  (\bibinfo {year} {2017}),\ \Eprint {https://arxiv.org/abs/1702.00786}
  {arXiv:1702.00786 [astro-ph.IM]} \BibitemShut {NoStop}%
\bibitem [{\citenamefont {Katz}\ \emph {et~al.}(2020)\citenamefont {Katz},
  \citenamefont {Kelley}, \citenamefont {Dosopoulou}, \citenamefont {Berry},
  \citenamefont {Blecha},\ and\ \citenamefont {Larson}}]{Katz:2019qlu}%
  \BibitemOpen
  \bibfield  {author} {\bibinfo {author} {\bibfnamefont {M.~L.}\ \bibnamefont
  {Katz}}, \bibinfo {author} {\bibfnamefont {L.~Z.}\ \bibnamefont {Kelley}},
  \bibinfo {author} {\bibfnamefont {F.}~\bibnamefont {Dosopoulou}}, \bibinfo
  {author} {\bibfnamefont {S.}~\bibnamefont {Berry}}, \bibinfo {author}
  {\bibfnamefont {L.}~\bibnamefont {Blecha}},\ and\ \bibinfo {author}
  {\bibfnamefont {S.~L.}\ \bibnamefont {Larson}},\ }\bibfield  {title}
  {\bibinfo {title} {{Probing Massive Black Hole Binary Populations with
  LISA}},\ }\href {https://doi.org/10.1093/mnras/stz3102} {\bibfield  {journal}
  {\bibinfo  {journal} {Mon. Not. Roy. Astron. Soc.}\ }\textbf {\bibinfo
  {volume} {491}},\ \bibinfo {pages} {2301} (\bibinfo {year} {2020})},\ \Eprint
  {https://arxiv.org/abs/1908.05779} {arXiv:1908.05779 [astro-ph.HE]}
  \BibitemShut {NoStop}%
\bibitem [{\citenamefont {Gair}\ \emph {et~al.}(2011)\citenamefont {Gair},
  \citenamefont {Sesana}, \citenamefont {Berti},\ and\ \citenamefont
  {Volonteri}}]{Gair:2010bx}%
  \BibitemOpen
  \bibfield  {author} {\bibinfo {author} {\bibfnamefont {J.~R.}\ \bibnamefont
  {Gair}}, \bibinfo {author} {\bibfnamefont {A.}~\bibnamefont {Sesana}},
  \bibinfo {author} {\bibfnamefont {E.}~\bibnamefont {Berti}},\ and\ \bibinfo
  {author} {\bibfnamefont {M.}~\bibnamefont {Volonteri}},\ }\bibfield  {title}
  {\bibinfo {title} {{Constraining properties of the black hole population
  using LISA}},\ }\href {https://doi.org/10.1088/0264-9381/28/9/094018}
  {\bibfield  {journal} {\bibinfo  {journal} {Class. Quant. Grav.}\ }\textbf
  {\bibinfo {volume} {28}},\ \bibinfo {pages} {094018} (\bibinfo {year}
  {2011})},\ \Eprint {https://arxiv.org/abs/1009.6172} {arXiv:1009.6172
  [gr-qc]} \BibitemShut {NoStop}%
\bibitem [{\citenamefont {Volonteri}\ \emph {et~al.}(2020)\citenamefont
  {Volonteri} \emph {et~al.}}]{Volonteri:2020wkx}%
  \BibitemOpen
  \bibfield  {author} {\bibinfo {author} {\bibfnamefont {M.}~\bibnamefont
  {Volonteri}} \emph {et~al.},\ }\bibfield  {title} {\bibinfo {title} {{Black
  hole mergers from dwarf to massive galaxies with the NewHorizon and
  Horizon-AGN simulations}},\ }\href {https://doi.org/10.1093/mnras/staa2384}
  {\bibfield  {journal} {\bibinfo  {journal} {Mon. Not. Roy. Astron. Soc.}\
  }\textbf {\bibinfo {volume} {498}},\ \bibinfo {pages} {2219} (\bibinfo {year}
  {2020})},\ \Eprint {https://arxiv.org/abs/2005.04902} {arXiv:2005.04902
  [astro-ph.GA]} \BibitemShut {NoStop}%
\bibitem [{\citenamefont {Baumgarte}\ and\ \citenamefont
  {Shapiro}(2010)}]{BauSha10}%
  \BibitemOpen
  \bibfield  {author} {\bibinfo {author} {\bibfnamefont {T.~W.}\ \bibnamefont
  {Baumgarte}}\ and\ \bibinfo {author} {\bibfnamefont {S.~L.}\ \bibnamefont
  {Shapiro}},\ }\href@noop {} {\emph {\bibinfo {title} {Numerical Relativity:
  Solving {E}instein's Equations on the Computer}}}\ (\bibinfo  {publisher}
  {Cambridge University Press},\ \bibinfo {address} {Cambridge},\ \bibinfo
  {year} {2010})\BibitemShut {NoStop}%
\bibitem [{\citenamefont {Boyle}\ \emph {et~al.}(2019)\citenamefont {Boyle}
  \emph {et~al.}}]{Boyle:2019kee}%
  \BibitemOpen
  \bibfield  {author} {\bibinfo {author} {\bibfnamefont {M.}~\bibnamefont
  {Boyle}} \emph {et~al.},\ }\bibfield  {title} {\bibinfo {title} {{The SXS
  Collaboration catalog of binary black hole simulations}},\ }\href
  {https://doi.org/10.1088/1361-6382/ab34e2} {\bibfield  {journal} {\bibinfo
  {journal} {Class. Quant. Grav.}\ }\textbf {\bibinfo {volume} {36}},\ \bibinfo
  {pages} {195006} (\bibinfo {year} {2019})},\ \Eprint
  {https://arxiv.org/abs/1904.04831} {arXiv:1904.04831 [gr-qc]} \BibitemShut
  {NoStop}%
\bibitem [{\citenamefont {Lousto}\ and\ \citenamefont
  {Healy}(2020)}]{Lousto:2020tnb}%
  \BibitemOpen
  \bibfield  {author} {\bibinfo {author} {\bibfnamefont {C.~O.}\ \bibnamefont
  {Lousto}}\ and\ \bibinfo {author} {\bibfnamefont {J.}~\bibnamefont {Healy}},\
  }\bibfield  {title} {\bibinfo {title} {{Exploring the Small Mass Ratio Binary
  Black Hole Merger via Zeno\textquoteright{}s Dichotomy Approach}},\ }\href
  {https://doi.org/10.1103/PhysRevLett.125.191102} {\bibfield  {journal}
  {\bibinfo  {journal} {Phys. Rev. Lett.}\ }\textbf {\bibinfo {volume} {125}},\
  \bibinfo {pages} {191102} (\bibinfo {year} {2020})},\ \Eprint
  {https://arxiv.org/abs/2006.04818} {arXiv:2006.04818 [gr-qc]} \BibitemShut
  {NoStop}%
\bibitem [{\citenamefont {Rosato}\ \emph {et~al.}(2021)\citenamefont {Rosato},
  \citenamefont {Healy},\ and\ \citenamefont {Lousto}}]{Rosato:2021jsq}%
  \BibitemOpen
  \bibfield  {author} {\bibinfo {author} {\bibfnamefont {N.}~\bibnamefont
  {Rosato}}, \bibinfo {author} {\bibfnamefont {J.}~\bibnamefont {Healy}},\ and\
  \bibinfo {author} {\bibfnamefont {C.~O.}\ \bibnamefont {Lousto}},\ }\bibfield
   {title} {\bibinfo {title} {{Adapted gauge to small mass ratio binary black
  hole evolutions}},\ }\href {https://doi.org/10.1103/PhysRevD.103.104068}
  {\bibfield  {journal} {\bibinfo  {journal} {Phys. Rev. D}\ }\textbf {\bibinfo
  {volume} {103}},\ \bibinfo {pages} {104068} (\bibinfo {year} {2021})},\
  \Eprint {https://arxiv.org/abs/2103.09326} {arXiv:2103.09326 [gr-qc]}
  \BibitemShut {NoStop}%
\bibitem [{\citenamefont {Sperhake}\ \emph {et~al.}(2011)\citenamefont
  {Sperhake}, \citenamefont {Cardoso}, \citenamefont {Ott}, \citenamefont
  {Schnetter},\ and\ \citenamefont {Witek}}]{Sperhake:2011ik}%
  \BibitemOpen
  \bibfield  {author} {\bibinfo {author} {\bibfnamefont {U.}~\bibnamefont
  {Sperhake}}, \bibinfo {author} {\bibfnamefont {V.}~\bibnamefont {Cardoso}},
  \bibinfo {author} {\bibfnamefont {C.~D.}\ \bibnamefont {Ott}}, \bibinfo
  {author} {\bibfnamefont {E.}~\bibnamefont {Schnetter}},\ and\ \bibinfo
  {author} {\bibfnamefont {H.}~\bibnamefont {Witek}},\ }\bibfield  {title}
  {\bibinfo {title} {{Extreme black hole simulations: collisions of unequal
  mass black holes and the point particle limit}},\ }\href
  {https://doi.org/10.1103/PhysRevD.84.084038} {\bibfield  {journal} {\bibinfo
  {journal} {Phys. Rev. D}\ }\textbf {\bibinfo {volume} {84}},\ \bibinfo
  {pages} {084038} (\bibinfo {year} {2011})},\ \Eprint
  {https://arxiv.org/abs/1105.5391} {arXiv:1105.5391 [gr-qc]} \BibitemShut
  {NoStop}%
\bibitem [{\citenamefont {Lousto}\ and\ \citenamefont
  {Healy}(2023)}]{Lousto:2022}%
  \BibitemOpen
  \bibfield  {author} {\bibinfo {author} {\bibfnamefont {C.~O.}\ \bibnamefont
  {Lousto}}\ and\ \bibinfo {author} {\bibfnamefont {J.}~\bibnamefont {Healy}},\
  }\bibfield  {title} {\bibinfo {title} {{Study of the intermediate mass ratio
  black hole binary merger up to 1000:1 with numerical relativity}},\ }\href
  {https://doi.org/10.1088/1361-6382/acc7ef} {\bibfield  {journal} {\bibinfo
  {journal} {Class. Quant. Grav.}\ }\textbf {\bibinfo {volume} {40}},\ \bibinfo
  {pages} {09LT01} (\bibinfo {year} {2023})},\ \Eprint
  {https://arxiv.org/abs/2203.08831} {arXiv:2203.08831 [gr-qc]} \BibitemShut
  {NoStop}%
\bibitem [{\citenamefont {Barack}\ and\ \citenamefont
  {Pound}(2019)}]{Barack:2018yvs}%
  \BibitemOpen
  \bibfield  {author} {\bibinfo {author} {\bibfnamefont {L.}~\bibnamefont
  {Barack}}\ and\ \bibinfo {author} {\bibfnamefont {A.}~\bibnamefont {Pound}},\
  }\bibfield  {title} {\bibinfo {title} {{Self-force and radiation reaction in
  general relativity}},\ }\href {https://doi.org/10.1088/1361-6633/aae552}
  {\bibfield  {journal} {\bibinfo  {journal} {Rept. Prog. Phys.}\ }\textbf
  {\bibinfo {volume} {82}},\ \bibinfo {pages} {016904} (\bibinfo {year}
  {2019})},\ \Eprint {https://arxiv.org/abs/1805.10385} {arXiv:1805.10385
  [gr-qc]} \BibitemShut {NoStop}%
\bibitem [{\citenamefont {Pound}\ and\ \citenamefont
  {Wardell}(2022)}]{Pound:2021qin}%
  \BibitemOpen
  \bibfield  {author} {\bibinfo {author} {\bibfnamefont {A.}~\bibnamefont
  {Pound}}\ and\ \bibinfo {author} {\bibfnamefont {B.}~\bibnamefont
  {Wardell}},\ }\bibinfo {title} {Black hole perturbation theory and
  gravitational self-force},\ in\ \href
  {https://doi.org/10.1007/978-981-16-4306-4_38} {\emph {\bibinfo {booktitle}
  {Handbook of Gravitational Wave Astronomy}}},\ \bibinfo {editor} {edited by\
  \bibinfo {editor} {\bibfnamefont {C.}~\bibnamefont {Bambi}}, \bibinfo
  {editor} {\bibfnamefont {S.}~\bibnamefont {Katsanevas}},\ and\ \bibinfo
  {editor} {\bibfnamefont {K.~D.}\ \bibnamefont {Kokkotas}}}\ (\bibinfo
  {publisher} {Springer Nature Singapore},\ \bibinfo {address} {Singapore},\
  \bibinfo {year} {2022})\ pp.\ \bibinfo {pages} {1411--1529}\BibitemShut
  {NoStop}%
\bibitem [{\citenamefont {van~de Meent}(2018)}]{vandeMeent:2017bcc}%
  \BibitemOpen
  \bibfield  {author} {\bibinfo {author} {\bibfnamefont {M.}~\bibnamefont
  {van~de Meent}},\ }\bibfield  {title} {\bibinfo {title} {{Gravitational
  self-force on generic bound geodesics in Kerr spacetime}},\ }\href
  {https://doi.org/10.1103/PhysRevD.97.104033} {\bibfield  {journal} {\bibinfo
  {journal} {Phys. Rev. D}\ }\textbf {\bibinfo {volume} {97}},\ \bibinfo
  {pages} {104033} (\bibinfo {year} {2018})},\ \Eprint
  {https://arxiv.org/abs/1711.09607} {arXiv:1711.09607 [gr-qc]} \BibitemShut
  {NoStop}%
\bibitem [{\citenamefont {Chua}\ \emph {et~al.}(2021)\citenamefont {Chua},
  \citenamefont {Katz}, \citenamefont {Warburton},\ and\ \citenamefont
  {Hughes}}]{Chua:2020stf}%
  \BibitemOpen
  \bibfield  {author} {\bibinfo {author} {\bibfnamefont {A.~J.~K.}\
  \bibnamefont {Chua}}, \bibinfo {author} {\bibfnamefont {M.~L.}\ \bibnamefont
  {Katz}}, \bibinfo {author} {\bibfnamefont {N.}~\bibnamefont {Warburton}},\
  and\ \bibinfo {author} {\bibfnamefont {S.~A.}\ \bibnamefont {Hughes}},\
  }\bibfield  {title} {\bibinfo {title} {{Rapid generation of fully
  relativistic extreme-mass-ratio-inspiral waveform templates for LISA data
  analysis}},\ }\href {https://doi.org/10.1103/PhysRevLett.126.051102}
  {\bibfield  {journal} {\bibinfo  {journal} {Phys. Rev. Lett.}\ }\textbf
  {\bibinfo {volume} {126}},\ \bibinfo {pages} {051102} (\bibinfo {year}
  {2021})},\ \Eprint {https://arxiv.org/abs/2008.06071} {arXiv:2008.06071
  [gr-qc]} \BibitemShut {NoStop}%
\bibitem [{\citenamefont {Hughes}\ \emph {et~al.}(2021)\citenamefont {Hughes},
  \citenamefont {Warburton}, \citenamefont {Khanna}, \citenamefont {Chua},\
  and\ \citenamefont {Katz}}]{Hughes:2021exa}%
  \BibitemOpen
  \bibfield  {author} {\bibinfo {author} {\bibfnamefont {S.~A.}\ \bibnamefont
  {Hughes}}, \bibinfo {author} {\bibfnamefont {N.}~\bibnamefont {Warburton}},
  \bibinfo {author} {\bibfnamefont {G.}~\bibnamefont {Khanna}}, \bibinfo
  {author} {\bibfnamefont {A.~J.~K.}\ \bibnamefont {Chua}},\ and\ \bibinfo
  {author} {\bibfnamefont {M.~L.}\ \bibnamefont {Katz}},\ }\bibfield  {title}
  {\bibinfo {title} {{Adiabatic waveforms for extreme mass-ratio inspirals via
  multivoice decomposition in time and frequency}},\ }\href
  {https://doi.org/10.1103/PhysRevD.103.104014} {\bibfield  {journal} {\bibinfo
   {journal} {Phys. Rev. D}\ }\textbf {\bibinfo {volume} {103}},\ \bibinfo
  {pages} {104014} (\bibinfo {year} {2021})},\ \Eprint
  {https://arxiv.org/abs/2102.02713} {arXiv:2102.02713 [gr-qc]} \BibitemShut
  {NoStop}%
\bibitem [{\citenamefont {Pound}\ \emph {et~al.}(2020)\citenamefont {Pound},
  \citenamefont {Wardell}, \citenamefont {Warburton},\ and\ \citenamefont
  {Miller}}]{Pound:2019lzj}%
  \BibitemOpen
  \bibfield  {author} {\bibinfo {author} {\bibfnamefont {A.}~\bibnamefont
  {Pound}}, \bibinfo {author} {\bibfnamefont {B.}~\bibnamefont {Wardell}},
  \bibinfo {author} {\bibfnamefont {N.}~\bibnamefont {Warburton}},\ and\
  \bibinfo {author} {\bibfnamefont {J.}~\bibnamefont {Miller}},\ }\bibfield
  {title} {\bibinfo {title} {{Second-Order Self-Force Calculation of
  Gravitational Binding Energy in Compact Binaries}},\ }\href
  {https://doi.org/10.1103/PhysRevLett.124.021101} {\bibfield  {journal}
  {\bibinfo  {journal} {Phys. Rev. Lett.}\ }\textbf {\bibinfo {volume} {124}},\
  \bibinfo {pages} {021101} (\bibinfo {year} {2020})},\ \Eprint
  {https://arxiv.org/abs/1908.07419} {arXiv:1908.07419 [gr-qc]} \BibitemShut
  {NoStop}%
\bibitem [{\citenamefont {Warburton}\ \emph {et~al.}(2021)\citenamefont
  {Warburton}, \citenamefont {Pound}, \citenamefont {Wardell}, \citenamefont
  {Miller},\ and\ \citenamefont {Durkan}}]{Warburton:2021kwk}%
  \BibitemOpen
  \bibfield  {author} {\bibinfo {author} {\bibfnamefont {N.}~\bibnamefont
  {Warburton}}, \bibinfo {author} {\bibfnamefont {A.}~\bibnamefont {Pound}},
  \bibinfo {author} {\bibfnamefont {B.}~\bibnamefont {Wardell}}, \bibinfo
  {author} {\bibfnamefont {J.}~\bibnamefont {Miller}},\ and\ \bibinfo {author}
  {\bibfnamefont {L.}~\bibnamefont {Durkan}},\ }\bibfield  {title} {\bibinfo
  {title} {{Gravitational-Wave Energy Flux for Compact Binaries through Second
  Order in the Mass Ratio}},\ }\href
  {https://doi.org/10.1103/PhysRevLett.127.151102} {\bibfield  {journal}
  {\bibinfo  {journal} {Phys. Rev. Lett.}\ }\textbf {\bibinfo {volume} {127}},\
  \bibinfo {pages} {151102} (\bibinfo {year} {2021})},\ \Eprint
  {https://arxiv.org/abs/2107.01298} {arXiv:2107.01298 [gr-qc]} \BibitemShut
  {NoStop}%
\bibitem [{\citenamefont {Wardell}\ \emph {et~al.}(2021)\citenamefont
  {Wardell}, \citenamefont {Pound}, \citenamefont {Warburton}, \citenamefont
  {Miller}, \citenamefont {Durkan},\ and\ \citenamefont
  {Le~Tiec}}]{Wardell:2021fyy}%
  \BibitemOpen
  \bibfield  {author} {\bibinfo {author} {\bibfnamefont {B.}~\bibnamefont
  {Wardell}}, \bibinfo {author} {\bibfnamefont {A.}~\bibnamefont {Pound}},
  \bibinfo {author} {\bibfnamefont {N.}~\bibnamefont {Warburton}}, \bibinfo
  {author} {\bibfnamefont {J.}~\bibnamefont {Miller}}, \bibinfo {author}
  {\bibfnamefont {L.}~\bibnamefont {Durkan}},\ and\ \bibinfo {author}
  {\bibfnamefont {A.}~\bibnamefont {Le~Tiec}},\ }\bibfield  {title} {\bibinfo
  {title} {{Gravitational waveforms for compact binaries from second-order
  self-force theory}},\ }\href@noop {} {\  (\bibinfo {year} {2021})},\ \Eprint
  {https://arxiv.org/abs/2112.12265} {arXiv:2112.12265 [gr-qc]} \BibitemShut
  {NoStop}%
\bibitem [{\citenamefont {van~de Meent}\ and\ \citenamefont
  {Pfeiffer}(2020)}]{vandeMeent:2020xgc}%
  \BibitemOpen
  \bibfield  {author} {\bibinfo {author} {\bibfnamefont {M.}~\bibnamefont
  {van~de Meent}}\ and\ \bibinfo {author} {\bibfnamefont {H.~P.}\ \bibnamefont
  {Pfeiffer}},\ }\bibfield  {title} {\bibinfo {title} {{Intermediate mass-ratio
  black hole binaries: Applicability of small mass-ratio perturbation
  theory}},\ }\href {https://doi.org/10.1103/PhysRevLett.125.181101} {\bibfield
   {journal} {\bibinfo  {journal} {Phys. Rev. Lett.}\ }\textbf {\bibinfo
  {volume} {125}},\ \bibinfo {pages} {181101} (\bibinfo {year} {2020})},\
  \Eprint {https://arxiv.org/abs/2006.12036} {arXiv:2006.12036 [gr-qc]}
  \BibitemShut {NoStop}%
\bibitem [{\citenamefont {Ramos-Buades}\ \emph {et~al.}(2022)\citenamefont
  {Ramos-Buades}, \citenamefont {van~de Meent}, \citenamefont {Pfeiffer},
  \citenamefont {R\"uter}, \citenamefont {Scheel}, \citenamefont {Boyle},\ and\
  \citenamefont {Kidder}}]{Ramos-Buades:2022lgf}%
  \BibitemOpen
  \bibfield  {author} {\bibinfo {author} {\bibfnamefont {A.}~\bibnamefont
  {Ramos-Buades}}, \bibinfo {author} {\bibfnamefont {M.}~\bibnamefont {van~de
  Meent}}, \bibinfo {author} {\bibfnamefont {H.~P.}\ \bibnamefont {Pfeiffer}},
  \bibinfo {author} {\bibfnamefont {H.~R.}\ \bibnamefont {R\"uter}}, \bibinfo
  {author} {\bibfnamefont {M.~A.}\ \bibnamefont {Scheel}}, \bibinfo {author}
  {\bibfnamefont {M.}~\bibnamefont {Boyle}},\ and\ \bibinfo {author}
  {\bibfnamefont {L.~E.}\ \bibnamefont {Kidder}},\ }\bibfield  {title}
  {\bibinfo {title} {{Eccentric binary black holes: Comparing numerical
  relativity and small mass-ratio perturbation theory}},\ }\href
  {https://doi.org/10.1103/PhysRevD.106.124040} {\bibfield  {journal} {\bibinfo
   {journal} {Phys. Rev. D}\ }\textbf {\bibinfo {volume} {106}},\ \bibinfo
  {pages} {124040} (\bibinfo {year} {2022})},\ \Eprint
  {https://arxiv.org/abs/2209.03390} {arXiv:2209.03390 [gr-qc]} \BibitemShut
  {NoStop}%
\bibitem [{\citenamefont {Albertini}\ \emph {et~al.}(2022)\citenamefont
  {Albertini}, \citenamefont {Nagar}, \citenamefont {Pound}, \citenamefont
  {Warburton}, \citenamefont {Wardell}, \citenamefont {Durkan},\ and\
  \citenamefont {Miller}}]{Albertini:2022rfe}%
  \BibitemOpen
  \bibfield  {author} {\bibinfo {author} {\bibfnamefont {A.}~\bibnamefont
  {Albertini}}, \bibinfo {author} {\bibfnamefont {A.}~\bibnamefont {Nagar}},
  \bibinfo {author} {\bibfnamefont {A.}~\bibnamefont {Pound}}, \bibinfo
  {author} {\bibfnamefont {N.}~\bibnamefont {Warburton}}, \bibinfo {author}
  {\bibfnamefont {B.}~\bibnamefont {Wardell}}, \bibinfo {author} {\bibfnamefont
  {L.}~\bibnamefont {Durkan}},\ and\ \bibinfo {author} {\bibfnamefont
  {J.}~\bibnamefont {Miller}},\ }\bibfield  {title} {\bibinfo {title}
  {{Comparing second-order gravitational self-force, numerical relativity, and
  effective one body waveforms from inspiralling, quasicircular, and
  nonspinning black hole binaries}},\ }\href
  {https://doi.org/10.1103/PhysRevD.106.084061} {\bibfield  {journal} {\bibinfo
   {journal} {Phys. Rev. D}\ }\textbf {\bibinfo {volume} {106}},\ \bibinfo
  {pages} {084061} (\bibinfo {year} {2022})},\ \Eprint
  {https://arxiv.org/abs/2208.01049} {arXiv:2208.01049 [gr-qc]} \BibitemShut
  {NoStop}%
\bibitem [{\citenamefont {Dhesi}\ \emph {et~al.}(2021)\citenamefont {Dhesi},
  \citenamefont {R\"uter}, \citenamefont {Pound}, \citenamefont {Barack},\ and\
  \citenamefont {Pfeiffer}}]{Dhesi:2021yje}%
  \BibitemOpen
  \bibfield  {author} {\bibinfo {author} {\bibfnamefont {M.}~\bibnamefont
  {Dhesi}}, \bibinfo {author} {\bibfnamefont {H.~R.}\ \bibnamefont {R\"uter}},
  \bibinfo {author} {\bibfnamefont {A.}~\bibnamefont {Pound}}, \bibinfo
  {author} {\bibfnamefont {L.}~\bibnamefont {Barack}},\ and\ \bibinfo {author}
  {\bibfnamefont {H.~P.}\ \bibnamefont {Pfeiffer}},\ }\bibfield  {title}
  {\bibinfo {title} {{Worldtube excision method for intermediate-mass-ratio
  inspirals: Scalar-field toy model}},\ }\href
  {https://doi.org/10.1103/PhysRevD.104.124002} {\bibfield  {journal} {\bibinfo
   {journal} {Phys. Rev. D}\ }\textbf {\bibinfo {volume} {104}},\ \bibinfo
  {pages} {124002} (\bibinfo {year} {2021})},\ \Eprint
  {https://arxiv.org/abs/2109.03531} {arXiv:2109.03531 [gr-qc]} \BibitemShut
  {NoStop}%
\bibitem [{\citenamefont {Deppe}\ \emph {et~al.}(2023)\citenamefont {Deppe},
  \citenamefont {Throwe}, \citenamefont {Kidder}, \citenamefont {Vu},
  \citenamefont {H\'ebert}, \citenamefont {Moxon}, \citenamefont {Armaza},
  \citenamefont {Bonilla}, \citenamefont {Kim}, \citenamefont {Kumar},
  \citenamefont {Lovelace}, \citenamefont {Macedo}, \citenamefont {Nelli},
  \citenamefont {O'Shea}, \citenamefont {Pfeiffer}, \citenamefont {Scheel},
  \citenamefont {Teukolsky}, \citenamefont {Wittek} \emph
  {et~al.}}]{spectrecode}%
  \BibitemOpen
  \bibfield  {author} {\bibinfo {author} {\bibfnamefont {N.}~\bibnamefont
  {Deppe}}, \bibinfo {author} {\bibfnamefont {W.}~\bibnamefont {Throwe}},
  \bibinfo {author} {\bibfnamefont {L.~E.}\ \bibnamefont {Kidder}}, \bibinfo
  {author} {\bibfnamefont {N.~L.}\ \bibnamefont {Vu}}, \bibinfo {author}
  {\bibfnamefont {F.}~\bibnamefont {H\'ebert}}, \bibinfo {author}
  {\bibfnamefont {J.}~\bibnamefont {Moxon}}, \bibinfo {author} {\bibfnamefont
  {C.}~\bibnamefont {Armaza}}, \bibinfo {author} {\bibfnamefont {G.~S.}\
  \bibnamefont {Bonilla}}, \bibinfo {author} {\bibfnamefont {Y.}~\bibnamefont
  {Kim}}, \bibinfo {author} {\bibfnamefont {P.}~\bibnamefont {Kumar}}, \bibinfo
  {author} {\bibfnamefont {G.}~\bibnamefont {Lovelace}}, \bibinfo {author}
  {\bibfnamefont {A.}~\bibnamefont {Macedo}}, \bibinfo {author} {\bibfnamefont
  {K.~C.}\ \bibnamefont {Nelli}}, \bibinfo {author} {\bibfnamefont
  {E.}~\bibnamefont {O'Shea}}, \bibinfo {author} {\bibfnamefont {H.~P.}\
  \bibnamefont {Pfeiffer}}, \bibinfo {author} {\bibfnamefont {M.~A.}\
  \bibnamefont {Scheel}}, \bibinfo {author} {\bibfnamefont {S.~A.}\
  \bibnamefont {Teukolsky}}, \bibinfo {author} {\bibfnamefont {N.~A.}\
  \bibnamefont {Wittek}}, \emph {et~al.},\ }\href
  {https://doi.org/10.5281/zenodo.7713393} {\bibinfo {title} {\texttt{SpECTRE
  v2023.03.09}}},\ \bibinfo {howpublished}
  {\href{https://doi.org/10.5281/zenodo.7713393}{10.5281/zenodo.7713393}}
  (\bibinfo {year} {2023})\BibitemShut {NoStop}%
\bibitem [{\citenamefont {Vega}\ \emph {et~al.}(2009)\citenamefont {Vega},
  \citenamefont {Diener}, \citenamefont {Tichy},\ and\ \citenamefont
  {Detweiler}}]{Vega:2009}%
  \BibitemOpen
  \bibfield  {author} {\bibinfo {author} {\bibfnamefont {I.}~\bibnamefont
  {Vega}}, \bibinfo {author} {\bibfnamefont {P.}~\bibnamefont {Diener}},
  \bibinfo {author} {\bibfnamefont {W.}~\bibnamefont {Tichy}},\ and\ \bibinfo
  {author} {\bibfnamefont {S.}~\bibnamefont {Detweiler}},\ }\bibfield  {title}
  {\bibinfo {title} {Self-force with ($3+1$) codes: A primer for numerical
  relativists},\ }\href {https://doi.org/10.1103/PhysRevD.80.084021} {\bibfield
   {journal} {\bibinfo  {journal} {Phys. Rev. D}\ }\textbf {\bibinfo {volume}
  {80}},\ \bibinfo {pages} {084021} (\bibinfo {year} {2009})}\BibitemShut
  {NoStop}%
\bibitem [{\citenamefont {Scheel}\ \emph {et~al.}(2004)\citenamefont {Scheel},
  \citenamefont {Erickcek}, \citenamefont {Burko}, \citenamefont {Kidder},
  \citenamefont {Pfeiffer},\ and\ \citenamefont {Teukolsky}}]{Scheel:2004}%
  \BibitemOpen
  \bibfield  {author} {\bibinfo {author} {\bibfnamefont {M.~A.}\ \bibnamefont
  {Scheel}}, \bibinfo {author} {\bibfnamefont {A.~L.}\ \bibnamefont
  {Erickcek}}, \bibinfo {author} {\bibfnamefont {L.~M.}\ \bibnamefont {Burko}},
  \bibinfo {author} {\bibfnamefont {L.~E.}\ \bibnamefont {Kidder}}, \bibinfo
  {author} {\bibfnamefont {H.~P.}\ \bibnamefont {Pfeiffer}},\ and\ \bibinfo
  {author} {\bibfnamefont {S.~A.}\ \bibnamefont {Teukolsky}},\ }\bibfield
  {title} {\bibinfo {title} {3d simulations of linearized scalar fields in kerr
  spacetime},\ }\bibfield  {journal} {\bibinfo  {journal} {Physical Review D}\
  }\textbf {\bibinfo {volume} {69}},\ \href
  {https://doi.org/10.1103/physrevd.69.104006} {10.1103/physrevd.69.104006}
  (\bibinfo {year} {2004})\BibitemShut {NoStop}%
\bibitem [{\citenamefont {Holst}\ \emph {et~al.}(2004)\citenamefont {Holst},
  \citenamefont {Lindblom}, \citenamefont {Owen}, \citenamefont {Pfeiffer},
  \citenamefont {Scheel},\ and\ \citenamefont {Kidder}}]{Holst:2004}%
  \BibitemOpen
  \bibfield  {author} {\bibinfo {author} {\bibfnamefont {M.}~\bibnamefont
  {Holst}}, \bibinfo {author} {\bibfnamefont {L.}~\bibnamefont {Lindblom}},
  \bibinfo {author} {\bibfnamefont {R.}~\bibnamefont {Owen}}, \bibinfo {author}
  {\bibfnamefont {H.~P.}\ \bibnamefont {Pfeiffer}}, \bibinfo {author}
  {\bibfnamefont {M.~A.}\ \bibnamefont {Scheel}},\ and\ \bibinfo {author}
  {\bibfnamefont {L.~E.}\ \bibnamefont {Kidder}},\ }\bibfield  {title}
  {\bibinfo {title} {Optimal constraint projection for hyperbolic evolution
  systems},\ }\bibfield  {journal} {\bibinfo  {journal} {Physical Review D}\
  }\textbf {\bibinfo {volume} {70}},\ \href
  {https://doi.org/10.1103/physrevd.70.084017} {10.1103/physrevd.70.084017}
  (\bibinfo {year} {2004})\BibitemShut {NoStop}%
\bibitem [{\citenamefont {Lindblom}\ \emph {et~al.}(2006)\citenamefont
  {Lindblom}, \citenamefont {Scheel}, \citenamefont {Kidder}, \citenamefont
  {Owen},\ and\ \citenamefont {Rinne}}]{Lindblom:2006}%
  \BibitemOpen
  \bibfield  {author} {\bibinfo {author} {\bibfnamefont {L.}~\bibnamefont
  {Lindblom}}, \bibinfo {author} {\bibfnamefont {M.~A.}\ \bibnamefont
  {Scheel}}, \bibinfo {author} {\bibfnamefont {L.~E.}\ \bibnamefont {Kidder}},
  \bibinfo {author} {\bibfnamefont {R.}~\bibnamefont {Owen}},\ and\ \bibinfo
  {author} {\bibfnamefont {O.}~\bibnamefont {Rinne}},\ }\bibfield  {title}
  {\bibinfo {title} {A new generalized harmonic evolution system},\ }\href
  {https://doi.org/10.1088/0264-9381/23/16/S09} {\bibfield  {journal} {\bibinfo
   {journal} {Classical and Quantum Gravity}\ }\textbf {\bibinfo {volume}
  {23}},\ \bibinfo {pages} {S447} (\bibinfo {year} {2006})}\BibitemShut
  {NoStop}%
\bibitem [{\citenamefont {Bayliss}\ and\ \citenamefont
  {Turkel}(1980)}]{Bayliss:1980}%
  \BibitemOpen
  \bibfield  {author} {\bibinfo {author} {\bibfnamefont {A.}~\bibnamefont
  {Bayliss}}\ and\ \bibinfo {author} {\bibfnamefont {E.}~\bibnamefont
  {Turkel}},\ }\bibfield  {title} {\bibinfo {title} {Radiation boundary
  conditions for wave-like equations},\ }\href
  {https://doi.org/https://doi.org/10.1002/cpa.3160330603} {\bibfield
  {journal} {\bibinfo  {journal} {Communications on Pure and Applied
  Mathematics}\ }\textbf {\bibinfo {volume} {33}},\ \bibinfo {pages} {707}
  (\bibinfo {year} {1980})},\ \Eprint
  {https://arxiv.org/abs/https://onlinelibrary.wiley.com/doi/pdf/10.1002/cpa.3160330603}
  {https://onlinelibrary.wiley.com/doi/pdf/10.1002/cpa.3160330603} \BibitemShut
  {NoStop}%
\bibitem [{\citenamefont {Bj\o{}rhus}(1995)}]{Bjorhus:1995}%
  \BibitemOpen
  \bibfield  {author} {\bibinfo {author} {\bibfnamefont {M.}~\bibnamefont
  {Bj\o{}rhus}},\ }\bibfield  {title} {\bibinfo {title} {The ode formulation of
  hyperbolic pdes discretized by the spectral collocation method},\ }\href
  {https://doi.org/10.1137/0916035} {\bibfield  {journal} {\bibinfo  {journal}
  {SIAM Journal on Scientific Computing}\ }\textbf {\bibinfo {volume} {16}},\
  \bibinfo {pages} {542} (\bibinfo {year} {1995})},\ \Eprint
  {https://arxiv.org/abs/https://doi.org/10.1137/0916035}
  {https://doi.org/10.1137/0916035} \BibitemShut {NoStop}%
\bibitem [{\citenamefont {Kidder}\ \emph {et~al.}(2005)\citenamefont {Kidder},
  \citenamefont {Lindblom}, \citenamefont {Scheel}, \citenamefont {Buchman},\
  and\ \citenamefont {Pfeiffer}}]{Kidder:2005}%
  \BibitemOpen
  \bibfield  {author} {\bibinfo {author} {\bibfnamefont {L.~E.}\ \bibnamefont
  {Kidder}}, \bibinfo {author} {\bibfnamefont {L.}~\bibnamefont {Lindblom}},
  \bibinfo {author} {\bibfnamefont {M.~A.}\ \bibnamefont {Scheel}}, \bibinfo
  {author} {\bibfnamefont {L.~T.}\ \bibnamefont {Buchman}},\ and\ \bibinfo
  {author} {\bibfnamefont {H.~P.}\ \bibnamefont {Pfeiffer}},\ }\bibfield
  {title} {\bibinfo {title} {Boundary conditions for the einstein evolution
  system},\ }\bibfield  {journal} {\bibinfo  {journal} {Physical Review D}\
  }\textbf {\bibinfo {volume} {71}},\ \href
  {https://doi.org/10.1103/physrevd.71.064020} {10.1103/physrevd.71.064020}
  (\bibinfo {year} {2005})\BibitemShut {NoStop}%
\bibitem [{\citenamefont {Hesthaven}\ and\ \citenamefont
  {Warburton}(2007)}]{Hesthaven:2007}%
  \BibitemOpen
  \bibfield  {author} {\bibinfo {author} {\bibfnamefont {J.}~\bibnamefont
  {Hesthaven}}\ and\ \bibinfo {author} {\bibfnamefont {T.}~\bibnamefont
  {Warburton}},\ }\bibinfo {title} {Nodal discontinuous galerkin methods:
  Algorithms, analysis, and applications}\ (\bibinfo  {publisher} {Springer New
  York, NY},\ \bibinfo {year} {2007})\BibitemShut {NoStop}%
\bibitem [{\citenamefont {Scheel}\ \emph {et~al.}(2006)\citenamefont {Scheel},
  \citenamefont {Pfeiffer}, \citenamefont {Lindblom}, \citenamefont {Kidder},
  \citenamefont {Rinne},\ and\ \citenamefont {Teukolsky}}]{Scheel:2006}%
  \BibitemOpen
  \bibfield  {author} {\bibinfo {author} {\bibfnamefont {M.~A.}\ \bibnamefont
  {Scheel}}, \bibinfo {author} {\bibfnamefont {H.~P.}\ \bibnamefont
  {Pfeiffer}}, \bibinfo {author} {\bibfnamefont {L.}~\bibnamefont {Lindblom}},
  \bibinfo {author} {\bibfnamefont {L.~E.}\ \bibnamefont {Kidder}}, \bibinfo
  {author} {\bibfnamefont {O.}~\bibnamefont {Rinne}},\ and\ \bibinfo {author}
  {\bibfnamefont {S.~A.}\ \bibnamefont {Teukolsky}},\ }\bibfield  {title}
  {\bibinfo {title} {Solving einstein's equations with dual coordinate
  frames},\ }\bibfield  {journal} {\bibinfo  {journal} {Physical Review D}\
  }\textbf {\bibinfo {volume} {74}},\ \href
  {https://doi.org/10.1103/physrevd.74.104006} {10.1103/physrevd.74.104006}
  (\bibinfo {year} {2006})\BibitemShut {NoStop}%
\bibitem [{\citenamefont {Hemberger}\ \emph {et~al.}(2013)\citenamefont
  {Hemberger}, \citenamefont {Scheel}, \citenamefont {Kidder}, \citenamefont
  {Szil\'agyi}, \citenamefont {Lovelace}, \citenamefont {Taylor},\ and\
  \citenamefont {Teukolsky}}]{Hemberger:2012jz}%
  \BibitemOpen
  \bibfield  {author} {\bibinfo {author} {\bibfnamefont {D.~A.}\ \bibnamefont
  {Hemberger}}, \bibinfo {author} {\bibfnamefont {M.~A.}\ \bibnamefont
  {Scheel}}, \bibinfo {author} {\bibfnamefont {L.~E.}\ \bibnamefont {Kidder}},
  \bibinfo {author} {\bibfnamefont {B.}~\bibnamefont {Szil\'agyi}}, \bibinfo
  {author} {\bibfnamefont {G.}~\bibnamefont {Lovelace}}, \bibinfo {author}
  {\bibfnamefont {N.~W.}\ \bibnamefont {Taylor}},\ and\ \bibinfo {author}
  {\bibfnamefont {S.~A.}\ \bibnamefont {Teukolsky}},\ }\bibfield  {title}
  {\bibinfo {title} {{Dynamical Excision Boundaries in Spectral Evolutions of
  Binary Black Hole Spacetimes}},\ }\href
  {https://doi.org/10.1088/0264-9381/30/11/115001} {\bibfield  {journal}
  {\bibinfo  {journal} {Class. Quant. Grav.}\ }\textbf {\bibinfo {volume}
  {30}},\ \bibinfo {pages} {115001} (\bibinfo {year} {2013})},\ \Eprint
  {https://arxiv.org/abs/1211.6079} {arXiv:1211.6079 [gr-qc]} \BibitemShut
  {NoStop}%
\bibitem [{\citenamefont {Scheel}\ \emph {et~al.}(2015)\citenamefont {Scheel},
  \citenamefont {Giesler}, \citenamefont {Hemberger}, \citenamefont {Lovelace},
  \citenamefont {Kuper}, \citenamefont {Boyle}, \citenamefont {Szil\'agyi},\
  and\ \citenamefont {Kidder}}]{Scheel:2014ina}%
  \BibitemOpen
  \bibfield  {author} {\bibinfo {author} {\bibfnamefont {M.~A.}\ \bibnamefont
  {Scheel}}, \bibinfo {author} {\bibfnamefont {M.}~\bibnamefont {Giesler}},
  \bibinfo {author} {\bibfnamefont {D.~A.}\ \bibnamefont {Hemberger}}, \bibinfo
  {author} {\bibfnamefont {G.}~\bibnamefont {Lovelace}}, \bibinfo {author}
  {\bibfnamefont {K.}~\bibnamefont {Kuper}}, \bibinfo {author} {\bibfnamefont
  {M.}~\bibnamefont {Boyle}}, \bibinfo {author} {\bibfnamefont
  {B.}~\bibnamefont {Szil\'agyi}},\ and\ \bibinfo {author} {\bibfnamefont
  {L.~E.}\ \bibnamefont {Kidder}},\ }\bibfield  {title} {\bibinfo {title}
  {{Improved methods for simulating nearly extremal binary black holes}},\
  }\href {https://doi.org/10.1088/0264-9381/32/10/105009} {\bibfield  {journal}
  {\bibinfo  {journal} {Class. Quant. Grav.}\ }\textbf {\bibinfo {volume}
  {32}},\ \bibinfo {pages} {105009} (\bibinfo {year} {2015})},\ \Eprint
  {https://arxiv.org/abs/1412.1803} {arXiv:1412.1803 [gr-qc]} \BibitemShut
  {NoStop}%
\bibitem [{\citenamefont {Kale}\ \emph
  {et~al.}(2021{\natexlab{a}})\citenamefont {Kale}, \citenamefont {Acun},
  \citenamefont {Bak}, \citenamefont {Becker}, \citenamefont {Bhandarkar},
  \citenamefont {Bhat}, \citenamefont {Bhatele}, \citenamefont {Bohm},
  \citenamefont {Bordage}, \citenamefont {Brunner}, \citenamefont {Buch},
  \citenamefont {Chakravorty}, \citenamefont {Chandrasekar}, \citenamefont
  {Choi}, \citenamefont {Denardo}, \citenamefont {DeSouza}, \citenamefont
  {Diener}, \citenamefont {Dokania}, \citenamefont {Dooley}, \citenamefont
  {Fenton}, \citenamefont {Fink}, \citenamefont {Galvez}, \citenamefont
  {Ghosh}, \citenamefont {Gioachin}, \citenamefont {Gupta}, \citenamefont
  {Gupta}, \citenamefont {Gupta}, \citenamefont {Gursoy}, \citenamefont
  {Harsh}, \citenamefont {Hu}, \citenamefont {Huang}, \citenamefont
  {Jagathesan}, \citenamefont {Jain}, \citenamefont {Jetley}, \citenamefont
  {Jindal}, \citenamefont {Kanakagiri}, \citenamefont {Koenig}, \citenamefont
  {Krishnan}, \citenamefont {Kumar}, \citenamefont {Kunzman}, \citenamefont
  {Lang}, \citenamefont {Langer}, \citenamefont {Lawlor}, \citenamefont {Lee},
  \citenamefont {Lifflander}, \citenamefont {Mahesh}, \citenamefont {Mendes},
  \citenamefont {Menon}, \citenamefont {Mei}, \citenamefont {Meneses},
  \citenamefont {Mikida}, \citenamefont {Miller}, \citenamefont {Mokos},
  \citenamefont {Narayanan}, \citenamefont {Ni}, \citenamefont {Nomura},
  \citenamefont {Paranjpye}, \citenamefont {Ramachandran}, \citenamefont
  {Ramkumar}, \citenamefont {Ramos}, \citenamefont {Robson}, \citenamefont
  {Saboo}, \citenamefont {Saletore}, \citenamefont {Sarood}, \citenamefont
  {Senthil}, \citenamefont {Shah}, \citenamefont {Shu}, \citenamefont {Sinha},
  \citenamefont {Sun}, \citenamefont {Sura}, \citenamefont {Szaday},
  \citenamefont {Totoni}, \citenamefont {Varadarajan}, \citenamefont
  {Venkataraman}, \citenamefont {Wang}, \citenamefont {Wesolowski},
  \citenamefont {White}, \citenamefont {Wilmarth}, \citenamefont {Wright},
  \citenamefont {Yelon},\ and\ \citenamefont {Zheng}}]{charm}%
  \BibitemOpen
  \bibfield  {author} {\bibinfo {author} {\bibfnamefont {L.}~\bibnamefont
  {Kale}}, \bibinfo {author} {\bibfnamefont {B.}~\bibnamefont {Acun}}, \bibinfo
  {author} {\bibfnamefont {S.}~\bibnamefont {Bak}}, \bibinfo {author}
  {\bibfnamefont {A.}~\bibnamefont {Becker}}, \bibinfo {author} {\bibfnamefont
  {M.}~\bibnamefont {Bhandarkar}}, \bibinfo {author} {\bibfnamefont
  {N.}~\bibnamefont {Bhat}}, \bibinfo {author} {\bibfnamefont {A.}~\bibnamefont
  {Bhatele}}, \bibinfo {author} {\bibfnamefont {E.}~\bibnamefont {Bohm}},
  \bibinfo {author} {\bibfnamefont {C.}~\bibnamefont {Bordage}}, \bibinfo
  {author} {\bibfnamefont {R.}~\bibnamefont {Brunner}}, \bibinfo {author}
  {\bibfnamefont {R.}~\bibnamefont {Buch}}, \bibinfo {author} {\bibfnamefont
  {S.}~\bibnamefont {Chakravorty}}, \bibinfo {author} {\bibfnamefont
  {K.}~\bibnamefont {Chandrasekar}}, \bibinfo {author} {\bibfnamefont
  {J.}~\bibnamefont {Choi}}, \bibinfo {author} {\bibfnamefont {M.}~\bibnamefont
  {Denardo}}, \bibinfo {author} {\bibfnamefont {J.}~\bibnamefont {DeSouza}},
  \bibinfo {author} {\bibfnamefont {M.}~\bibnamefont {Diener}}, \bibinfo
  {author} {\bibfnamefont {H.}~\bibnamefont {Dokania}}, \bibinfo {author}
  {\bibfnamefont {I.}~\bibnamefont {Dooley}}, \bibinfo {author} {\bibfnamefont
  {W.}~\bibnamefont {Fenton}}, \bibinfo {author} {\bibfnamefont
  {Z.}~\bibnamefont {Fink}}, \bibinfo {author} {\bibfnamefont {J.}~\bibnamefont
  {Galvez}}, \bibinfo {author} {\bibfnamefont {P.}~\bibnamefont {Ghosh}},
  \bibinfo {author} {\bibfnamefont {F.}~\bibnamefont {Gioachin}}, \bibinfo
  {author} {\bibfnamefont {A.}~\bibnamefont {Gupta}}, \bibinfo {author}
  {\bibfnamefont {G.}~\bibnamefont {Gupta}}, \bibinfo {author} {\bibfnamefont
  {M.}~\bibnamefont {Gupta}}, \bibinfo {author} {\bibfnamefont
  {A.}~\bibnamefont {Gursoy}}, \bibinfo {author} {\bibfnamefont
  {V.}~\bibnamefont {Harsh}}, \bibinfo {author} {\bibfnamefont
  {F.}~\bibnamefont {Hu}}, \bibinfo {author} {\bibfnamefont {C.}~\bibnamefont
  {Huang}}, \bibinfo {author} {\bibfnamefont {N.}~\bibnamefont {Jagathesan}},
  \bibinfo {author} {\bibfnamefont {N.}~\bibnamefont {Jain}}, \bibinfo {author}
  {\bibfnamefont {P.}~\bibnamefont {Jetley}}, \bibinfo {author} {\bibfnamefont
  {P.}~\bibnamefont {Jindal}}, \bibinfo {author} {\bibfnamefont
  {R.}~\bibnamefont {Kanakagiri}}, \bibinfo {author} {\bibfnamefont
  {G.}~\bibnamefont {Koenig}}, \bibinfo {author} {\bibfnamefont
  {S.}~\bibnamefont {Krishnan}}, \bibinfo {author} {\bibfnamefont
  {S.}~\bibnamefont {Kumar}}, \bibinfo {author} {\bibfnamefont
  {D.}~\bibnamefont {Kunzman}}, \bibinfo {author} {\bibfnamefont
  {M.}~\bibnamefont {Lang}}, \bibinfo {author} {\bibfnamefont {A.}~\bibnamefont
  {Langer}}, \bibinfo {author} {\bibfnamefont {O.}~\bibnamefont {Lawlor}},
  \bibinfo {author} {\bibfnamefont {C.~W.}\ \bibnamefont {Lee}}, \bibinfo
  {author} {\bibfnamefont {J.}~\bibnamefont {Lifflander}}, \bibinfo {author}
  {\bibfnamefont {K.}~\bibnamefont {Mahesh}}, \bibinfo {author} {\bibfnamefont
  {C.}~\bibnamefont {Mendes}}, \bibinfo {author} {\bibfnamefont
  {H.}~\bibnamefont {Menon}}, \bibinfo {author} {\bibfnamefont
  {C.}~\bibnamefont {Mei}}, \bibinfo {author} {\bibfnamefont {E.}~\bibnamefont
  {Meneses}}, \bibinfo {author} {\bibfnamefont {E.}~\bibnamefont {Mikida}},
  \bibinfo {author} {\bibfnamefont {P.}~\bibnamefont {Miller}}, \bibinfo
  {author} {\bibfnamefont {R.}~\bibnamefont {Mokos}}, \bibinfo {author}
  {\bibfnamefont {V.}~\bibnamefont {Narayanan}}, \bibinfo {author}
  {\bibfnamefont {X.}~\bibnamefont {Ni}}, \bibinfo {author} {\bibfnamefont
  {K.}~\bibnamefont {Nomura}}, \bibinfo {author} {\bibfnamefont
  {S.}~\bibnamefont {Paranjpye}}, \bibinfo {author} {\bibfnamefont
  {P.}~\bibnamefont {Ramachandran}}, \bibinfo {author} {\bibfnamefont
  {B.}~\bibnamefont {Ramkumar}}, \bibinfo {author} {\bibfnamefont
  {E.}~\bibnamefont {Ramos}}, \bibinfo {author} {\bibfnamefont
  {M.}~\bibnamefont {Robson}}, \bibinfo {author} {\bibfnamefont
  {N.}~\bibnamefont {Saboo}}, \bibinfo {author} {\bibfnamefont
  {V.}~\bibnamefont {Saletore}}, \bibinfo {author} {\bibfnamefont
  {O.}~\bibnamefont {Sarood}}, \bibinfo {author} {\bibfnamefont
  {K.}~\bibnamefont {Senthil}}, \bibinfo {author} {\bibfnamefont
  {N.}~\bibnamefont {Shah}}, \bibinfo {author} {\bibfnamefont {W.}~\bibnamefont
  {Shu}}, \bibinfo {author} {\bibfnamefont {A.~B.}\ \bibnamefont {Sinha}},
  \bibinfo {author} {\bibfnamefont {Y.}~\bibnamefont {Sun}}, \bibinfo {author}
  {\bibfnamefont {Z.}~\bibnamefont {Sura}}, \bibinfo {author} {\bibfnamefont
  {J.}~\bibnamefont {Szaday}}, \bibinfo {author} {\bibfnamefont
  {E.}~\bibnamefont {Totoni}}, \bibinfo {author} {\bibfnamefont
  {K.}~\bibnamefont {Varadarajan}}, \bibinfo {author} {\bibfnamefont
  {R.}~\bibnamefont {Venkataraman}}, \bibinfo {author} {\bibfnamefont
  {J.}~\bibnamefont {Wang}}, \bibinfo {author} {\bibfnamefont {L.}~\bibnamefont
  {Wesolowski}}, \bibinfo {author} {\bibfnamefont {S.}~\bibnamefont {White}},
  \bibinfo {author} {\bibfnamefont {T.}~\bibnamefont {Wilmarth}}, \bibinfo
  {author} {\bibfnamefont {J.}~\bibnamefont {Wright}}, \bibinfo {author}
  {\bibfnamefont {J.}~\bibnamefont {Yelon}},\ and\ \bibinfo {author}
  {\bibfnamefont {G.}~\bibnamefont {Zheng}},\ }\href
  {https://doi.org/10.5281/zenodo.5597907} {\bibinfo {title} {Uiuc-ppl/charm:
  Charm++ version 7.0.0}} (\bibinfo {year} {2021}{\natexlab{a}})\BibitemShut
  {NoStop}%
\bibitem [{\citenamefont {Detweiler}\ and\ \citenamefont
  {Whiting}(2003)}]{Detweiler:2002mi}%
  \BibitemOpen
  \bibfield  {author} {\bibinfo {author} {\bibfnamefont {S.~L.}\ \bibnamefont
  {Detweiler}}\ and\ \bibinfo {author} {\bibfnamefont {B.~F.}\ \bibnamefont
  {Whiting}},\ }\bibfield  {title} {\bibinfo {title} {{Selfforce via a Green's
  functsion decomposition}},\ }\href
  {https://doi.org/10.1103/PhysRevD.67.024025} {\bibfield  {journal} {\bibinfo
  {journal} {Phys. Rev. D}\ }\textbf {\bibinfo {volume} {67}},\ \bibinfo
  {pages} {024025} (\bibinfo {year} {2003})},\ \Eprint
  {https://arxiv.org/abs/gr-qc/0202086} {arXiv:gr-qc/0202086} \BibitemShut
  {NoStop}%
\bibitem [{\citenamefont {Heffernan}\ \emph {et~al.}(2012)\citenamefont
  {Heffernan}, \citenamefont {Ottewill},\ and\ \citenamefont
  {Wardell}}]{Heffernan:2012su}%
  \BibitemOpen
  \bibfield  {author} {\bibinfo {author} {\bibfnamefont {A.}~\bibnamefont
  {Heffernan}}, \bibinfo {author} {\bibfnamefont {A.}~\bibnamefont
  {Ottewill}},\ and\ \bibinfo {author} {\bibfnamefont {B.}~\bibnamefont
  {Wardell}},\ }\bibfield  {title} {\bibinfo {title} {{High-order expansions of
  the Detweiler-Whiting singular field in Schwarzschild spacetime}},\ }\href
  {https://doi.org/10.1103/PhysRevD.86.104023} {\bibfield  {journal} {\bibinfo
  {journal} {Phys. Rev. D}\ }\textbf {\bibinfo {volume} {86}},\ \bibinfo
  {pages} {104023} (\bibinfo {year} {2012})},\ \Eprint
  {https://arxiv.org/abs/1204.0794} {arXiv:1204.0794 [gr-qc]} \BibitemShut
  {NoStop}%
\bibitem [{\citenamefont {Haas}\ and\ \citenamefont
  {Poisson}(2006)}]{Haas:2006ne}%
  \BibitemOpen
  \bibfield  {author} {\bibinfo {author} {\bibfnamefont {R.}~\bibnamefont
  {Haas}}\ and\ \bibinfo {author} {\bibfnamefont {E.}~\bibnamefont {Poisson}},\
  }\bibfield  {title} {\bibinfo {title} {{Mode-sum regularization of the scalar
  self-force: Formulation in terms of a tetrad decomposition of the singular
  field}},\ }\href {https://doi.org/10.1103/PhysRevD.74.044009} {\bibfield
  {journal} {\bibinfo  {journal} {Phys. Rev. D}\ }\textbf {\bibinfo {volume}
  {74}},\ \bibinfo {pages} {044009} (\bibinfo {year} {2006})},\ \Eprint
  {https://arxiv.org/abs/gr-qc/0605077} {arXiv:gr-qc/0605077} \BibitemShut
  {NoStop}%
\bibitem [{\citenamefont {Wardell}\ \emph {et~al.}(2012)\citenamefont
  {Wardell}, \citenamefont {Vega}, \citenamefont {Thornburg},\ and\
  \citenamefont {Diener}}]{Wardell:2011gb}%
  \BibitemOpen
  \bibfield  {author} {\bibinfo {author} {\bibfnamefont {B.}~\bibnamefont
  {Wardell}}, \bibinfo {author} {\bibfnamefont {I.}~\bibnamefont {Vega}},
  \bibinfo {author} {\bibfnamefont {J.}~\bibnamefont {Thornburg}},\ and\
  \bibinfo {author} {\bibfnamefont {P.}~\bibnamefont {Diener}},\ }\bibfield
  {title} {\bibinfo {title} {{A Generic effective source for scalar self-force
  calculations}},\ }\href {https://doi.org/10.1103/PhysRevD.85.104044}
  {\bibfield  {journal} {\bibinfo  {journal} {Phys. Rev. D}\ }\textbf {\bibinfo
  {volume} {85}},\ \bibinfo {pages} {104044} (\bibinfo {year} {2012})},\
  \Eprint {https://arxiv.org/abs/1112.6355} {arXiv:1112.6355 [gr-qc]}
  \BibitemShut {NoStop}%
\bibitem [{\citenamefont {Synge}(1960)}]{J.L.Synge:1960zz}%
  \BibitemOpen
  \bibfield  {author} {\bibinfo {author} {\bibfnamefont {J.~L.}\ \bibnamefont
  {Synge}},\ }\href@noop {} {\emph {\bibinfo {title} {{Relativity: The General
  theory}}}}\ (\bibinfo  {publisher} {North-Holland Publishing Company},\
  \bibinfo {year} {1960})\BibitemShut {NoStop}%
\bibitem [{\citenamefont {Barack}\ and\ \citenamefont
  {Ori}(2002)}]{Barack:2002mha}%
  \BibitemOpen
  \bibfield  {author} {\bibinfo {author} {\bibfnamefont {L.}~\bibnamefont
  {Barack}}\ and\ \bibinfo {author} {\bibfnamefont {A.}~\bibnamefont {Ori}},\
  }\bibfield  {title} {\bibinfo {title} {{Regularization parameters for the
  self-force in Schwarzschild space-time. 1. Scalar case}},\ }\href
  {https://doi.org/10.1103/PhysRevD.66.084022} {\bibfield  {journal} {\bibinfo
  {journal} {Phys. Rev. D}\ }\textbf {\bibinfo {volume} {66}},\ \bibinfo
  {pages} {084022} (\bibinfo {year} {2002})},\ \Eprint
  {https://arxiv.org/abs/gr-qc/0204093} {arXiv:gr-qc/0204093} \BibitemShut
  {NoStop}%
\bibitem [{\citenamefont {Pound}(2012)}]{Pound:2012dk}%
  \BibitemOpen
  \bibfield  {author} {\bibinfo {author} {\bibfnamefont {A.}~\bibnamefont
  {Pound}},\ }\bibfield  {title} {\bibinfo {title} {{Nonlinear gravitational
  self-force. I. Field outside a small body}},\ }\href
  {https://doi.org/10.1103/PhysRevD.86.084019} {\bibfield  {journal} {\bibinfo
  {journal} {Phys. Rev. D}\ }\textbf {\bibinfo {volume} {86}},\ \bibinfo
  {pages} {084019} (\bibinfo {year} {2012})},\ \Eprint
  {https://arxiv.org/abs/1206.6538} {arXiv:1206.6538 [gr-qc]} \BibitemShut
  {NoStop}%
\bibitem [{\citenamefont {Mino}(2003)}]{Mino:2003yg}%
  \BibitemOpen
  \bibfield  {author} {\bibinfo {author} {\bibfnamefont {Y.}~\bibnamefont
  {Mino}},\ }\bibfield  {title} {\bibinfo {title} {{Perturbative approach to an
  orbital evolution around a supermassive black hole}},\ }\href
  {https://doi.org/10.1103/PhysRevD.67.084027} {\bibfield  {journal} {\bibinfo
  {journal} {Phys. Rev. D}\ }\textbf {\bibinfo {volume} {67}},\ \bibinfo
  {pages} {084027} (\bibinfo {year} {2003})},\ \Eprint
  {https://arxiv.org/abs/gr-qc/0302075} {arXiv:gr-qc/0302075} \BibitemShut
  {NoStop}%
\bibitem [{\citenamefont {Hinderer}\ and\ \citenamefont
  {Flanagan}(2008)}]{Hinderer:2008dm}%
  \BibitemOpen
  \bibfield  {author} {\bibinfo {author} {\bibfnamefont {T.}~\bibnamefont
  {Hinderer}}\ and\ \bibinfo {author} {\bibfnamefont {E.~E.}\ \bibnamefont
  {Flanagan}},\ }\bibfield  {title} {\bibinfo {title} {{Two timescale analysis
  of extreme mass ratio inspirals in Kerr. I. Orbital Motion}},\ }\href
  {https://doi.org/10.1103/PhysRevD.78.064028} {\bibfield  {journal} {\bibinfo
  {journal} {Phys. Rev. D}\ }\textbf {\bibinfo {volume} {78}},\ \bibinfo
  {pages} {064028} (\bibinfo {year} {2008})},\ \Eprint
  {https://arxiv.org/abs/0805.3337} {arXiv:0805.3337 [gr-qc]} \BibitemShut
  {NoStop}%
\bibitem [{\citenamefont {Diaz-Rivera}\ \emph {et~al.}(2004)\citenamefont
  {Diaz-Rivera}, \citenamefont {Messaritaki}, \citenamefont {Whiting},\ and\
  \citenamefont {Detweiler}}]{Diaz_Rivera:2004}%
  \BibitemOpen
  \bibfield  {author} {\bibinfo {author} {\bibfnamefont {L.}~\bibnamefont
  {Diaz-Rivera}}, \bibinfo {author} {\bibfnamefont {E.}~\bibnamefont
  {Messaritaki}}, \bibinfo {author} {\bibfnamefont {B.}~\bibnamefont
  {Whiting}},\ and\ \bibinfo {author} {\bibfnamefont {S.}~\bibnamefont
  {Detweiler}},\ }\bibfield  {title} {\bibinfo {title} {Scalar field self-force
  effects on orbits about a schwarzschild black hole},\ }\bibfield  {journal}
  {\bibinfo  {journal} {Physical Review D}\ }\textbf {\bibinfo {volume} {70}},\
  \href {https://doi.org/10.1103/physrevd.70.124018}
  {10.1103/physrevd.70.124018} (\bibinfo {year} {2004})\BibitemShut {NoStop}%
\bibitem [{\citenamefont {Macedo}\ \emph {et~al.}(2022)\citenamefont {Macedo},
  \citenamefont {Leather}, \citenamefont {Warburton}, \citenamefont {Wardell},\
  and\ \citenamefont {Zengino{\u{g} }lu}}]{Macedo_2022}%
  \BibitemOpen
  \bibfield  {author} {\bibinfo {author} {\bibfnamefont {R.~P.}\ \bibnamefont
  {Macedo}}, \bibinfo {author} {\bibfnamefont {B.}~\bibnamefont {Leather}},
  \bibinfo {author} {\bibfnamefont {N.}~\bibnamefont {Warburton}}, \bibinfo
  {author} {\bibfnamefont {B.}~\bibnamefont {Wardell}},\ and\ \bibinfo {author}
  {\bibfnamefont {A.}~\bibnamefont {Zengino{\u{g} }lu}},\ }\bibfield  {title}
  {\bibinfo {title} {Hyperboloidal method for frequency-domain self-force
  calculations},\ }\bibfield  {journal} {\bibinfo  {journal} {Physical Review
  D}\ }\textbf {\bibinfo {volume} {105}},\ \href
  {https://doi.org/10.1103/physrevd.105.104033} {10.1103/physrevd.105.104033}
  (\bibinfo {year} {2022})\BibitemShut {NoStop}%
\bibitem [{\citenamefont {Diener}\ \emph {et~al.}(2012)\citenamefont {Diener},
  \citenamefont {Vega}, \citenamefont {Wardell},\ and\ \citenamefont
  {Detweiler}}]{Diener:2012}%
  \BibitemOpen
  \bibfield  {author} {\bibinfo {author} {\bibfnamefont {P.}~\bibnamefont
  {Diener}}, \bibinfo {author} {\bibfnamefont {I.}~\bibnamefont {Vega}},
  \bibinfo {author} {\bibfnamefont {B.}~\bibnamefont {Wardell}},\ and\ \bibinfo
  {author} {\bibfnamefont {S.}~\bibnamefont {Detweiler}},\ }\bibfield  {title}
  {\bibinfo {title} {Self-consistent orbital evolution of a particle around a
  schwarzschild black hole},\ }\bibfield  {journal} {\bibinfo  {journal}
  {Physical Review Letters}\ }\textbf {\bibinfo {volume} {108}},\ \href
  {https://doi.org/10.1103/physrevlett.108.191102}
  {10.1103/physrevlett.108.191102} (\bibinfo {year} {2012})\BibitemShut
  {NoStop}%
\bibitem [{\citenamefont {Quinn}(2000)}]{Quinn_2000}%
  \BibitemOpen
  \bibfield  {author} {\bibinfo {author} {\bibfnamefont {T.~C.}\ \bibnamefont
  {Quinn}},\ }\bibfield  {title} {\bibinfo {title} {Axiomatic approach to
  radiation reaction of scalar point particles in curved spacetime},\
  }\bibfield  {journal} {\bibinfo  {journal} {Physical Review D}\ }\textbf
  {\bibinfo {volume} {62}},\ \href {https://doi.org/10.1103/physrevd.62.064029}
  {10.1103/physrevd.62.064029} (\bibinfo {year} {2000})\BibitemShut {NoStop}%
\bibitem [{\citenamefont {Kale}\ \emph
  {et~al.}(2021{\natexlab{b}})\citenamefont {Kale}, \citenamefont {Acun},
  \citenamefont {Bak}, \citenamefont {Becker}, \citenamefont {Bhandarkar},
  \citenamefont {Bhat}, \citenamefont {Bhatele}, \citenamefont {Bohm},
  \citenamefont {Bordage}, \citenamefont {Brunner}, \citenamefont {Buch},
  \citenamefont {Chakravorty}, \citenamefont {Chandrasekar}, \citenamefont
  {Choi}, \citenamefont {Denardo}, \citenamefont {DeSouza}, \citenamefont
  {Diener}, \citenamefont {Dokania}, \citenamefont {Dooley}, \citenamefont
  {Fenton}, \citenamefont {Fink}, \citenamefont {Galvez}, \citenamefont
  {Ghosh}, \citenamefont {Gioachin}, \citenamefont {Gupta}, \citenamefont
  {Gupta}, \citenamefont {Gupta}, \citenamefont {Gursoy}, \citenamefont
  {Harsh}, \citenamefont {Hu}, \citenamefont {Huang}, \citenamefont
  {Jagathesan}, \citenamefont {Jain}, \citenamefont {Jetley}, \citenamefont
  {Jindal}, \citenamefont {Kanakagiri}, \citenamefont {Koenig}, \citenamefont
  {Krishnan}, \citenamefont {Kumar}, \citenamefont {Kunzman}, \citenamefont
  {Lang}, \citenamefont {Langer}, \citenamefont {Lawlor}, \citenamefont {Lee},
  \citenamefont {Lifflander}, \citenamefont {Mahesh}, \citenamefont {Mendes},
  \citenamefont {Menon}, \citenamefont {Mei}, \citenamefont {Meneses},
  \citenamefont {Mikida}, \citenamefont {Miller}, \citenamefont {Mokos},
  \citenamefont {Narayanan}, \citenamefont {Ni}, \citenamefont {Nomura},
  \citenamefont {Paranjpye}, \citenamefont {Ramachandran}, \citenamefont
  {Ramkumar}, \citenamefont {Ramos}, \citenamefont {Robson}, \citenamefont
  {Saboo}, \citenamefont {Saletore}, \citenamefont {Sarood}, \citenamefont
  {Senthil}, \citenamefont {Shah}, \citenamefont {Shu}, \citenamefont {Sinha},
  \citenamefont {Sun}, \citenamefont {Sura}, \citenamefont {Szaday},
  \citenamefont {Totoni}, \citenamefont {Varadarajan}, \citenamefont
  {Venkataraman}, \citenamefont {Wang}, \citenamefont {Wesolowski},
  \citenamefont {White}, \citenamefont {Wilmarth}, \citenamefont {Wright},
  \citenamefont {Yelon},\ and\ \citenamefont
  {Zheng}}]{laxmikant_kale_2021_5597907}%
  \BibitemOpen
  \bibfield  {author} {\bibinfo {author} {\bibfnamefont {L.}~\bibnamefont
  {Kale}}, \bibinfo {author} {\bibfnamefont {B.}~\bibnamefont {Acun}}, \bibinfo
  {author} {\bibfnamefont {S.}~\bibnamefont {Bak}}, \bibinfo {author}
  {\bibfnamefont {A.}~\bibnamefont {Becker}}, \bibinfo {author} {\bibfnamefont
  {M.}~\bibnamefont {Bhandarkar}}, \bibinfo {author} {\bibfnamefont
  {N.}~\bibnamefont {Bhat}}, \bibinfo {author} {\bibfnamefont {A.}~\bibnamefont
  {Bhatele}}, \bibinfo {author} {\bibfnamefont {E.}~\bibnamefont {Bohm}},
  \bibinfo {author} {\bibfnamefont {C.}~\bibnamefont {Bordage}}, \bibinfo
  {author} {\bibfnamefont {R.}~\bibnamefont {Brunner}}, \bibinfo {author}
  {\bibfnamefont {R.}~\bibnamefont {Buch}}, \bibinfo {author} {\bibfnamefont
  {S.}~\bibnamefont {Chakravorty}}, \bibinfo {author} {\bibfnamefont
  {K.}~\bibnamefont {Chandrasekar}}, \bibinfo {author} {\bibfnamefont
  {J.}~\bibnamefont {Choi}}, \bibinfo {author} {\bibfnamefont {M.}~\bibnamefont
  {Denardo}}, \bibinfo {author} {\bibfnamefont {J.}~\bibnamefont {DeSouza}},
  \bibinfo {author} {\bibfnamefont {M.}~\bibnamefont {Diener}}, \bibinfo
  {author} {\bibfnamefont {H.}~\bibnamefont {Dokania}}, \bibinfo {author}
  {\bibfnamefont {I.}~\bibnamefont {Dooley}}, \bibinfo {author} {\bibfnamefont
  {W.}~\bibnamefont {Fenton}}, \bibinfo {author} {\bibfnamefont
  {Z.}~\bibnamefont {Fink}}, \bibinfo {author} {\bibfnamefont {J.}~\bibnamefont
  {Galvez}}, \bibinfo {author} {\bibfnamefont {P.}~\bibnamefont {Ghosh}},
  \bibinfo {author} {\bibfnamefont {F.}~\bibnamefont {Gioachin}}, \bibinfo
  {author} {\bibfnamefont {A.}~\bibnamefont {Gupta}}, \bibinfo {author}
  {\bibfnamefont {G.}~\bibnamefont {Gupta}}, \bibinfo {author} {\bibfnamefont
  {M.}~\bibnamefont {Gupta}}, \bibinfo {author} {\bibfnamefont
  {A.}~\bibnamefont {Gursoy}}, \bibinfo {author} {\bibfnamefont
  {V.}~\bibnamefont {Harsh}}, \bibinfo {author} {\bibfnamefont
  {F.}~\bibnamefont {Hu}}, \bibinfo {author} {\bibfnamefont {C.}~\bibnamefont
  {Huang}}, \bibinfo {author} {\bibfnamefont {N.}~\bibnamefont {Jagathesan}},
  \bibinfo {author} {\bibfnamefont {N.}~\bibnamefont {Jain}}, \bibinfo {author}
  {\bibfnamefont {P.}~\bibnamefont {Jetley}}, \bibinfo {author} {\bibfnamefont
  {P.}~\bibnamefont {Jindal}}, \bibinfo {author} {\bibfnamefont
  {R.}~\bibnamefont {Kanakagiri}}, \bibinfo {author} {\bibfnamefont
  {G.}~\bibnamefont {Koenig}}, \bibinfo {author} {\bibfnamefont
  {S.}~\bibnamefont {Krishnan}}, \bibinfo {author} {\bibfnamefont
  {S.}~\bibnamefont {Kumar}}, \bibinfo {author} {\bibfnamefont
  {D.}~\bibnamefont {Kunzman}}, \bibinfo {author} {\bibfnamefont
  {M.}~\bibnamefont {Lang}}, \bibinfo {author} {\bibfnamefont {A.}~\bibnamefont
  {Langer}}, \bibinfo {author} {\bibfnamefont {O.}~\bibnamefont {Lawlor}},
  \bibinfo {author} {\bibfnamefont {C.~W.}\ \bibnamefont {Lee}}, \bibinfo
  {author} {\bibfnamefont {J.}~\bibnamefont {Lifflander}}, \bibinfo {author}
  {\bibfnamefont {K.}~\bibnamefont {Mahesh}}, \bibinfo {author} {\bibfnamefont
  {C.}~\bibnamefont {Mendes}}, \bibinfo {author} {\bibfnamefont
  {H.}~\bibnamefont {Menon}}, \bibinfo {author} {\bibfnamefont
  {C.}~\bibnamefont {Mei}}, \bibinfo {author} {\bibfnamefont {E.}~\bibnamefont
  {Meneses}}, \bibinfo {author} {\bibfnamefont {E.}~\bibnamefont {Mikida}},
  \bibinfo {author} {\bibfnamefont {P.}~\bibnamefont {Miller}}, \bibinfo
  {author} {\bibfnamefont {R.}~\bibnamefont {Mokos}}, \bibinfo {author}
  {\bibfnamefont {V.}~\bibnamefont {Narayanan}}, \bibinfo {author}
  {\bibfnamefont {X.}~\bibnamefont {Ni}}, \bibinfo {author} {\bibfnamefont
  {K.}~\bibnamefont {Nomura}}, \bibinfo {author} {\bibfnamefont
  {S.}~\bibnamefont {Paranjpye}}, \bibinfo {author} {\bibfnamefont
  {P.}~\bibnamefont {Ramachandran}}, \bibinfo {author} {\bibfnamefont
  {B.}~\bibnamefont {Ramkumar}}, \bibinfo {author} {\bibfnamefont
  {E.}~\bibnamefont {Ramos}}, \bibinfo {author} {\bibfnamefont
  {M.}~\bibnamefont {Robson}}, \bibinfo {author} {\bibfnamefont
  {N.}~\bibnamefont {Saboo}}, \bibinfo {author} {\bibfnamefont
  {V.}~\bibnamefont {Saletore}}, \bibinfo {author} {\bibfnamefont
  {O.}~\bibnamefont {Sarood}}, \bibinfo {author} {\bibfnamefont
  {K.}~\bibnamefont {Senthil}}, \bibinfo {author} {\bibfnamefont
  {N.}~\bibnamefont {Shah}}, \bibinfo {author} {\bibfnamefont {W.}~\bibnamefont
  {Shu}}, \bibinfo {author} {\bibfnamefont {A.~B.}\ \bibnamefont {Sinha}},
  \bibinfo {author} {\bibfnamefont {Y.}~\bibnamefont {Sun}}, \bibinfo {author}
  {\bibfnamefont {Z.}~\bibnamefont {Sura}}, \bibinfo {author} {\bibfnamefont
  {J.}~\bibnamefont {Szaday}}, \bibinfo {author} {\bibfnamefont
  {E.}~\bibnamefont {Totoni}}, \bibinfo {author} {\bibfnamefont
  {K.}~\bibnamefont {Varadarajan}}, \bibinfo {author} {\bibfnamefont
  {R.}~\bibnamefont {Venkataraman}}, \bibinfo {author} {\bibfnamefont
  {J.}~\bibnamefont {Wang}}, \bibinfo {author} {\bibfnamefont {L.}~\bibnamefont
  {Wesolowski}}, \bibinfo {author} {\bibfnamefont {S.}~\bibnamefont {White}},
  \bibinfo {author} {\bibfnamefont {T.}~\bibnamefont {Wilmarth}}, \bibinfo
  {author} {\bibfnamefont {J.}~\bibnamefont {Wright}}, \bibinfo {author}
  {\bibfnamefont {J.}~\bibnamefont {Yelon}},\ and\ \bibinfo {author}
  {\bibfnamefont {G.}~\bibnamefont {Zheng}},\ }\href
  {https://doi.org/10.5281/zenodo.5597907} {\bibinfo {title} {{UIUC-PPL}/charm:
  {Charm++} version 7.0.0}} (\bibinfo {year} {2021}{\natexlab{b}})\BibitemShut
  {NoStop}%
\bibitem [{\citenamefont {Kale}\ and\ \citenamefont
  {Krishnan}(1996)}]{kale1996charm++}%
  \BibitemOpen
  \bibfield  {author} {\bibinfo {author} {\bibfnamefont {L.~V.}\ \bibnamefont
  {Kale}}\ and\ \bibinfo {author} {\bibfnamefont {S.}~\bibnamefont
  {Krishnan}},\ }\bibfield  {title} {\bibinfo {title} {Charm++: Parallel
  programming with message-driven objects},\ }in\ \href@noop {} {\emph
  {\bibinfo {booktitle} {Parallel programming using C++}}},\ \bibinfo {editor}
  {edited by\ \bibinfo {editor} {\bibfnamefont {G.~V.}\ \bibnamefont {Wilson}}\
  and\ \bibinfo {editor} {\bibfnamefont {P.}~\bibnamefont {Lu}}}\ (\bibinfo
  {publisher} {The MIT Press},\ \bibinfo {year} {1996})\ pp.\ \bibinfo {pages}
  {175--213}\BibitemShut {NoStop}%
\bibitem [{\citenamefont {Iglberger}\ \emph
  {et~al.}(2012{\natexlab{a}})\citenamefont {Iglberger}, \citenamefont {Hager},
  \citenamefont {Treibig},\ and\ \citenamefont {Rüde}}]{Blaze1}%
  \BibitemOpen
  \bibfield  {author} {\bibinfo {author} {\bibfnamefont {K.}~\bibnamefont
  {Iglberger}}, \bibinfo {author} {\bibfnamefont {G.}~\bibnamefont {Hager}},
  \bibinfo {author} {\bibfnamefont {J.}~\bibnamefont {Treibig}},\ and\ \bibinfo
  {author} {\bibfnamefont {U.}~\bibnamefont {Rüde}},\ }\bibfield  {title}
  {\bibinfo {title} {High performance smart expression template math
  libraries},\ }in\ \href {https://doi.org/10.1109/HPCSim.2012.6266939} {\emph
  {\bibinfo {booktitle} {2012 International Conference on High Performance
  Computing \& Simulation (HPCS)}}}\ (\bibinfo  {publisher} {IEEE},\ \bibinfo
  {year} {2012})\ pp.\ \bibinfo {pages} {367--373}\BibitemShut {NoStop}%
\bibitem [{\citenamefont {Iglberger}\ \emph
  {et~al.}(2012{\natexlab{b}})\citenamefont {Iglberger}, \citenamefont {Hager},
  \citenamefont {Treibig},\ and\ \citenamefont {R\"{u}de}}]{Blaze2}%
  \BibitemOpen
  \bibfield  {author} {\bibinfo {author} {\bibfnamefont {K.}~\bibnamefont
  {Iglberger}}, \bibinfo {author} {\bibfnamefont {G.}~\bibnamefont {Hager}},
  \bibinfo {author} {\bibfnamefont {J.}~\bibnamefont {Treibig}},\ and\ \bibinfo
  {author} {\bibfnamefont {U.}~\bibnamefont {R\"{u}de}},\ }\bibfield  {title}
  {\bibinfo {title} {Expression templates revisited: A performance analysis of
  current methodologies},\ }\href {https://doi.org/10.1137/110830125}
  {\bibfield  {journal} {\bibinfo  {journal} {SIAM Journal on Scientific
  Computing}\ }\textbf {\bibinfo {volume} {34}},\ \bibinfo {pages} {C42}
  (\bibinfo {year} {2012}{\natexlab{b}})},\ \Eprint
  {https://arxiv.org/abs/https://doi.org/10.1137/110830125}
  {https://doi.org/10.1137/110830125} \BibitemShut {NoStop}%
\bibitem [{\citenamefont {{The HDF Group}}(2023)}]{hdf5}%
  \BibitemOpen
  \bibfield  {author} {\bibinfo {author} {\bibnamefont {{The HDF Group}}},\
  }\href@noop {} {\bibinfo {title} {{Hierarchical Data Format, version 5}}}
  (\bibinfo {year} {1997-2023}),\ \bibinfo {note}
  {https://www.hdfgroup.org/HDF5/}\BibitemShut {NoStop}%
\bibitem [{\citenamefont {Galassi}\ \emph {et~al.}(2009)\citenamefont {Galassi}
  \emph {et~al.}}]{gsl}%
  \BibitemOpen
  \bibfield  {author} {\bibinfo {author} {\bibfnamefont {M.}~\bibnamefont
  {Galassi}} \emph {et~al.},\ }\href@noop {} {\emph {\bibinfo {title} {{GNU}
  Scientific Library Reference Manual}}},\ \bibinfo {edition} {3rd}\ ed.\
  (\bibinfo  {publisher} {Network Theory Ltd., Surrey, UK},\ \bibinfo {year}
  {2009})\BibitemShut {NoStop}%
\bibitem [{\citenamefont {Beder}\ \emph {et~al.}(2009)\citenamefont {Beder},
  \citenamefont {Woehlke}, \citenamefont {Breitbart}, \citenamefont {Wolchok},
  \citenamefont {Hackimov}, \citenamefont {Snape}, \citenamefont {Hamlet},
  \citenamefont {Novotny}, \citenamefont {Tambre}, \citenamefont {Reinhold},
  \citenamefont {Vaucher}, \citenamefont {Zaitsev}, \citenamefont {Anokhin},
  \citenamefont {Karatarakis}, \citenamefont {Maloney}, \citenamefont
  {Polukhin}, \citenamefont {Brandenburg}, \citenamefont {Ibanez},
  \citenamefont {Gladkikh}, \citenamefont {Eich}, \citenamefont {Dumont},
  \citenamefont {Trigg}, \citenamefont {King}, \citenamefont {Frederico},
  \citenamefont {Hamilton}, \citenamefont {Langley}, \citenamefont
  {Hämäläinen}, \citenamefont {Blair}, \citenamefont {Duggan}, \citenamefont
  {Wang}, \citenamefont {Stotko}, \citenamefont {Levine}, \citenamefont {Bena},
  \citenamefont {Cordoba}, \citenamefont {Schmidt}, \citenamefont {Gottlieb},
  \citenamefont {Prilmeier}, \citenamefont {Zhang}, \citenamefont {Ishi},
  \citenamefont {Lyngmo}, \citenamefont {Mataré}, \citenamefont {Konečný},\
  and\ \citenamefont {USDOE}}]{yamlcpp}%
  \BibitemOpen
  \bibfield  {author} {\bibinfo {author} {\bibfnamefont {J.}~\bibnamefont
  {Beder}}, \bibinfo {author} {\bibfnamefont {M.}~\bibnamefont {Woehlke}},
  \bibinfo {author} {\bibfnamefont {J.}~\bibnamefont {Breitbart}}, \bibinfo
  {author} {\bibfnamefont {S.}~\bibnamefont {Wolchok}}, \bibinfo {author}
  {\bibfnamefont {A.~H.}\ \bibnamefont {Hackimov}}, \bibinfo {author}
  {\bibfnamefont {J.}~\bibnamefont {Snape}}, \bibinfo {author} {\bibfnamefont
  {O.}~\bibnamefont {Hamlet}}, \bibinfo {author} {\bibfnamefont
  {P.}~\bibnamefont {Novotny}}, \bibinfo {author} {\bibfnamefont
  {R.}~\bibnamefont {Tambre}}, \bibinfo {author} {\bibfnamefont
  {S.}~\bibnamefont {Reinhold}}, \bibinfo {author} {\bibfnamefont
  {A.}~\bibnamefont {Vaucher}}, \bibinfo {author} {\bibfnamefont
  {A.}~\bibnamefont {Zaitsev}}, \bibinfo {author} {\bibfnamefont
  {A.}~\bibnamefont {Anokhin}}, \bibinfo {author} {\bibfnamefont
  {A.}~\bibnamefont {Karatarakis}}, \bibinfo {author} {\bibfnamefont
  {A.}~\bibnamefont {Maloney}}, \bibinfo {author} {\bibfnamefont
  {A.}~\bibnamefont {Polukhin}}, \bibinfo {author} {\bibfnamefont {C.~M.}\
  \bibnamefont {Brandenburg}}, \bibinfo {author} {\bibfnamefont
  {D.}~\bibnamefont {Ibanez}}, \bibinfo {author} {\bibfnamefont
  {D.}~\bibnamefont {Gladkikh}}, \bibinfo {author} {\bibfnamefont
  {F.}~\bibnamefont {Eich}}, \bibinfo {author} {\bibfnamefont {G.}~\bibnamefont
  {Dumont}}, \bibinfo {author} {\bibfnamefont {H.}~\bibnamefont {Trigg}},
  \bibinfo {author} {\bibfnamefont {J.}~\bibnamefont {King}}, \bibinfo {author}
  {\bibfnamefont {J.}~\bibnamefont {Frederico}}, \bibinfo {author}
  {\bibfnamefont {J.}~\bibnamefont {Hamilton}}, \bibinfo {author}
  {\bibfnamefont {J.}~\bibnamefont {Langley}}, \bibinfo {author} {\bibfnamefont
  {L.}~\bibnamefont {Hämäläinen}}, \bibinfo {author} {\bibfnamefont
  {M.}~\bibnamefont {Blair}}, \bibinfo {author} {\bibfnamefont {M.~W.}\
  \bibnamefont {Duggan}}, \bibinfo {author} {\bibfnamefont {O.}~\bibnamefont
  {Wang}}, \bibinfo {author} {\bibfnamefont {P.}~\bibnamefont {Stotko}},
  \bibinfo {author} {\bibfnamefont {P.}~\bibnamefont {Levine}}, \bibinfo
  {author} {\bibfnamefont {P.}~\bibnamefont {Bena}}, \bibinfo {author}
  {\bibfnamefont {R.~H.}\ \bibnamefont {Cordoba}}, \bibinfo {author}
  {\bibfnamefont {R.}~\bibnamefont {Schmidt}}, \bibinfo {author} {\bibfnamefont
  {S.~G.}\ \bibnamefont {Gottlieb}}, \bibinfo {author} {\bibfnamefont
  {F.}~\bibnamefont {Prilmeier}}, \bibinfo {author} {\bibfnamefont
  {T.}~\bibnamefont {Zhang}}, \bibinfo {author} {\bibfnamefont
  {T.}~\bibnamefont {Ishi}}, \bibinfo {author} {\bibfnamefont {T.}~\bibnamefont
  {Lyngmo}}, \bibinfo {author} {\bibfnamefont {V.}~\bibnamefont {Mataré}},
  \bibinfo {author} {\bibfnamefont {M.}~\bibnamefont {Konečný}},\ and\
  \bibinfo {author} {\bibnamefont {USDOE}},\ }\href
  {https://doi.org/10.11578/dc.20220817.13} {\bibinfo {title} {yaml-cpp}}
  (\bibinfo {year} {2009})\BibitemShut {NoStop}%
\bibitem [{\citenamefont {Jakob}\ \emph {et~al.}(2017)\citenamefont {Jakob},
  \citenamefont {Rhinelander},\ and\ \citenamefont {Moldovan}}]{pybind11}%
  \BibitemOpen
  \bibfield  {author} {\bibinfo {author} {\bibfnamefont {W.}~\bibnamefont
  {Jakob}}, \bibinfo {author} {\bibfnamefont {J.}~\bibnamefont {Rhinelander}},\
  and\ \bibinfo {author} {\bibfnamefont {D.}~\bibnamefont {Moldovan}},\
  }\href@noop {} {\bibinfo {title} {pybind11 -- seamless operability between
  c++11 and python}} (\bibinfo {year} {2017}),\ \bibinfo {note}
  {https://github.com/pybind/pybind11}\BibitemShut {NoStop}%
\bibitem [{\citenamefont {{Reinecke}}\ and\ \citenamefont
  {{Seljebotn}}(2013)}]{libsharp}%
  \BibitemOpen
  \bibfield  {author} {\bibinfo {author} {\bibfnamefont {M.}~\bibnamefont
  {{Reinecke}}}\ and\ \bibinfo {author} {\bibfnamefont {D.~S.}\ \bibnamefont
  {{Seljebotn}}},\ }\bibfield  {title} {\bibinfo {title} {{Libsharp - spherical
  harmonic transforms revisited}},\ }\href
  {https://doi.org/10.1051/0004-6361/201321494} {\bibfield  {journal} {\bibinfo
   {journal} {Astronomy and Astrophysics}\ }\textbf {\bibinfo {volume} {554}},\
  \bibinfo {eid} {A112} (\bibinfo {year} {2013})},\ \Eprint
  {https://arxiv.org/abs/1303.4945} {arXiv:1303.4945 [physics.comp-ph]}
  \BibitemShut {NoStop}%
\bibitem [{\citenamefont {Heinecke}\ \emph {et~al.}(2016)\citenamefont
  {Heinecke}, \citenamefont {Henry}, \citenamefont {Hutchinson},\ and\
  \citenamefont {Pabst}}]{libxsmm}%
  \BibitemOpen
  \bibfield  {author} {\bibinfo {author} {\bibfnamefont {A.}~\bibnamefont
  {Heinecke}}, \bibinfo {author} {\bibfnamefont {G.}~\bibnamefont {Henry}},
  \bibinfo {author} {\bibfnamefont {M.}~\bibnamefont {Hutchinson}},\ and\
  \bibinfo {author} {\bibfnamefont {H.}~\bibnamefont {Pabst}},\ }\bibfield
  {title} {\bibinfo {title} {{LIBXSMM}: {Accelerating} small matrix
  multiplications by runtime code generation},\ }in\ \href@noop {} {\emph
  {\bibinfo {booktitle} {Proceedings of the International Conference for High
  Performance Computing, Networking, Storage and Analysis}}},\ \bibinfo {series
  and number} {SC '16}\ (\bibinfo  {publisher} {IEEE Press},\ \bibinfo {year}
  {2016})\ pp.\ \bibinfo {pages} {1--11}\BibitemShut {NoStop}%
\bibitem [{\citenamefont {Hunter}(2007)}]{Hunter:2007}%
  \BibitemOpen
  \bibfield  {author} {\bibinfo {author} {\bibfnamefont {J.~D.}\ \bibnamefont
  {Hunter}},\ }\bibfield  {title} {\bibinfo {title} {Matplotlib: A 2d graphics
  environment},\ }\href {https://doi.org/10.1109/MCSE.2007.55} {\bibfield
  {journal} {\bibinfo  {journal} {Computing in Science \& Engineering}\
  }\textbf {\bibinfo {volume} {9}},\ \bibinfo {pages} {90} (\bibinfo {year}
  {2007})}\BibitemShut {NoStop}%
\bibitem [{\citenamefont {Caswell}\ \emph {et~al.}(2020)\citenamefont
  {Caswell}, \citenamefont {Droettboom}, \citenamefont {Lee}, \citenamefont
  {Hunter}, \citenamefont {Firing}, \citenamefont {de~Andrade}, \citenamefont
  {Hoffmann}, \citenamefont {Stansby}, \citenamefont {Klymak}, \citenamefont
  {Varoquaux}, \citenamefont {Nielsen}, \citenamefont {Root}, \citenamefont
  {May}, \citenamefont {Elson}, \citenamefont {Dale}, \citenamefont {Lee},
  \citenamefont {Seppänen}, \citenamefont {McDougall}, \citenamefont {Straw},
  \citenamefont {Hobson}, \citenamefont {Gohlke}, \citenamefont {Yu},
  \citenamefont {Ma}, \citenamefont {Vincent}, \citenamefont {Silvester},
  \citenamefont {Moad}, \citenamefont {Kniazev}, \citenamefont {hannah},
  \citenamefont {Ernest},\ and\ \citenamefont
  {Ivanov}}]{thomas_a_caswell_2020_3948793}%
  \BibitemOpen
  \bibfield  {author} {\bibinfo {author} {\bibfnamefont {T.~A.}\ \bibnamefont
  {Caswell}}, \bibinfo {author} {\bibfnamefont {M.}~\bibnamefont {Droettboom}},
  \bibinfo {author} {\bibfnamefont {A.}~\bibnamefont {Lee}}, \bibinfo {author}
  {\bibfnamefont {J.}~\bibnamefont {Hunter}}, \bibinfo {author} {\bibfnamefont
  {E.}~\bibnamefont {Firing}}, \bibinfo {author} {\bibfnamefont {E.~S.}\
  \bibnamefont {de~Andrade}}, \bibinfo {author} {\bibfnamefont
  {T.}~\bibnamefont {Hoffmann}}, \bibinfo {author} {\bibfnamefont
  {D.}~\bibnamefont {Stansby}}, \bibinfo {author} {\bibfnamefont
  {J.}~\bibnamefont {Klymak}}, \bibinfo {author} {\bibfnamefont
  {N.}~\bibnamefont {Varoquaux}}, \bibinfo {author} {\bibfnamefont {J.~H.}\
  \bibnamefont {Nielsen}}, \bibinfo {author} {\bibfnamefont {B.}~\bibnamefont
  {Root}}, \bibinfo {author} {\bibfnamefont {R.}~\bibnamefont {May}}, \bibinfo
  {author} {\bibfnamefont {P.}~\bibnamefont {Elson}}, \bibinfo {author}
  {\bibfnamefont {D.}~\bibnamefont {Dale}}, \bibinfo {author} {\bibfnamefont
  {J.-J.}\ \bibnamefont {Lee}}, \bibinfo {author} {\bibfnamefont {J.~K.}\
  \bibnamefont {Seppänen}}, \bibinfo {author} {\bibfnamefont {D.}~\bibnamefont
  {McDougall}}, \bibinfo {author} {\bibfnamefont {A.}~\bibnamefont {Straw}},
  \bibinfo {author} {\bibfnamefont {P.}~\bibnamefont {Hobson}}, \bibinfo
  {author} {\bibfnamefont {C.}~\bibnamefont {Gohlke}}, \bibinfo {author}
  {\bibfnamefont {T.~S.}\ \bibnamefont {Yu}}, \bibinfo {author} {\bibfnamefont
  {E.}~\bibnamefont {Ma}}, \bibinfo {author} {\bibfnamefont {A.~F.}\
  \bibnamefont {Vincent}}, \bibinfo {author} {\bibfnamefont {S.}~\bibnamefont
  {Silvester}}, \bibinfo {author} {\bibfnamefont {C.}~\bibnamefont {Moad}},
  \bibinfo {author} {\bibfnamefont {N.}~\bibnamefont {Kniazev}}, \bibinfo
  {author} {\bibnamefont {hannah}}, \bibinfo {author} {\bibfnamefont
  {E.}~\bibnamefont {Ernest}},\ and\ \bibinfo {author} {\bibfnamefont
  {P.}~\bibnamefont {Ivanov}},\ }\href {https://doi.org/10.5281/zenodo.3948793}
  {\bibinfo {title} {matplotlib/matplotlib: {REL:} v3.3.0}} (\bibinfo {year}
  {2020})\BibitemShut {NoStop}%
\bibitem [{\citenamefont {Harris}\ \emph {et~al.}(2020)\citenamefont {Harris},
  \citenamefont {Millman}, \citenamefont {van~der Walt}, \citenamefont
  {Gommers}, \citenamefont {Virtanen}, \citenamefont {Cournapeau},
  \citenamefont {Wieser}, \citenamefont {Taylor}, \citenamefont {Berg},
  \citenamefont {Smith}, \citenamefont {Kern}, \citenamefont {Picus},
  \citenamefont {Hoyer}, \citenamefont {van Kerkwijk}, \citenamefont {Brett},
  \citenamefont {Haldane}, \citenamefont {del R{\'{i}}o}, \citenamefont
  {Wiebe}, \citenamefont {Peterson}, \citenamefont {G{\'{e}}rard-Marchant},
  \citenamefont {Sheppard}, \citenamefont {Reddy}, \citenamefont {Weckesser},
  \citenamefont {Abbasi}, \citenamefont {Gohlke},\ and\ \citenamefont
  {Oliphant}}]{harris2020array}%
  \BibitemOpen
  \bibfield  {author} {\bibinfo {author} {\bibfnamefont {C.~R.}\ \bibnamefont
  {Harris}}, \bibinfo {author} {\bibfnamefont {K.~J.}\ \bibnamefont {Millman}},
  \bibinfo {author} {\bibfnamefont {S.~J.}\ \bibnamefont {van~der Walt}},
  \bibinfo {author} {\bibfnamefont {R.}~\bibnamefont {Gommers}}, \bibinfo
  {author} {\bibfnamefont {P.}~\bibnamefont {Virtanen}}, \bibinfo {author}
  {\bibfnamefont {D.}~\bibnamefont {Cournapeau}}, \bibinfo {author}
  {\bibfnamefont {E.}~\bibnamefont {Wieser}}, \bibinfo {author} {\bibfnamefont
  {J.}~\bibnamefont {Taylor}}, \bibinfo {author} {\bibfnamefont
  {S.}~\bibnamefont {Berg}}, \bibinfo {author} {\bibfnamefont {N.~J.}\
  \bibnamefont {Smith}}, \bibinfo {author} {\bibfnamefont {R.}~\bibnamefont
  {Kern}}, \bibinfo {author} {\bibfnamefont {M.}~\bibnamefont {Picus}},
  \bibinfo {author} {\bibfnamefont {S.}~\bibnamefont {Hoyer}}, \bibinfo
  {author} {\bibfnamefont {M.~H.}\ \bibnamefont {van Kerkwijk}}, \bibinfo
  {author} {\bibfnamefont {M.}~\bibnamefont {Brett}}, \bibinfo {author}
  {\bibfnamefont {A.}~\bibnamefont {Haldane}}, \bibinfo {author} {\bibfnamefont
  {J.~F.}\ \bibnamefont {del R{\'{i}}o}}, \bibinfo {author} {\bibfnamefont
  {M.}~\bibnamefont {Wiebe}}, \bibinfo {author} {\bibfnamefont
  {P.}~\bibnamefont {Peterson}}, \bibinfo {author} {\bibfnamefont
  {P.}~\bibnamefont {G{\'{e}}rard-Marchant}}, \bibinfo {author} {\bibfnamefont
  {K.}~\bibnamefont {Sheppard}}, \bibinfo {author} {\bibfnamefont
  {T.}~\bibnamefont {Reddy}}, \bibinfo {author} {\bibfnamefont
  {W.}~\bibnamefont {Weckesser}}, \bibinfo {author} {\bibfnamefont
  {H.}~\bibnamefont {Abbasi}}, \bibinfo {author} {\bibfnamefont
  {C.}~\bibnamefont {Gohlke}},\ and\ \bibinfo {author} {\bibfnamefont {T.~E.}\
  \bibnamefont {Oliphant}},\ }\bibfield  {title} {\bibinfo {title} {Array
  programming with {NumPy}},\ }\href
  {https://doi.org/10.1038/s41586-020-2649-2} {\bibfield  {journal} {\bibinfo
  {journal} {Nature}\ }\textbf {\bibinfo {volume} {585}},\ \bibinfo {pages}
  {357} (\bibinfo {year} {2020})}\BibitemShut {NoStop}%
\bibitem [{\citenamefont {Ayachit}(2015)}]{paraview}%
  \BibitemOpen
  \bibfield  {author} {\bibinfo {author} {\bibfnamefont {U.}~\bibnamefont
  {Ayachit}},\ }\href@noop {} {\emph {\bibinfo {title} {The ParaView Guide: A
  Parallel Visualization Application}}}\ (\bibinfo  {publisher} {Kitware,
  Inc.},\ \bibinfo {address} {Clifton Park, NY, USA},\ \bibinfo {year}
  {2015})\BibitemShut {NoStop}%
\bibitem [{\citenamefont {Ahrens}\ \emph {et~al.}(2005)\citenamefont {Ahrens},
  \citenamefont {Geveci},\ and\ \citenamefont {Law}}]{paraview2}%
  \BibitemOpen
  \bibfield  {author} {\bibinfo {author} {\bibfnamefont {J.}~\bibnamefont
  {Ahrens}}, \bibinfo {author} {\bibfnamefont {B.}~\bibnamefont {Geveci}},\
  and\ \bibinfo {author} {\bibfnamefont {C.}~\bibnamefont {Law}},\ }\href@noop
  {} {\emph {\bibinfo {title} {{ParaView}: {An} end-user tool for large-data
  visualization}}}\ (\bibinfo  {publisher} {Elsevier},\ \bibinfo {year}
  {2005})\BibitemShut {NoStop}%
\bibitem [{\citenamefont {Poisson}\ and\ \citenamefont
  {Will}(2014)}]{Poisson:2014}%
  \BibitemOpen
  \bibfield  {author} {\bibinfo {author} {\bibfnamefont {E.}~\bibnamefont
  {Poisson}}\ and\ \bibinfo {author} {\bibfnamefont {C.~M.}\ \bibnamefont
  {Will}},\ }\href {https://doi.org/10.1017/CBO9781139507486} {\emph {\bibinfo
  {title} {Gravity: Newtonian, Post-Newtonian, Relativistic}}}\ (\bibinfo
  {publisher} {Cambridge University Press},\ \bibinfo {year}
  {2014})\BibitemShut {NoStop}%
\bibitem [{\citenamefont {Thorne}(1980)}]{Thorne:1980}%
  \BibitemOpen
  \bibfield  {author} {\bibinfo {author} {\bibfnamefont {K.~S.}\ \bibnamefont
  {Thorne}},\ }\bibfield  {title} {\bibinfo {title} {Multipole expansions of
  gravitational radiation},\ }\href {https://doi.org/10.1103/RevModPhys.52.299}
  {\bibfield  {journal} {\bibinfo  {journal} {Rev. Mod. Phys.}\ }\textbf
  {\bibinfo {volume} {52}},\ \bibinfo {pages} {299} (\bibinfo {year}
  {1980})}\BibitemShut {NoStop}%
\bibitem [{\citenamefont {{Blanchet}}\ and\ \citenamefont
  {{Damour}}(1986)}]{Blanchet:1986}%
  \BibitemOpen
  \bibfield  {author} {\bibinfo {author} {\bibfnamefont {L.}~\bibnamefont
  {{Blanchet}}}\ and\ \bibinfo {author} {\bibfnamefont {T.}~\bibnamefont
  {{Damour}}},\ }\bibfield  {title} {\bibinfo {title} {{Radiative gravitational
  fields in general relativity. I - General structure of the field outside the
  source}},\ }\href {https://doi.org/10.1098/rsta.1986.0125} {\bibfield
  {journal} {\bibinfo  {journal} {Philosophical Transactions of the Royal
  Society of London Series A}\ }\textbf {\bibinfo {volume} {320}},\ \bibinfo
  {pages} {379} (\bibinfo {year} {1986})}\BibitemShut {NoStop}%
\end{thebibliography}%

\end{document}